\documentclass[10pt]{iopart}

\usepackage{graphicx}

\usepackage{hyperref,bm,bbm}
\hypersetup{colorlinks,allcolors=blue}
\usepackage[normalem]{ulem}

\renewcommand{\lsim}{{\;\raise0.3ex\hbox{$<$\kern-0.75em\raise-1.1ex\hbox{$\sim$}}\;}}
\newcommand{\gsim}{{\;\raise0.3ex\hbox{$>$\kern-0.75em\raise-1.1ex\hbox{$\sim$}}\;}}

\begin{document}

\topical[Polymer gels]{Swelling thermodynamics and phase transitions of polymer gels}

\author{Michael S Dimitriyev$^1$, Ya-Wen Chang$^2$, Paul M Goldbart$^3$ and Alberto Fern\'andez-Nieves$^{1,4,5}$}

\address{$^1$ School of Physics, Georgia Institute of Technology,
Atlanta, Georgia 30332}
\address{$^2$ Department of Chemical Engineering, Texas Tech University, Lubbock, Texas 79409}
\address{$^3$ Department of Physics, The University of Texas, Austin, Texas 78712}
\address{$^4$ Department of Condensed Matter Physics, University of Barcelona, 08028 Barcelona, Spain}
\address{$^5$ ICREA-Instituci\'o Catalana de Recerca i Estudis Avan\c{c}ats, 08010 Barcelona, Spain}

\ead{\mailto{alberto.fernandez@physics.gatech.edu}, \mailto{michael.dimitriyev@physics.gatech.edu}}
\vspace{10pt}
\begin{indented}
\item[] \today
\end{indented}

\begin{abstract}
We present a pedagogical review of the swelling thermodynamics and phase transitions of polymer gels.
In particular, we discuss how features of the \emph{volume phase transition} of the gel's osmotic equilibrium is analogous to other transitions described by mean-field models of binary mixtures, and the failure of this analogy at the critical point due to shear rigidity.
We then consider the \emph{phase transition at fixed volume}, a relatively unexplored paradigm for polymer gels that results in a phase-separated equilibrium consisting of coexisting solvent-rich and solvent-poor regions of gel.
Again, the gel's shear rigidity is found to have a profound effect on the phase transition, here resulting in macroscopic shape change at constant volume of the sample, exemplified by the tunable buckling of toroidal samples of polymer gel.
By drawing analogies with extreme mechanics, where large shape changes are achieved via \emph{mechanical} instabilities, we formulate the notion of \emph{extreme thermodynamics}, where large shape changes are achieved via \emph{thermodynamic} instabilities, i.e.~phase transitions.
\end{abstract}

%
%
\submitto{\NT}
%
\maketitle
%
\ioptwocol

\section{Introduction} 


Within the realm of amorphous, rigid materials without crystalline symmetries, polymer gels possess an interesting duality, having a rubber-like elasticity whilst being able to undergo \emph{large} volume changes due to mixing with a solvent.
These materials are soft, being composed of large macromolecules whose interactions are often governed by thermal fluctuations.
There are three essential ingredients:~polymers, solvent, and cross-links.

\begin{figure}
\includegraphics[width=8.3cm]{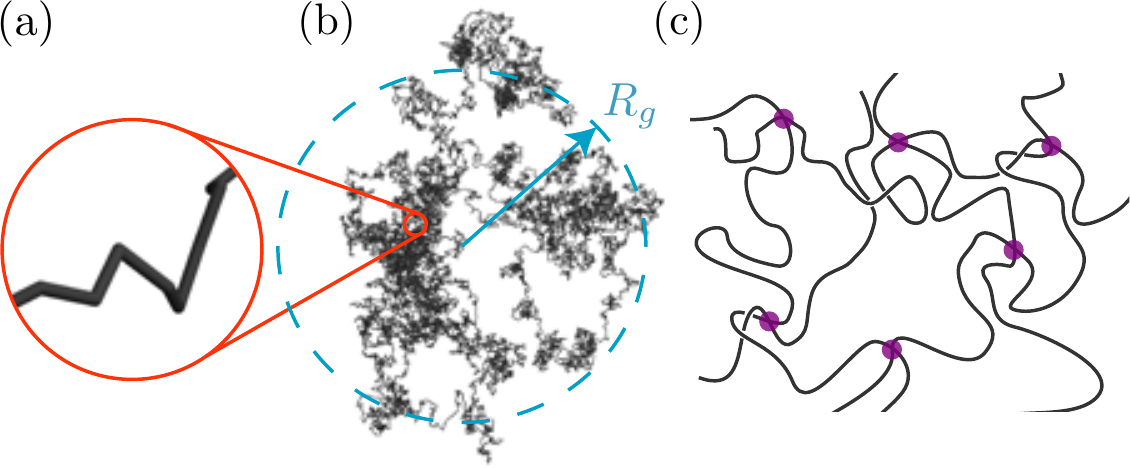}
\caption{\label{fig:polymer} (a) A sequence of monomer units joined with randomly sampled bond angles. (b) Example of an ``ideal chain'' consisting of $\mathcal{N}=10^4$ monomers that form a static random walk conformation with radius of gyration $R_g$. (c) Pairs of polymers (gray curves) are permanently joined by molecular cross-links (purple dots) to form a polymer network.}
\end{figure}

Polymers are composed of many short segments (i.e., monomers) that are typically chemically bonded end-to-end, as shown in figure \ref{fig:polymer}(a).
Whilst there are energetically favorable bond angles between successive monomers, there often are several monomer-monomer bond conformations that are at comparable energies \cite{deGennes1979}.
As the number $\mathcal{N}$ of monomers that constitute a polymer is typically on the order of $10^{3}-10^{4}$, there are many different mutually accessible polymer conformations, within a small energy window, that a polymer may be found in.
Thus, polymers are said to have \emph{static flexibility} \cite{Brereton1976}.
Furthermore, there are modest energy barriers between bond angles, enabling frequent transitions that are driven by thermal fluctuations, causing the polymer to explore many different conformations over time.
As such, polymers are also said to have \emph{dynamic flexibility}.

Quite generally, as a consequence of such static and dynamic flexibility, correlations between monomer-monomer bond angles decay with distance along the backbone of the polymer.
Beyond a certain $\emph{persistence length}$, bond angles are barely correlated.
Therefore, conformations of polymers that span many persistence lengths have the form of a ``random walk,'' an example of which is shown in figure \ref{fig:polymer}(b).
The radius of gyration $R_g$ specifies the characteristic size of the polymer, as also illustrated in figure \ref{fig:polymer}(b); for large $\mathcal{N}$, $R_g$ scales with the number of monomers $\mathcal{N}$ as $\mathcal{N}^\nu$, where $\nu = 1/2$ for a random-walk polymer, or ``ideal chain,'' for which the excluded volume interaction between polymer segments is neglected.
More realistically, an isolated polymer does not intersect with itself, resulting in statistics of a self-avoiding (as opposed to ideal) random walk, for which $\nu \approx 3/5$ in three dimensions, reflecting the swelling of the polymer sequence.
This is indeed the situation when the polymer is immersed in a ``good'' solvent, one in which the polymer is miscible, as opposed to the case of immiscibility in a ``poor'' solvent, where the polymer radius is decreased and a compact structure is formed due to the high energetic penalty for the mixing of the polymer and solvent.
An intermediate case is the $\vartheta$-solvent, where the radius-decrease due to a mildly poor solvent counteracts the radius-increase due to the self-repulsion of the polymer, resulting in ideal-chain scaling, for which $\nu = 1/2$ (see, e.g., \cite{deGennes1979}).

Now consider a solution of many such polymers.
Focusing on the case where all polymers are composed of roughly the same number $\mathcal{N}$ monomers that are chemically identical, the solution can be brought to a polymer concentration for which individual polymer coils overlap spatially in equilibrium.
In this case, application of a static stress induces steady state flow, since after a short-time elastic response, where the polymer coils deform, they are able to re-arrange in space continually.
Rigidity results from the introduction of cross-links between neighboring polymers, since cross-linked molecules can no longer individually undergo substantial re-arrangement relative to cross-linked partners.
If there is sufficient linking of different polymers then a container-spanning, percolating, network of linked polymer coils forms; this constitutes a gel \cite{deGennes1979}.
In a gel, the cross-linked clusters of polymers are thus \emph{localized} in space relative to one another, unable to explore the volume of their container via Brownian motion; such ergodicity breaking of the polymers due to the formation of a percolating polymer network is a hallmark of the onset of rigidity \cite{Goldbart1996}, at least in spaces of dimension $d > 2$.

Despite the large variety of cross-links that can be formed, they are generally classified according to two categories: physical and chemical \footnote{Note that we do not consider here polymer rings that may be \emph{topologically} linked to form rigid ``Olympic'' gels \cite{deGennes1979,Raphael1997} or other knotted polymer networks \cite{MacArthur1995,Horner2016}.}.
Examples of physical cross-links include entangled polymers as well as non-covalent bonds, such as ionic (i.e., electrostatic) bonds.
Physical cross-links enable an elastic response over potentially extended timescales; however, they are not truly rigid, as they allow flow at sufficiently long times due to the reversible nature of the cross-linking process.
Instead, we shall focus our attention on gels formed from \emph{chemical cross-links}, which are typically induced by the introduction of small--when compared to the typical polymer size--cross-linking molecules that form covalent bonds between polymers, as depicted in figure \ref{fig:polymer}(c).
Chemical cross-links may be regarded as permanent so that the network topology of the gel is frozen in after cross-linking, much like the cross-links in rubber, resulting in a thermodynamically rigid material \cite{Deam1976}.
Moreover, unlike amorphous elastic materials, such as glasses, these gels are in a well-defined, equilibrium solid phase; unlike conventional solids, gels lack long-range order.
Thus, gels are \emph{equilibrium amorphous solids} \cite{Goldbart1996,Panyukov1996}.

However, unlike rubber, polymer gels are typically cross-linked in the presence of a solvent, which permeates the polymer network of the gel.
The magnitude of the osmotic pressure $\Pi$ due to the mixing of polymer with the solvent is set by the thermal energy scale $k_B T$, and is thus on the same scale as the entropic elastic stresses of the polymer network.
As a result, both are important in determining the \emph{macroscopic} equilibrium state of the gel.
This is especially true for gels having low cross-link densities, which have the signature ability to undergo large macroscopic volume changes in response to varying solvent conditions.
In the presence of a \emph{good solvent}, the polymer network is well mixed with solvent molecules, and the gel incorporates a large volume of solvent and is said to be \emph{swollen}; in the presence of a \emph{poor solvent}, the polymer network is essentially segregated from the solvent molecules and is said to be \emph{deswollen}.
Suitably prepared gels have a remarkably large volume response, capable of swelling to an equilibrium volume on the order of $10^3$ times their deswollen volume by absorbing solvent \cite{Matsuo1988}.
Whilst a variety of different swollen volumes can be achieved by continuous changes in solubility (as induced, e.g., via temperature), certain gels can exhibit a \emph{discontinuous} change in volume.
For example, poly(N-isopropylacrylamide) (pNIPAM) gels in water gradually deswell under \emph{heating} until $\sim\mkern-4mu 32^{\circ}{\rm C}$.
Beyond this, due to a change in solvent nature from good to poor, they abruptly expel most of their solvent \cite{Matsuo1988}.
In fact, there is a \emph{first-order phase transition} between well-defined swollen and deswollen phases of gel.
This phase transition is confirmed by varying the osmotic pressure, leading to a phase diagram having a first-order transition region separating the two phases, terminating at a critical point \cite{Hirotsu1988,Shibayama1993}.

In addition to swelling, polymer gels can undergo shape changes.
Mechanical constraints, such as attachment to a stiff substrate, can frustrate \emph{homogeneous} deswelling, resulting in inhomogeneous deswelling of the gel, which can lead to the formation of surface ripples \cite{Tanaka1987mechanical,Kang2010}.
Gels that are subject to inhomogeneous swelling are of particular interest, as the resulting deformations typically cannot be realized in flat space \cite{Efrati2013}, resulting in a variety of buckled shapes \cite{Pezzulla2015,Holmes2011}, some of which mirror patterns found in nature \cite{Trujillo2008,Hohlfeld2011,Breid2011}.
This has led to origami-inspired \cite{Zhang2011} and biology-inspired \cite{Gladman2016} searches for ways to program certain shapes that are actuated upon swelling.

In this Topical Review, we first discuss in Section \ref{sec:swelling_thermo} the thermodynamic description of polymer gel swelling.
We give a brief outline of the statistical mechanical treatment due to Flory and Rehner \cite{FloryRehner1943_1,FloryRehner1943_2} and mention some subtleties that arise in describing the rubber-like elasticity of the gel \cite{Deam1976}.
We then present a pedagogical review of phase transitions in Section~\ref{sec:phase_transitions}, to develop an intuition for the volume phase transition and the critical behavior of gels in analogy with the Van der Waals theory of the liquid-vapor phase transition.
Continuing with this analogy, we consider in Section~\ref{sec:arrested_deswelling} how a transition to \emph{phase coexistence} between swollen and deswollen phases can be achieved by arresting the deswelling transition of a swollen gel.
In Section~\ref{sec:shape_change}, we show how the equilibrium phase-coexistent gel is characterized by a large deformation of the macroscopic gel shape that is distinct from the usual volume phase transition.
Drawing on analogies with the \emph{extreme mechanics} of shape-changing materials through programmed mechanical instability \cite{Reis2015,Holmes2019}, we propose that phase-coexistent gels provide a route to large deformation via \emph{thermodynamic instability}.
To provide an illustrative example of this \emph{extreme thermodynamics}, we give a detailed description of the phase-coexistent equilibrium of gel \emph{toroids} and the accompanying shape-buckling transition, which has been realized in experiments.
Finally, in Section~\ref{sec:conclusion}, we summarize some of the open problems in the field and highlight some of the gaps in the understanding of polymer gels that have yet to be filled.

\section{\label{sec:swelling_thermo} Swelling thermodynamics} 

The thermodynamic description of polymer gels is based on that of non-ideal fluids, in which interactions between particles give rise to equations of state, such as the Van der Waals equation, that differ from the universal, ideal gas description.
To begin, we consider the state functions that are required to describe the macroscopic state of the gel.
Then we turn to the microscopic description and outline the Flory-Rehner \cite{FloryRehner1943_1,FloryRehner1943_2} mean-field theory which yields approximate equations of state of the gel that are analogous to the Van der Waals equation of state.
Next, just as the Van der Waals equation predicts that a fluid expands with increasing temperature at constant pressure, we show how the Flory-Rehner equation of state predicts gel deswelling with increasing temperature.
Finally, we examine the breakdown of thermodynamic stability predicted by the Flory-Rehner theory, highlighting analogies and differences with phase transitions in fluids.

\subsection{State functions and thermodynamic potentials}


Macroscopic materials, such as polymer gels, are composed of a vast number of microscopic degrees of freedom that are in continual flux, i.e., are thermally fluctuating.
In a thermodynamic description of such materials, these many fluctuating microscopic degrees of freedom are averaged over time and space to yield \emph{state functions} (see, e.g., \cite{Callen1985}).
For example, in a fluid consisting of $N$ identical particles occupying a fixed volume $V$, both $V$ and $N$ are state functions. 
Whilst the spacings between particles are not fixed, there are, on average, $\rho \equiv N/V$ particles per unit volume.
If the fluid is isolated then its total energy $E$ is fixed, as are the total number $N$ of particles in the fluid as well as its total volume $V$.
These three state functions are sufficient for characterizing the macroscopic equilibrium state of the system.
In order to quantify what happens when the macroscopic degrees of freedom $(E,V,N)$ are changed, one employs a thermodynamic potential.
The entropy $S(E,V,N)$ is one example of a thermodynamic potential, which has the fundamental property that if the energy, volume, or number constraints are relaxed, the equilibrium state that the system eventually attains corresponds to one of maximum entropy.
Note that entropy is also a state function, corresponding to the number of microstates of the fluid that give rise to a fixed macrostate $(E,V,N)$.
At times, it is useful to use the entropy as a state-characterizing function, exchanging it with the total energy $E$, which can then take the role of the thermodynamic potential, corresponding to a macroscopic description $(S,V,N)$.

It is often convenient to consider interactions between the fluid and a much larger ``bath,'' whose state is not affected by the presence of the fluid.
If we imagine that the fluid is kept in a container that allows heat to flow between the fluid and the surrounding bath then the energy of the fluid and that of the bath are allowed to change.
The total entropy $S = S_{\rm fluid} + S_{\rm bath}$ is maximized when the temperature $T$ of the fluid matches that of the bath, which characterizes a state of \emph{thermal equilibrium}.
This container can either maintain a fixed volume $V$ of the fluid or be flexible, in which case \emph{mechanical equilibrium} is reached when the pressure $P$ of particles in the fluid is balanced by a similar pressure from the bath.
Similarly, the container can either be impermeable, maintaining a constant number $N$ of particles, or it can be permeable, so that \emph{chemical equilibrium} is reached when the chemical potential $\mu$ of the fluid matches that of the surrounding bath.
In this way, the paired state functions $(S,T)$, $(V,P)$, and $(N,\mu)$ are considered conjugate to one another.
Much like the density $\rho$ of the fluid, $(T,P,\mu)$ are intensive state functions that characterize material properties of the fluid, whereas $(S,V,N)$ are extensive state functions.
Whilst there is freedom in choosing the three state functions that describe the macroscopic state of the fluid,\footnote{Recall, however, that due to the Gibbs-Duhem equation, which provides a link between the intensive parameters, at least one of the state functions must be extensive.} let us consider the temperature $T$ and the number of particles $N$ as specified properties, i.e., constraints imposed on the fluid.
There are two representations that can be considered: the Gibbs representation $(T,P,N)$ and the Helmholtz representation $(T,V,N)$, with corresponding thermodynamic potentials $G$, the Gibbs free energy, and $F$, the Helmholtz free energy.
Changes in constraints lead to changes in the thermodynamic potential, described by a Gibbs equation for each representation, namely
\numparts\begin{eqnarray}
{\rm d}G &=& -S\,{\rm d}T + V\,{\rm d}P + \mu\,{\rm d}N \, ,\\
{\rm d}F &=& -S\,{\rm d}T - P\,{\rm d}V + \mu\,{\rm d}N \, ,
\end{eqnarray}\endnumparts
from which we see that the two potentials are related via a Legendre transform, resulting in the relation $G = F + PV$.

Now consider a sample of gel that is allowed to exchange solvent with its surroundings but contains a constant number of monomers (i.e., polymer segments).
The gel is composed of $n_s$ solvent molecules, $n_m$ monomers, and $n_c$ cross-linking molecules.
Thus, it is natural to assume that in the Gibbs representation the state of the gel is characterized by the state functions $(T,P,n_s,n_m,n_c)$.
However, we will assume that each molecule and monomer occupy volumes $v_s$ and $v_m$, respectively, so that the volume $V$ of the system is approximately given by
\begin{equation} \label{eq:volume_relation}
V \approx n_s v_s + n_m v_m\, ,
\end{equation}
i.e., we have neglected the very small contribution due to the cross-linking molecules since the number of cross-links is typically orders of magnitude smaller than the number of solvent molecules and monomers.
Therefore, changing the pressure $P$ acts to change the volume per solvent molecule ($v_s$) and the volume per monomer ($v_m$).
This, however, only happens at very high pressures and is not of significance in the situations of interest here.
We will instead focus on the effect that mixing these two chemical species has on the macroscopic properties of polymer gels, and treat $v_s$ and $v_m$ as constants; for simplicity, we assume that they have the same value, namely $v_s \approx v_m \equiv v$.
Furthermore, the number of monomers $n_m$ and the number of cross-links $n_c$ are imposed at the formation of the gel, and are also assumed constant.
Thus, we are left with the state functions $(T,n_s)$, where the volume $V$ is determined as a function of $n_s$ via equation (\ref{eq:volume_relation}); this state characterization then amounts to a Helmholtz representation of the gel.

\begin{figure}
	\includegraphics[width=8.3cm]{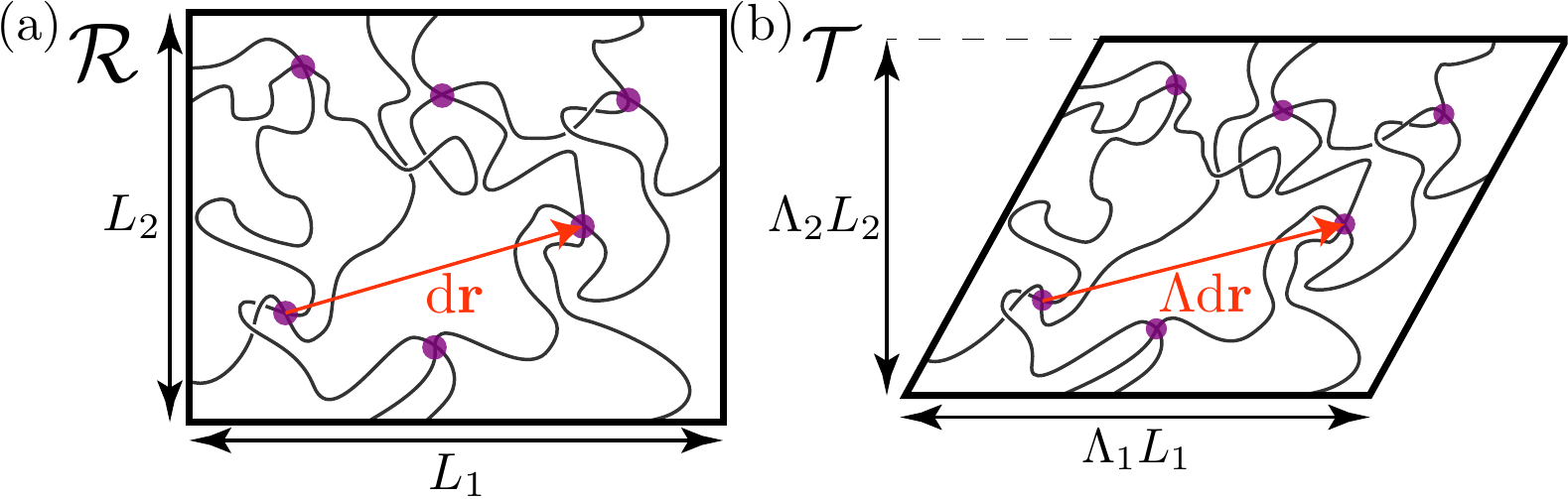}
	\caption{Representative sample of a cross-linked polymer network with linear dimensions $L_1$ and $L_2$ and distance between two arbitrary cross-links given by ${\rm d}\mathbf{r}$ (a) prior to deformation, in reference configuration $\mathcal{R}$, and (b) after affine deformation, in target configuration $\mathcal{T}$, prescribed by deformation matrix $\Lambda$.}
	\label{fig:gel_deformation}
\end{figure}

Unlike fluids, however, polymer gels possess a nonzero rigidity with respect to elastic deformations.
Therefore, in addition to occupying a volume $V$, the gel is able to maintain a deformed shape indefinitely when subjected to stress.
We therefore require additional state functions to account for this fact.
One such deformation is the change in the three side-lengths $\{L_1,L_2,L_3\}$ of the box-shaped sample of gel shown in figure \ref{fig:gel_deformation} to lengths $L_i' = \Lambda_i L_i$.
The deformation of the gel at constant volume is therefore set by the dimensionless ratios of length $\{\Lambda_1,\Lambda_2,\Lambda_3\}$ such that $\Lambda_1\Lambda_2\Lambda_3 = 1$.
Note that specifying the side-lengths of a parallelepiped region of gel is but one example deformation that can be achieved.
For general gel shapes, it is more appropriate to examine how the distance between any two points $\mathbf{r}$ and $\mathbf{r} + {\rm d}\mathbf{r}$ is altered upon deformation of the gel, which takes ${\rm d}\mathbf{r}$ to ${\rm d}\mathbf{R}$.
For example, it is useful to imagine $\mathbf{r}$ and $\mathbf{r}+{\rm d}\mathbf{r}$ as two neighboring cross-links.
Assuming \emph{affine} deformations, for which the changes in lengths between representative points in the gel are independent of position, all lengths are transformed by a deformation matrix $\Lambda$, such that ${\rm d} R_i = \Lambda_{ij} {\rm d} r_j$, where we use and adopt Einstein's summation convention over repeated indices.
Deformed volume elements ${\rm d}^3 R$ are related to the undeformed ones via ${\rm d}^3 R =  ({\rm det}\, \Lambda) {\rm d}^3 r$, so ${\rm det}\, \Lambda$ is the ratio of the deformed volume to the undeformed volume.
Thus, deformations that maintain the gel volume are characterized by ${\rm det}\, \Lambda = 1$.
Alternatively, we are free to choose a reference state, which we shall refer to as a \emph{reference configuration} $\mathcal{R}$, where the volume of the gel is given by $V_0$, such that after a deformation of the gel, the volume of the deformed state, which we shall refer to as a \emph{target configuration} $\mathcal{T}$, is given by 
\begin{equation}\label{eq:volume-deformation_relation}
V = V_0\,({\rm det}\, \Lambda) \; .
\end{equation}
Therefore, the determinant $({\rm det}\, \Lambda)$ may be expressed in terms of the amount of solvent $n_s$ in configuration $\mathcal{T}$ via equations (\ref{eq:volume_relation}) and (\ref{eq:volume-deformation_relation}).
As cross-links undergo Brownian motion, some care has to be taken in relating macroscopic affine deformation to a corresponding microscopic deformation.
However, it has been found \cite{Panyukov1996} that \emph{average} cross-link positions indeed undergo affine deformation.

In order to account for the effect of deformation on the equilibrium thermodynamics of the gel, it is necessary to introduce the deformation matrix $\Lambda$ as a state function.
However, by equation \ref{eq:volume-deformation_relation}, the deformation matrix determines the volume of the gel. 
To account for this redundancy in state functions, we may express the Helmholtz free energy as
\begin{equation}\label{eq:free energy-with-LM}
F(T,\Lambda, n_s; \lambda) = F(T,\Lambda,n_s) + \lambda\left[V_0\,{\rm det}\,\Lambda - V(n_s)\right] \, ,
\end{equation}
where $\lambda$ a Lagrange multiplier accounting for the constraint associated to equation (\ref{eq:volume-deformation_relation}).
In this form, the Lagrange multiplier $\lambda$ is an additional state function and the constraint is an equation of state. 

It is useful to define the \emph{polymer volume fraction} $\phi$ via
\begin{equation}
\phi \equiv \frac{n_m v_m}{n_sv_s + n_mv_m} \approx  1 - \frac{v_s n_s}{V}\, , 
\end{equation}
i.e., the fraction of the gel volume that is occupied by polymer; $\phi = 1$ corresponds to a gel that is completely devoid of solvent, whereas $\phi=0$ is the limit of an infinitely dilute gel.
Noting that since $F$ is a homogeneous first-order function in its extensive parameters \cite{Callen1985}, we can define $F(T,\Lambda,n_s) = V\mathcal{F}(T,\Lambda,n_s/V)$, where $\mathcal{F}$ is a free energy density.
Therefore, in terms of the polymer volume fraction $\phi$, we have
\begin{equation}\label{eq:free energy_density}
F = \frac{v n_m}{\phi}\mathcal{F}(T,\Lambda,\phi) \, ,
\end{equation}
where we have used the assumption $v_s \approx v_m \equiv v$.

Inasmuch as $P$ and $V$ are conjugate to each other for a fluid, for a gel there is a state function that is paired with the polymer volume fraction $\phi$.
This is the \emph{osmotic pressure} $\Pi$.
If the gel is in equilibrium with a solvent bath, the chemical potential of the solvent in the gel, $\mu(T,P)$, equals the chemical potential of the solvent in the bath, $\mu_0(T,P)$, plus a contribution $\Delta \mu$ accounting for the presence of the polymer network.
Then $\mu(T,P) = \mu_0(T,P) + \Delta \mu$.
In addition, we may regard the boundary of the polymer network as a semipermeable membrane.
Equilibrium then requires an additional pressure in order to maintain the imbalance in solvent concentration in and out of the gel, $\mu(T,P + \Pi) = \mu_0(T,P)$; this additional pressure $\Pi$ is, by definition, the osmotic pressure.
It can be shown (see, e.g., \cite{LandauLifshitzSP1}) that the osmotic pressure $\Pi$ is related to $\Delta \mu$ via $\Pi = -\Delta\mu/v$, where $v$ is the solvent particle volume.

The thermodynamics of the polymer gel is determined by how polymer mixes with solvent.
Just as $\Delta \mu$ is the change in the chemical potential of the solvent due to the presence of the polymer network, we can decompose the total free energy $F$ as
\begin{equation}
F = F_{\rm sol} + \Delta F \, ,
\end{equation}
where $F_{\rm sol}$ is the part of the free energy due to solvent, without the effect of the polymer network.
Therefore, $\Delta \mu = (\partial \Delta F/\partial n_s)_{T,\Lambda}$.
Then, using the relation between $n_s$ and the volume fraction $\phi$ as well as equation (\ref{eq:free energy_density}), the osmotic pressure $\Pi$ is given by
\begin{equation}
\Pi(T,\phi,\Lambda) = -\left(\frac{\partial (\phi^{-1}\Delta\mathcal{F})}{\partial(\phi^{-1})}\right)_{T,\Lambda} \, ,
\end{equation}
where it is evident that if $\Pi$ is analogous to a pressure then $1/\phi$ plays the role of volume.
Note that, since the determinant of the deformation matrix $\Lambda$ depends on the polymer volume fraction via the volume-deformation relation (\ref{eq:volume-deformation_relation}), it is important to enforce the Lagrange multiplier constraint in equation (\ref{eq:free energy-with-LM}) when computing the osmotic pressure.
As $\Delta F$ is the part of the free energy that describes gel deformation, such as swelling, we shall refer to it as the deformation free energy.

\subsection{Flory-Rehner equation of state}

Just as we enumerated a set of macroscopic descriptors of the gel, let us consider some microscopic ones.
The gel is a mixture of $n_s$ solvent molecules, $n_m$ monomers, and $n_c$ cross-links.
Whilst the solvent molecules may have multiple internal degrees of freedom, e.g., rotational and vibrational, let us focus only on the center-of-mass degrees of freedom and treat them as point particles at positions $\{\bm{\sigma}_i\}$, each occupying a volume $v$, where $i$ runs from 1 to $n_s$.
Rather than treating each of the $n_m$ monomers as individual particles, we will group them into polymers.
For simplicity, assume that (i) we can ignore any ``free-ends'' or ``loops'' of polymers in the polymer network and consider only segments whose endpoints are cross-linked to other segments, and (ii) each of these segments, which we shall refer to as ``chains,'' are composed of $\mathcal{N}$ monomers.
Much like the simple representation of solvent molecules, we opt for a simple representation of chains as one dimensional curves $\{\mathbf{R}_j(s)\}$, where $s$ is the arclength parameter, running from 0 to chain length $L_{ch} \approx v^{1/3}\mathcal{N}$, and $j$ runs from 1 to the number of chains $n_{ch} \equiv 2n_c \approx n_m/\mathcal{N}$.

In order to link these microscopic degrees of freedom to the macroscopic properties of the system, one approach is to fix temperature $T [\equiv 1/(k_B\beta)]$ and the number of particles of each species in the system, and determine the canonical partition function $Z$, given by
\begin{equation}\label{eq:partition_function}
Z = \int[{\rm d}^3\bm{\sigma}_i]_{i=1}^{n_s}\int [\mathcal{D}\mathbf{R}_j(s)]_{j=1}^{n_{ch}}e^{-\beta E}\prod_{k=1}^{3n_c}\delta(\mathbf{f}^{\rm network}_k),
\end{equation}
where $E$ is the total potential energy of the system.
The network topology is set by a collection of constraints on the $2n_{ch} = 4n_c$ ends of the chains $\{\mathbf{R}_i(L_{ch}),\mathbf{R}_i(0)\}$.
At each cross-link there are 4 ends that coincide; we may choose these ends such that 2 are at $s = 0$ and 2 are at $s = L_{ch}$.
However for each cross-link there are only three independent constraints; the fourth is automatically satisfied.
For example, if a cross-link consists of the chains ends $\{\mathbf{R}_1(L_{ch}),\mathbf{R}_2(L_{ch}),\mathbf{R}_3(0),\mathbf{R}_4(0)\}$ then enforcing the constraints $\mathbf{R}_1(L_{ch}) = \mathbf{R}_3(0)$, $\mathbf{R}_1(L_{ch}) = \mathbf{R}_4(0)$, and $\mathbf{R}_2(L_{ch}) = \mathbf{R}_3(0)$ automatically implies that the fourth constraint $\mathbf{R}_2(L_{ch}) = \mathbf{R}_4(0)$ is satisfied.
Therefore, there are $3n_{c}$ vectors that are constrained, yielding exactly $n_c$ independent vectors, describing the positions of cross-links in space.
Thus, for each of the $n_c$ cross-links, there are 3 constraint equations that can be written as
\begin{equation}
\mathbf{f}^{\rm network}_k \equiv \sum^{n_{ch}}_{i\neq j}a_k^{ij}[\mathbf{R}_i(L_{ch}) - \mathbf{R}_j(0)] = \mathbf{0} \, ,
\end{equation}
for $k = 1\dots 3n_c$, where $a_k^{ij}$ is an adjacency matrix that is $1$ when the two polymer ends are joined by a cross-link and is $0$ otherwise.
These constraints are enforced by including a product of Dirac delta functions $\Pi_k \delta(\mathbf{f}^{\rm network}_k)$ in the integrand of the partition function, ensuring that the only contributions to the sum over states are those where $\mathbf{f}^{\rm network}_k = 0$ for all $k$.
Note that these topological constraints pose a considerable technical difficulty in the evaluation of the partition function $Z$ and the free energy $F = -k_BT\,{\rm ln}\, Z$, due to the lack of a periodic structure.
A mesoscopic representation of such a network with  ``quenched disorder'' is shown in figure \ref{fig:polymer}(c).
However, for sufficiently large gels, there are many different mesoscopic network structures.
Thus, instead of summing over polymer configurations with a certain fixed network topology, one can instead sample from a distribution of mesoscopic network structures, with the idea that they all appear somewhere in the gel; this is known as \emph{self-averaging}.
The partition function is then evaluated via the \emph{replica trick}, where many copies or ``replicas'' of the gel are treated as new interacting degrees of freedom to be integrated over \cite{Deam1976,Goldbart1987,Goldbart1996,Panyukov1996}.

However, instead of seeking a direct evaluation of the partition function $Z$ in equation (\ref{eq:partition_function}), we consider the classical construction of Flory and Rehner \cite{FloryRehner1943_1,FloryRehner1943_2}, which amounts to a mean-field approximation.
In particular, we seek a description of the deformation free energy $\Delta F = \Delta E - T\Delta S$, where $\Delta E$ is the change in the energy and $\Delta S$ is the change in the entropy due to the mixing of the solvent and polymer.
The total potential energy $E$ describes the microscopic interaction energy and is approximated by the sum of three contributions: $U_{m-m}$ is the energy of monomer-monomer interactions, $U_{s-s}$ is the energy of solvent-solvent interactions, and $U_{m-s}$ is the energy of monomer-solvent interactions (see \cite{deGennes1979}), with each of these terms depending on particle positions.
For example, $U_{m-m}$ contains an excluded-volume interaction $(v_m/2)\int{\rm d}s_i\int{\rm d}s_j\delta(\mathbf{R}_i(s_i) - \mathbf{R}_j(s_j))$ between chains $i$ and $j$ (including self-interactions, corresponding to the case $i = j$).
In the mean-field approximation, the interaction energy depends only on local densities of solvent and monomer, resulting in a simple form for the mean energy density $\overline{\mathcal{E}}$, namely
\begin{equation}
\overline{\mathcal{E}} \approx \frac{v k_B T}{2}\left[\chi_{m-m}\,\rho_m^2 + 2 \chi_{m-s}\, \rho_m\, \rho_s + \chi_{s-s}\, \rho_s^2\right]\, , 
\end{equation}
where $\rho_m [\equiv n_m/V]$ and $\rho_s [\equiv n_s/V]$ are number-densities of monomer and solvent and $\{\chi_{m-m}$, $\chi_{m-s}$, $\chi_{s-s}\}$ are the various interaction strengths, relative to $k_B T$, associated to Van der Waals, excluded volume,  and hydrophobic interactions \cite{deGennes1979}.
Re-writing in terms of $\phi$, the monomer density is $\rho_m = \phi/v$ and the solvent density is $\rho_s = (1-\phi)/v$ so that
\begin{equation} \eqalign{
\overline{\mathcal{E}} \approx \frac{k_B T}{2v}\big[&\chi_{m-m}\, \phi^2 \\
& + 2 \chi_{m-s}\, \phi(1 - \phi) + \chi_{s-s}\, (1 - \phi)^2\big] \, .
}\end{equation}
Each particle, independent of identity, shares the same mean energy density; the total energy $\overline{E}$ is therefore $V\overline{\mathcal{E}}$.
The quantity of interest is the change in energy $\Delta E$ due to mixing, which is $\overline{E} - \overline{E}_{s} - \overline{E}_{m}$, where $\overline{E}_s = n_s v\overline{\mathcal{E}}(\phi = 0)$ is the the total energy of a fictitious system having the same number $n_s$ of solvent molecules but without any monomers, so that $\phi = 0$; similarly, $\overline{E}_m = n_m v\overline{\mathcal{E}}(\phi = 1)$ is a system of monomers alone.
Introducing the total number $N = n_s + n_m$ of solvent molecules and monomers, the mixing energy $\Delta E$ is simply
\begin{equation}
\Delta E = N k_B T \chi \phi(1 - \phi)\, ,
\end{equation}
where $\chi [\equiv \chi_{m-s} - \chi_{m-m}/2 - \chi_{s-s}/2]$ is the so-called Flory parameter \cite{deGennes1979}.
If $\chi < 0$ the interaction energy is minimized when $\phi = 1/2$, corresponding to equal parts of solvent and polymer.
Since this case occurs when $\chi_{m-s} < (\chi_{m-m} + \chi_{s-s})/2$, it describes a regime in which the energetic cost of monomer-solvent interactions is less the average cost of pure monomer-monomer and pure solvent-solvent interactions.

Deformations of the polymer network generally result in changes in the contact interactions between chains.
At the mean-field level, these interactions are incorporated in the mixing energy $\Delta E$ through $\phi$ alone.
For anisotropic deformations at fixed $\phi$, however, we will assume that interactions between polymers are somewhat less important and approximate the polymer network via a \emph{phantom chain model} where the chain conformations are allowed to overlap one-another, leading to random-walk ``ideal'' polymers.
We thus focus on  single-chain deformations, implicitly assuming that this is the main contribution to the entropic cost of stretching the polymer network $\Delta S_{\rm net}$.
Although some degree of realism is lost, the problem gains tractability whilst retaining the essential physics -- the free energy cost of elastic deformations is simple to derive and has a form that reduces to the classical rubber elasticity model (see e.g., \cite{Treloar1975}) in the unswollen limit.
With this assumption, the elastic free energy of the gel is approximated as proportional to the \emph{net} conformational entropy change due to deforming $n_{ch}$ \emph{independent} polymer chains.
Thus, we require knowledge of (i) how deformations affect the conformational entropy of a single chain and (ii) how to determine the effect on an \emph{ensemble} of many such chains.
It should be noted, however, that this construction is limited in scope and fails to accurately capture the elastic free energy in the large shear-strain regime, where correlations between polymer fluctuations gain importance \cite{Xing2007}.

\begin{figure}
	\centering
	\includegraphics[width=8.3cm]{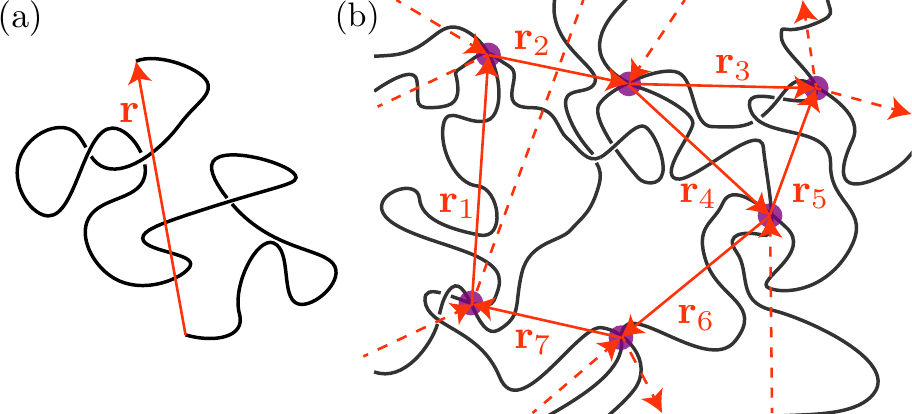}
	\caption{(a) Example of a single ideal chain with end-to-end vector $\mathbf{r}$. (b) Collection of chains in the polymer network with end-to-end vectors $\{\mathbf{r}_i\}$.} 
	\label{fig:network_vectors}
\end{figure}

To begin, consider a single chain of length $L$ with one terminal end at position $\mathbf{r}$ and the other at the origin, as shown in figure \ref{fig:network_vectors}(a).
In order to determine the entropy of the chain, first note that within the phantom chain model, all random paths of fixed length $L$ have the same energy.
Therefore, the entropy $S_1(\mathbf{r})$ of a single chain with end-to-end vector $\mathbf{r}$ is given by $S_1(\mathbf{r}) = k_B{\rm ln}\,\Omega_1(\mathbf{r})$, where $\Omega_1(\mathbf{r})$ is the total number of microstates available to the chain.
We can model chain conformations by considering a lattice model in which each lattice site has length $a$ and each component $r_i$ of the vector $\mathbf{r}$ can be expressed as $r_i/a = 2n_i - N_i$, where $n_i$ is the displacement along the lattice along one axis and $N_i$ is the total number of steps taken along that axis.
The number of microstates $\omega(r_i)$ for this one dimensional random walk is given by $N_i!/(n_i!(N - n_i)!)$.
Then $\Omega_1(\mathbf{r})$ is simply the product $(V/v)\omega(r_1)\omega(r_2)\omega(r_3)$, where $V/v$ is the number of possible locations for $\mathbf{r}=0$, and is given by
\begin{equation}
\Omega_1(\mathbf{r}) = \frac{V}{v}\prod_{i=1}^3\frac{N_i!}{\left(\frac{N_i a + r_i}{2a}\right)!\left(\frac{N_i a - r_i}{2a}\right)!} \, ,
\end{equation}
which can be simplified in the limit of large $N_i$ by taking advantage of Stirling's approximation, yielding
\begin{equation}\eqalign{
{\rm ln}\,\Omega_1 &\approx {\rm ln}\,\frac{V}{v} + \sum_i \bigg[N_i{\rm ln}\, N_i - \frac{N_i a + r_i}{2 a}{\rm ln}\, \frac{N_i a + r_i}{2 a}\\
&\mkern+124mu - \frac{N_i a - r_i}{2 a}{\rm ln}\, \frac{N_i a - r_i}{2 a} \bigg] \, .
}\end{equation}
Assuming that the polymer explores three-dimensional space isotropically, $N_1 = N_2 = N_3 = N/3$, where $N$ is the total number of monomer units.
Furthermore, we note that the polymer is most likely to be found with end-to-end distance $|\mathbf{r}|$ to be much smaller than its extreme maximum length $Na$.
Expanding ${\rm ln}\,\Omega_1$ in powers of $|\mathbf{r}|/(Na)$, the leading order contribution is
\begin{equation}
{\rm ln}\,\Omega_1 \approx {\rm ln}\,\frac{V}{v} + N{\rm ln}\,2 - \frac{3 |\mathbf{r}|^2}{2 R_0^2} \, ,
\end{equation}
where $R_0 = \sqrt{N}a$ \cite{deGennes1979}.
The entropy $S_1(\mathbf{r})$ for a single chain is therefore given by
\begin{equation}
S_1(\mathbf{r}) = S_0 - \frac{3k_B|\mathbf{r}|^2}{2 R_0^2} + k_B {\rm ln}\frac{V}{v}
\end{equation}
where $S_0$ is a constant that depends on the number of monomers in each chain.
Now consider a process that stretches the chain, resulting in a new end-to-end vector $\mathbf{r'}$, and hence a new entropy $S'(\mathbf{r'})$.
Writing the displaced vector as $r^{'i} = \Lambda_{ij}r^j$, where $\Lambda_{ij}$ is a deformation matrix, the change in entropy $\Delta S_1(\mathbf{r})$ for a single chain is given by
\begin{equation}\label{eq:entropy1}
\Delta S_1(\mathbf{r}) = -\frac{3k_B}{2 R_0^2}\left(\Lambda_{ki}\Lambda_{kj} - \delta_{ij}\right)r^ir^j + k_B {\rm ln}(V'/V)\, .
\end{equation}

Next, consider a collection of $n_{ch}$ chains that are cross-linked to form a polymer network.
The macrostate of these chains, prior to deformation, is specified by the collection of end-to-end vectors $\{\mathbf{r_1},\mathbf{r_2},\dots,\mathbf{r_{n_{ch}}}\}$, corresponding to vectors between cross-links as shown in figure \ref{fig:network_vectors}(b).
If this network undergoes \emph{affine deformation}, as illustrated in figure \ref{fig:gel_deformation}, then all end-to-end vectors are transformed by the same deformation matrix $\Lambda$.
Therefore, the change in entropy for this collection of chains, $\Delta S_{n_{ch}}$, is simply given by a sum over independent entropy contributions (\ref{eq:entropy1}), namely
\begin{equation}\eqalign{\label{eq:entropy2}
\Delta S_{n_{ch}} &= -\frac{3k_B}{2 R_0^2}\left(\Lambda_{ki}\Lambda_{kj} - \delta_{ij}\right)\big(r_1^ir_1^j +  \cdots \\
&\mkern+206mu \cdots + r_{n_{ch}}^ir_{n_{ch}}^j\big) \\
&\mkern+32mu +n_{ch}k_B{\rm ln}\, {\rm det}\, \Lambda \, ,
}\end{equation}
where we have assumed that all chains have the same length and thus the same value of $R_0$ and used the relation $V = V'({\rm det}\,\Lambda)$ for affine deformations.
This is the change in entropy for a given network, represented by the collection of end-to-end vectors.
Since we restrict our attention to chemical gels, the network topology is set upon cross-linking, and the same collection of end-to-end vectors describes the polymer network for all processes.
While each network has a distinct topology, we assume that (i) the polymer networks are sufficiently large that the same collection of end-to-end vectors is represented throughout every network, albeit in a possibly different arrangement (by the self-averaging property), and (ii) cross-linking occurs when a collection of polymers in solution are brought to a concentration where they overlap but not to the point where they would deform due to steric repulsion.
Then the change in entropy $\Delta S_{\rm net}$ representative of a polymer network composed of $n_{ch}$ chains of fixed $R_0$ is found by averaging equation (\ref{eq:entropy2}) over the equilibrium values of $\mathbf{r_1}\dots\mathbf{r_{n_{ch}}}$ \emph{before deformation}.
In order to do this, note that the probability $P_1(\mathbf{r})$ that a single polymer will have an end-to-end vector $\mathbf{r}$ is given by $P_1(\mathbf{r}) = \Omega_1(\mathbf{r})/Z_1$, where $Z_1 = \int {\rm d}^3r \Omega_1(\mathbf{r})$.
Note that in the phantom chain model, the energy is then equal to zero.
Using the result that
\begin{equation}
\left<r_i r_j\right> = \int {\rm d}^3 r\, P_1(\mathbf{r})r_i r_j = \frac{R_0^2}{3}\delta_{ij}\, ,
\end{equation}
the change in entropy $\Delta S_{\rm net} \equiv \left<\Delta S_{n_{ch}}(\mathbf{r_1},\dots,\mathbf{r_{n_{ch}}})\right>$ due to deforming a polymer network is given by
\begin{equation}\label{eq:entropy_net}
\Delta S_{\rm net} = -\frac{1}{2}n_{ch} k_B\left[{\rm tr}\,\Lambda^T\Lambda - 3 - 2\,{\rm ln}\, {\rm det}\, \Lambda\right] \, .
\end{equation}

However, we still have not arrived at our final result for $\Delta S_{\rm net}$.
Since chemically cross-linked chains share a common endpoint, some of the chain degrees of freedom must be eliminated \cite{FloryRehner1943_2,Flory1953}.
This will cause $\Delta S_{\rm net}$, as expressed in equation (\ref{eq:entropy_net}), to decrease.
To estimate this reduction, note that within the phantom chain assumption, the endpoint $\mathbf{r}$ of a chain is free to lie within any point in the volume $V$ of the gel, irrespective of the location of where the polymer is based.
After deformation, $V \rightarrow ({\rm det}\, \Lambda) V$ so the change in entropy due to the change in the volume of the gel that is accessible to the endpoint is given by $n_{ch} k_B {\rm ln}\,{\rm det}\,\Lambda$; this exactly cancels the last term in Eq.~(\ref{eq:entropy_net}).
However, each cross-link between two chains, say chain $i$ and chain $i+1$, constrains the endpoint motion of the chains; as a result, there is a constraint function $f_i(\mathbf{r}_{i},\mathbf{r}_{i+1}) = 0$ for these two chains.
Since there are $n_c = n_{ch}/2$ such constraints, there is an additional \emph{reduction} of the total entropy by $(n_{ch}k_B /2) {\rm ln}\,{\rm det}\,\Lambda$. 
Thus, the overall entropy change due to deformations of the polymer network is
\begin{equation} \label{eq:FW_entropy}
\Delta S_{\rm net} = -\frac{1}{2}n_{ch} k_B \left[{\rm tr}\,\Lambda^T\Lambda - 3 - {\rm ln} \, {\rm det}\, \Lambda\right] \; ,
\end{equation}
which is attributed to Flory and Wall \cite{Treloar1975}.
Note $\Delta S_{\rm net} \propto n_{ch}$ and is independent of chain length.

We emphasize, however, that the argument for reduction in entropy due to cross-linking, as presented above, is somewhat flawed.
In the seminal paper of Deam and Edwards \cite{Deam1976}, it was shown that this argument relies on the assumption that the cross-linked ends of the chains are free to explore the entire volume $V$ of the gel.
However, the cross-linked ends of the chains, whilst able to undergo thermal motion, are localized to a much smaller volume $\omega$ when the gel is formed.
Furthermore, whereas the volume $V$ of the gel depends on the affine deformation $\Lambda$, the volume $\omega$ of the localization is a much weaker function of $\Lambda$ owing to non-affine fluctuations of the cross-linked endpoints.
Therefore, the ${\rm ln} \, {\rm det}\, \Lambda$ term in the Flory-Wall entropy (\ref{eq:FW_entropy}) is not completely justified.
In addition, since the term can be re-written as ${\rm ln} (\phi_0/\phi)$, it only depends on the polymer volume fraction.
Since the mixing of solvent and polymer result in a similar contribution to the total entropy, it is nevertheless difficult to assess the validity of the inclusion of this term in the Flory-Wall entropy.

\begin{figure}
	\includegraphics[width=8.3cm]{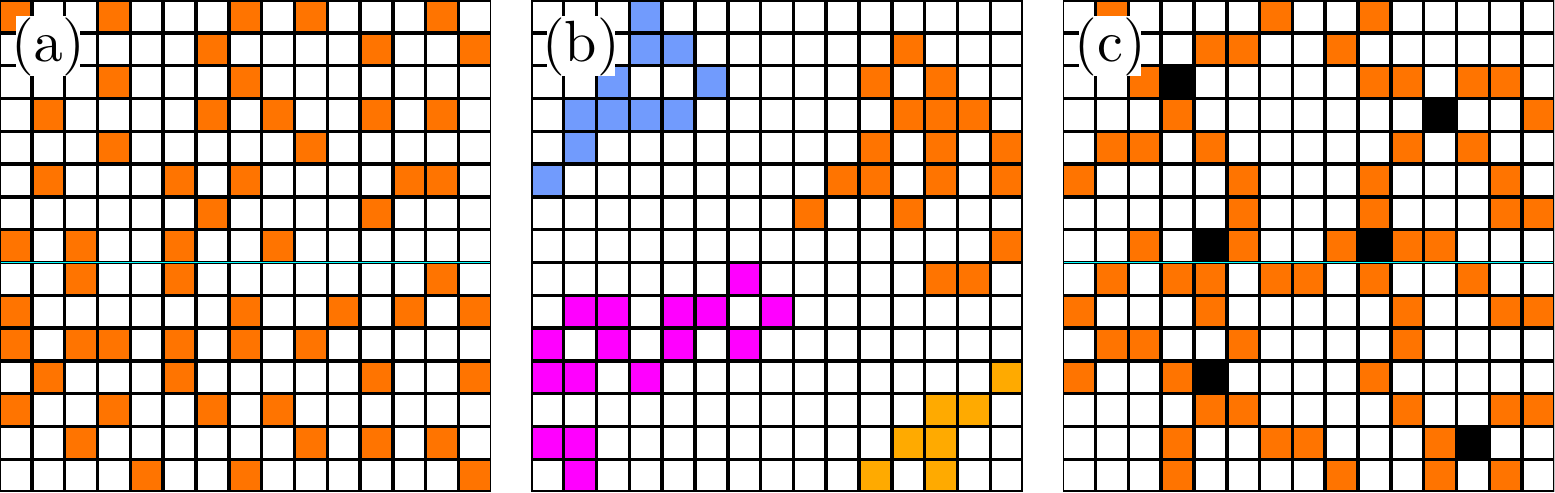}
	\caption{Lattice calculation of the mixing entropy. Each cell occupies a volume $v$. Solvent is represented by white cells. (a) Bragg-Williams case in which the monomer units (orange) are uncorrelated. (b) Flory-Huggins case where monomers are identified with mobile polymers of degree $\mathcal{N}$. (c) Flory-Rehner case where monomer translational freedom is frozen; cross-links between independent polymers are shown in black.} 
	\label{fig:flory_huggins}
\end{figure}

There is an additional contribution to the entropy coming from the mixing of solvent and the polymer network: $\Delta S_{\rm mix}$.
To estimate $\Delta S_{\rm mix}$, we model the space occupied by the gel by a lattice, as shown in figure \ref{fig:flory_huggins}, of $N$ sites, each occupied by either a solvent molecule or a monomer; because the system is densely filled, there are $n_s$ solvent molecules and $n_m = N - n_s$ monomers.
By fixing the total number of monomer and solvent molecules, the entropy $S_{\rm latt}$ of arranging monomers and solvent into the lattice is given by $S_{\rm latt} = k_B {\rm ln}\, \Omega$, where $\Omega$ is the number of possible lattice arrangements.
We must therefore count the number of ways that the lattice can be filled with solvent and monomers, where the monomers (i) are arranged into polymers that (ii) belong to a cross-linked network that spans space.
We will start with the simple case of ``free'' monomers that are unassociated into larger polymer molecules, all able to explore space independently and recover the entropy of Bragg-Williams theory \cite{ChaikinLubensky}.
Subsequently, we will progressively introduce the necessary constraints by associating the monomers into polymers and then introducing the cross-linking constraints.

Consider a binary system, consisting of $n_A$ and $n_B$ particles of species `A' and `B', respectively; `A' could, for example, represent solvent and `B' could represent free monomers, as shown in figure \ref{fig:flory_huggins}(a).
The total number of microstates of the lattice is given by
\begin{equation*}
\Omega = \frac{N!}{n_A!n_B!} \, ,
\end{equation*}
so that the entropy, using Stirling's approximation is
\begin{equation*}
S_{\rm latt} \approx k_B\,\left[N\, {\rm ln}\, N - n_A\, {\rm ln}\, n_A - n_B\, {\rm ln}\, n_B\right] \,.
\end{equation*}
Recalling that the volume fraction $\phi = n_B/N$,
\begin{equation*}
S_{\rm latt} \approx - N k_B\, \left[(1 - \phi)\, {\rm ln}\, (1 - \phi) + \phi\, {\rm ln}\, \phi\right]\, ,
\end{equation*}
from which we find that the state of maximum entropy corresponds to $\phi = 1/2$, which further corresponds to a mixed state composed equally of both species of particles.

We now associate monomers into polymer that are free to explore the entire space, whilst localizing individual monomers to much smaller volumes around the centers of mass of the polymers, as illustrated in figure \ref{fig:flory_huggins}(b).
Let each polymer consist of $\mathcal{N}$ monomers such that $n_{p} = n_m/\mathcal{N}$ is the total number of polymers.
Whilst the individual monomers have the adjacency condition, polymers are allowed full translational freedom on the lattice.
Thus, the entropy of the localized monomers is negligible compared with the translational entropy of the polymers.
The entropy $S_{\rm latt}$ is therefore \emph{dominated} by the translational entropy of the solvent and the polymers such that
\begin{equation*}
S_{\rm latt} \approx - N k_B \left[(1 - \phi)\, {\rm ln}\, (1 - \phi) + \frac{\phi}{\mathcal{N}}\, {\rm ln}\, \frac{\phi}{\mathcal{N}}\right] \, .
\end{equation*}
To obtain the mixing entropy $\Delta S_{{\rm mix}}$, first define an entropy density $\mathcal{S} = S/V$.
Following the definition of the mixing energy $\Delta E$, the mixing entropy is given by
\begin{equation}\eqalign{
\Delta S_{{\rm mix}} &= V \mathcal{S}(\phi) - vn_s \mathcal{S}(\phi = 0) - vn_m \mathcal{S}(\phi = 1) \\
&\approx -Nk_B\left[(1 - \phi)\, {\rm ln}\, (1 - \phi) + \frac{\phi}{\mathcal{N}}\, {\rm ln}\, \phi \right] \; , 
}\label{eq:mixing_entropy_FH}\end{equation}
which corresponds to the Flory-Huggins result \cite{Flory1942,deGennes1979} for polymer solutions.

Finally, we consider the case in which permanent cross-links are introduced, localizing polymers to small regions about the cross-link sites [see figure \ref{fig:flory_huggins}(c)].
In this case, the polymers have constraints that reach all the way to the sample boundary, resulting in rigidity.
Therefore, the translational entropy of polymers is negligible compared with the entropy of the solvent.
The result may be found by considering the limit of the Flory-Huggins theory for infinitely long polymer, i.e., taking $\mathcal{N}\rightarrow \infty$.
The result is
\begin{equation} \label{eq:mixing_entropy_FR}
\Delta S_{{\rm mix}} \approx -N k_B\,(1 - \phi)\,{\rm ln}\,(1 - \phi) \, ,
\end{equation}
which is independent of network details \cite{FloryRehner1943_2,deGennes1979,Flory1953}.

The deformation free energy $\Delta F$ can finally be decomposed as
\begin{equation}\label{eq:gel_fe}
\Delta F = \Delta F_{\rm elastic} + \Delta F_{\rm mix} + \lambda\left[{\rm det}\,\Lambda - \frac{\phi_0}{\phi}\right]\, ,
\end{equation}
where the elastic deformation free energy 
\begin{equation}\eqalign{
\Delta F_{\rm elastic} &= -T\Delta S_{\rm net} \\
&= \frac{1}{2}n_{ch} k_BT\left[{\rm tr}\,\Lambda^T\Lambda - 3 - {\rm ln} \, {\rm det}\, \Lambda\right]\, ,
}\end{equation}
arises from the entropy change due to deformation of the polymer network, and the mixing free energy,
\begin{equation}\label{eq:mixing_free energy}\eqalign{
\Delta F_{\rm mix} &= \Delta E - T\Delta S_{\rm mix} \\
&= Nk_B T\left[(1 - \phi)\,{\rm ln}\,(1 - \phi) + \chi\phi(1-\phi)\right] \, ,
}\end{equation}
is the net change in the free energy due to mixing polymer and solvent.
In the last term of equation (\ref{eq:gel_fe}), the constant $\phi_0$ corresponds to the volume fraction in the reference state of the gel, usually taken to be the volume fraction at which cross-linking is performed, or occasionally the volume fraction of a completely dry gel, namely $\phi_0 = 1$.
Notice that both the elastic and mixing free-energies scale with the thermal energy $k_B T$---the only term that presents a non-linear scaling with temperature is the Flory parameter term since $\chi$ is a function of temperature $T$.
It is therefore convenient to re-scale the total free energy $\Delta F$ by the thermal energy, i.e., $\Delta F/k_BT$, from which we find that the equilibrium state of polymer gels is determined by $\chi(T)$ alone.

A more careful and detailed look at the theory of gel elasticity confirms that the affine-deformation picture of classical rubber elasticity is inaccurate \cite{Panyukov1996}.
While the \emph{average} cross-link positions in space undergo affine transformation under a homogeneous deformation of the gel at its boundaries, there are in fact large fluctuations in cross-link positions due to thermal motion as well as network inhomogeneities.
In fact, these fluctuations are on the order of the mean cross-link spacing, which would melt ordinary solids, according to the Lindemann criterion, further highlighting the strangeness of these materials.
Additionally, the separation of the total free energy into a contribution due to the network elasticity and a contribution due to solvent-polymer mixing ultimately fails due to these large fluctuations, which renormalize both contributions.
Thus, while we will use the Flory-Rehner to illustrate the thermodynamics of polymer gels, it should be regarded as a semi-empirical model, that over-simplifies the true microscopic state of the gel.

\subsection{Isotropic swelling}


\begin{figure}
	\centering
	\includegraphics[width=8.3cm]{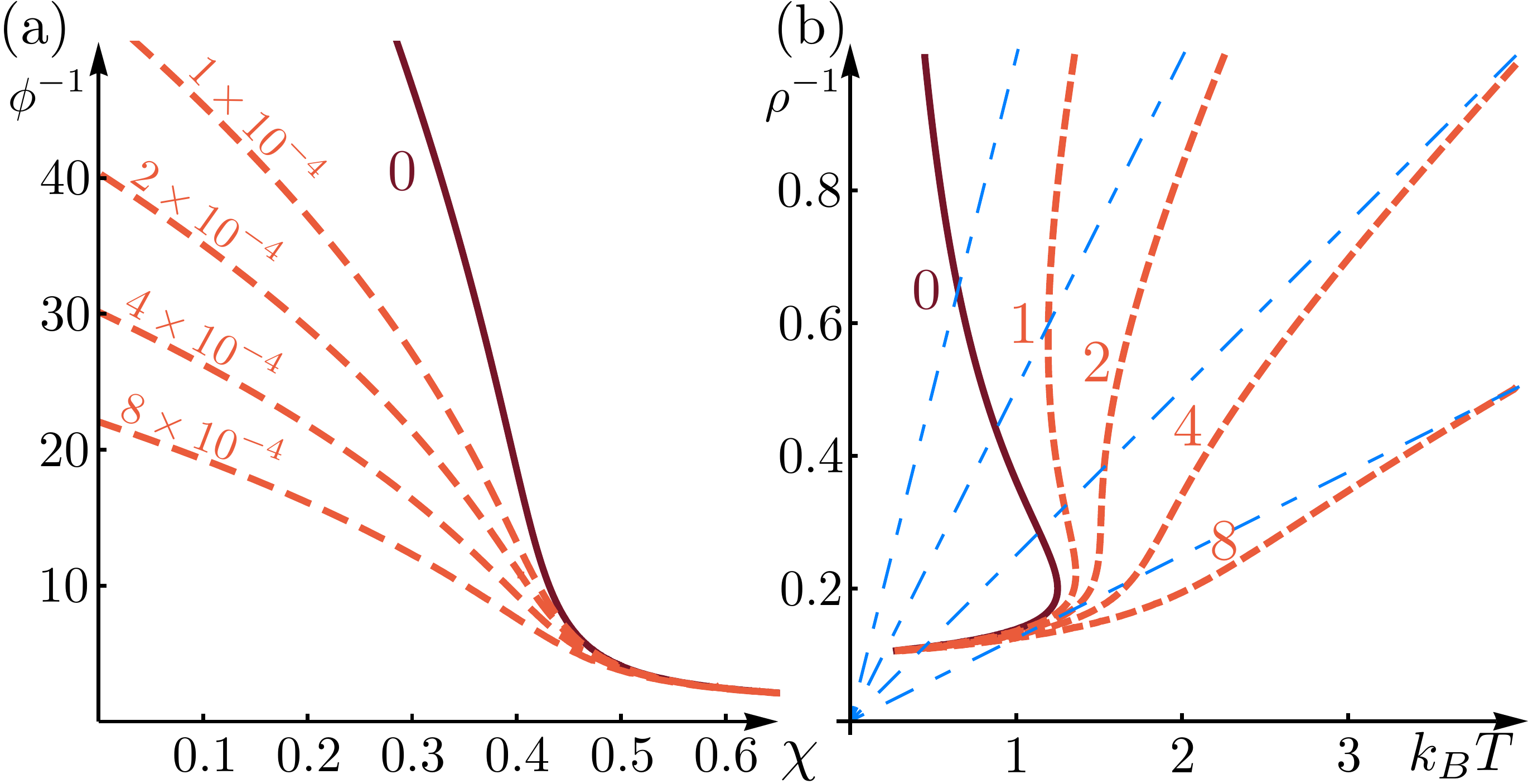}
	\caption{(a) Osmotic pressure ``isobars'' from the Flory-Rehner equation of state (\ref{eq:osmotic_pressure}) in units of thermal energy density $k_B T/v$ with chain fraction $\nu_0 v = 10^{-4}$ characterizing the connectivity of the polymer network and $\phi_0 = 10^{-1}$. (b) Isobars, corresponding to $P = 0,1,2,4,8$ in arbitrary units, from the Van der Waals equation of state with $\rho_0 = 10$ and $a = 0.5$. Teal dashed lines represent the ideal gas limit, $P = \rho k_B T$, realized when $\rho$ is small.}
	\label{fig:vol_chi}
\end{figure}

When allowed to equilibrate with a solvent bath, the amount of solvent in a polymer gel balances the osmotic pressure due to the thermal motion of the polymer network with the entropic cost of stretching this network.
Changes in this equilibrium state can be brought about by changing the solvent quality, as characterized by the Flory parameter $\chi$.
With the Flory-Rehner free energy (\ref{eq:gel_fe}) in hand, let us determine the equilibrium volume fraction $\phi(T)$ in the case of an isotropic gel.
Applying the volume constraint, viz., $\partial \Delta F/\partial \lambda = 0$, we readily obtain that the deformation matrix $\Lambda$ is given by
\begin{equation*}
\Lambda = \left(\frac{\phi_0}{\phi}\right)^{1/3}\mathbbm{1} \, .
\end{equation*}
The free energy density $\Delta \mathcal{F}$ of the gel is therefore
\begin{equation}\eqalign{
\Delta \mathcal{F} &= \frac{\nu_0\phi}{2\phi_0} k_BT\left[3\left(\frac{\phi_0}{\phi}\right)^{2/3} - 3 - {\rm ln} \, \left(\frac{\phi_0}{\phi}\right)\right] \\
&\mkern+32mu+ \frac{k_B T}{v}\left[(1 - \phi)\,{\rm ln}\,(1 - \phi) + \chi\phi(1-\phi)\right] \, ,
}\end{equation}
where $\nu_0 \equiv n_{ch}/V_0$ is the density of chains in the reference state of the gel.
The osmotic pressure follows from $\Delta F$ and is given by
\begin{equation}\eqalign{
\Pi(T,\phi) = \frac{k_B T}{v}\bigg[&\nu_0 v\left(\frac{\phi}{2\phi_0} - \left(\frac{\phi}{\phi_0}\right)^{1/3}\right) \\
& - \phi - {\rm ln}\,(1 - \phi) - \chi(T) \phi^2\bigg]\, .
}\label{eq:osmotic_pressure}
\end{equation}
where we have taken the Flory parameter $\chi$ to be a function only of temperature $T$.
Equation (\ref{eq:osmotic_pressure}) is the Flory-Rehner equation of state relating the osmotic pressure $\Pi$ to the volume fraction $\phi$ and temperature (via $\chi$).
Contours of variable volume fraction $\phi$ and Flory parameter $\chi$ at constant osmotic pressure $\Pi$ are shown in figure \ref{fig:vol_chi}(a).
For $\Pi = 0$, the volume fraction $\phi$ increases with decreasing $\chi$, corresponding to a gel that is swollen (low $\phi$) for a good solvent ($\chi < 0.5$) and that deswells as the solvent becomes poor.
Positive values of osmotic pressure can be obtained through the addition of a solute to the surrounding solvent; equilibrium osmotic isobars for positive values of $\Pi$ are also shown in figure \ref{fig:vol_chi}(a).

We can understand the behavior of the osmotic isobars in analogy with isobars from the Van der Waals equation of state
\begin{equation}\label{eq:vdw}
P = \frac{\rho k_B T}{1 - \rho/\rho_0} - a \rho^2,
\end{equation}
which relates the pressure $P$ to the density $\rho$ of particles, as shown in figure \ref{fig:vol_chi}(b).
Note that for low density, these curves asymptotically approach their ideal gas form $\rho^{-1} \sim P^{-1}k_B T$.
For sufficiently large positive pressure, the density $\rho$ decreases with increasing temperature, corresponding to an expanding gas.
However, for low pressures and temperatures, the density is a multivalued function of temperature.
In the case of the Van der Waals fluid, the emergence of the multivalued region is indicative of a loss of thermodynamic stability and the development of distinct liquid and gas phases.
Therefore, we might expect that the Flory-Rehner theory of gels has a similar phase transition separating a distinct low $\phi$ swollen phase and a high $\phi$ deswollen phase.
Such a phase transition indeed exists for gels, even though the Flory-Rehner equation requires a slight alteration to correctly capture it \cite{ErmanFlory1986}.

\section{\label{sec:phase_transitions} Phase transitions} 

\begin{figure}
	\includegraphics[width=8.3cm]{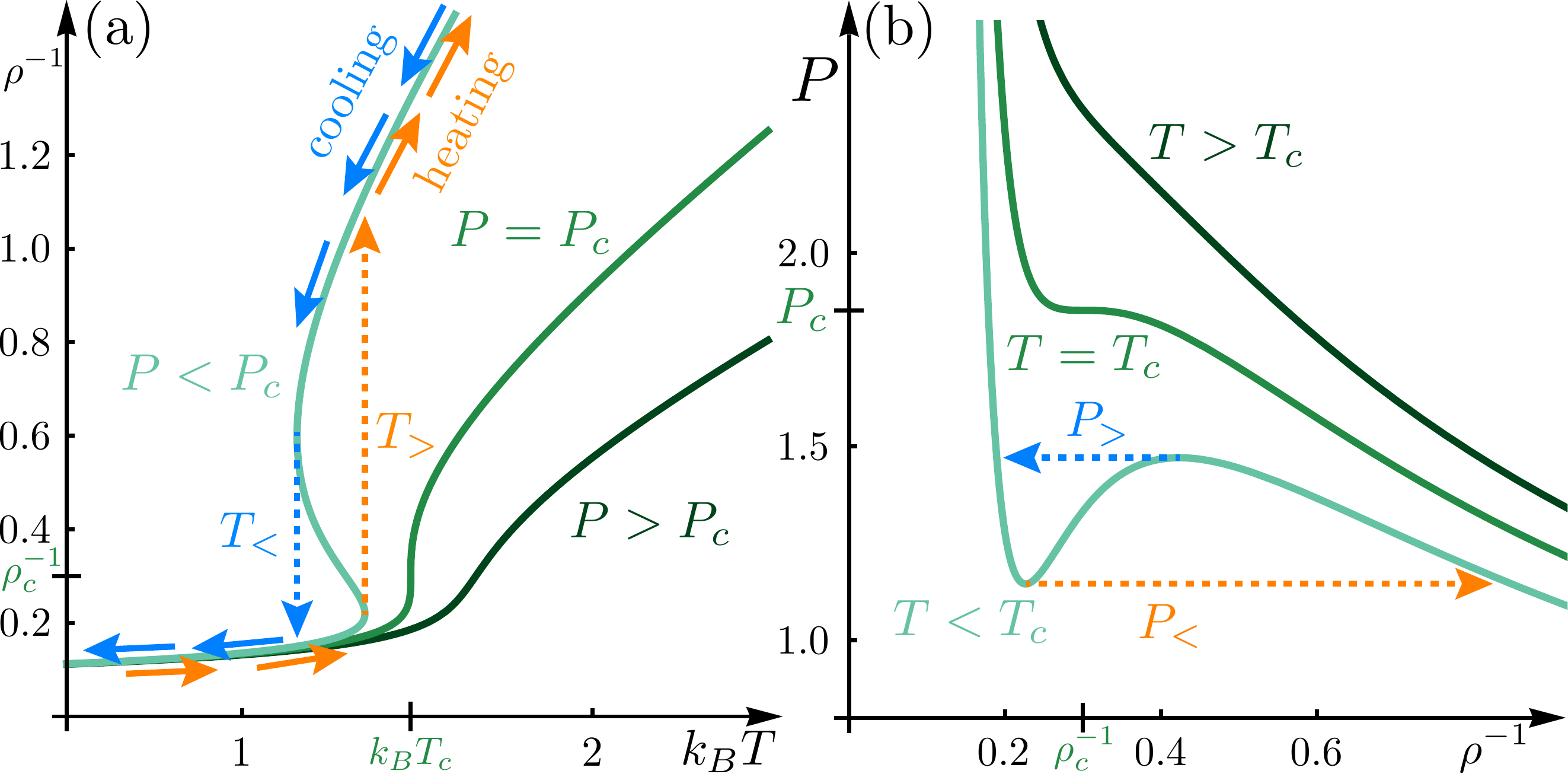}
	\caption{(a) Isobars of the Van der Waals equation of state for three values of pressure $P$ greater than, equal to, and less than the critical pressure $P_c$. The value of temperature for which the isobar becomes multivalued, $T_c$, is shown, along with the corresponding density $\rho_c^{-1}$. Also shown, a heating process starting at high density, which jumps to low density at a temperature $T_>$ and the reverse process with a density jump at $T_<$. (b) Isotherms for three temperatures $T$ greater than, equal to, and less than $T_c$. Two processes are shown in which pressure is increased and decreased, leading to density jumps at $P_>$ and $P_<$.}
	\label{fig:vdw_discontinuous}
\end{figure}

While there is a useful analogy that may be drawn between the isotropic swelling of polymer gels, as modeled by the Flory-Rehner theory, and the thermal expansion of a fluid, as modeled by the Van der Waals equation of state, there is also a key difference.
Examining the isobars in figure \ref{fig:vol_chi}(b), one finds that for sufficiently low temperature and pressure $P$, the $\rho^{-1} = V/N$ versus $T$ plot is multi-valued.
The value of pressure for which the well-defined, single-valued expansion curve becomes multi-valued is called the \emph{critical pressure}.
We highlight the situation in figure \ref{fig:vdw_discontinuous}(a), which shows three different isobars: $P<P_c$, $P=P_c$, and $P>P_c$.
For $P>P_c$, a fluid that is quasistatically heated from high-density (low $\rho^{-1}$) becomes lower-density (higher $\rho^{-1}$); this process is easily reversed upon cooling.
For $P=P_c$, while the equilibrium heating and cooling paths remain the same, the change in density with temperature diverges at a certain \emph{critical temperature} $T_c$.
However, for $P<P_c$, a high-density fluid can be quasistatically heated so that the density traces the lower part of the curve shown in figure \ref{fig:vdw_discontinuous}(a) until the curve folds back on itself.
Heating beyond this transition temperature $T_>$ results in a \emph{discontinuous} jump to a much lower density and the ensuing thermal expansion follows the upper branch of the isobar.
Cooling the fluid from low density and high temperature, however, traces the upper branch, until the discontinuity is encountered at a lower transition temperature $T_<$.
The low-density and high-density values of the fluid for $P<P_c$ distinguish separate fluid phases, which we recognize as the gas phase and the liquid phase, respectively.
Thus, this appearance of (i) a discontinuous jump in fluid density that (ii) depends on the heating path is not a failure of the Van der Waals model but rather a successful description of a \emph{first-order phase transition}.

Interestingly, experiments on certain polymer gels, notably pNIPAM, reveal similar discontinuous behavior in the equilibrium swelling curves \cite{Tanaka1978,Tanaka1980}.
For low temperatures, when the polymer network is miscible in the solvent, the gel is swollen.
Slowly increasing temperature increases the cost of polymer-solvent interaction, i.e., increases $\chi$, leading to gradual deswelling.
Above a certain temperature, roughly $32^{\circ}{\rm C}$ for pNIPAM, the gel suddenly expels most of its solvent into the surrounding bath, reducing its volume by orders of magnitude, and becomes opaque.
This discontinuity hints at a similar first-order phase transition of polymer gels and distinguishable $\emph{swollen}$ and \emph{deswollen phases}.
However, the osmotic pressure of the Flory-Rehner model, equation (\ref{eq:osmotic_pressure}), does not exhibit the multi-valued behavior of the Van der Waals model.
We will discuss the Erman-Flory extension of the Flory-Rehner model that allows for such a phase transition.
However, we will first briefly discuss the theory of phase transitions and critical phenomena more broadly.

\subsection{Preliminaries: general aspects of phase transitions}

In order to understand phase transitions, let us first consider thermodynamic stability.
While this discussion is generalizable, we will continue to use the example of the Van der Waals model of fluids.
We plot the constitutive relation between pressure $P$ and inverse density $\rho^{-1}$ for fixed temperature in figure \ref{fig:vdw_discontinuous}(b).
Evidently, if we are able to fix the temperature $T$ of a fluid and specify the total number of particles $N$ and volume $V$, then equation (\ref{eq:vdw}) tells us the pressure of the fluid, assuming that the density $\rho$ of particles is uniform everywhere.
Of course, at finite temperature, particles in the fluid undergo thermal fluctuations and the density $\rho$ varies in space and time.
The microscopic length scale over which spatial variations in particle density can be resolved is the correlation length $\xi$.
To connect to macroscopic physics, state functions like $\rho$ are found by \emph{coarse-graining}, or averaging over many particles in a certain region of space, the size of which is set by a coarse-graining length scale $\ell$.
By taking $\ell\gsim\xi$, thermal fluctuations are averaged out and we may approximate the state functions of each coarse-grained region by their thermodynamic limit.
For example, if we label a certain region by a position $\mathbf{x}$ so that the local coarse-grained density is $\rho(\mathbf{x})$ then the local pressure $P(\mathbf{x})$ can be approximated by the Van der Waals equation of state (\ref{eq:vdw}).

Now consider a disturbance to the fluid, such as a vibration or incident sound wave.
The result of such an external influence is a spatial modulation in density, typically of a longer length scale than $\xi$.
One coarse-grained region may have a slightly lower density of particles than its surroundings, which may have a slightly higher density of particles.
Consulting figure \ref{fig:vdw_discontinuous}(b), we find that for the higher temperature isotherms, pressure increases monotonically with density.
Therefore, assuming that the fluid is maintained at fixed temperature, the higher density regions are at higher pressure and the lower density regions are at lower pressure.
Subsequently, due to this pressure difference, particles will migrate from higher density to lower density in order to re-establish equilibrium.
The higher density region and the lower density regions eventually settle to a uniform density, namely $N/V$.
This resilience to perturbations is known as \emph{thermodynamic stability}.
Per this argument, the essence of \emph{Le Chatelier's Principle}, thermodynamic stability requires
\begin{equation}
\left(\frac{\partial P}{\partial \rho}\right)_{T} = \rho^{-1} K(T,\rho) > 0 \, ,
\end{equation}
where $K$ is the bulk modulus, which is the inverse of the isothermal compressibility $\kappa_T$ \cite{Callen1985}.
As long as the temperature $T > T_c$, where $T_c$ is the critical temperature, the constitutive relation $P(\rho)$ obeys this stability requirement.
However, at $T_c$, there is a critical pressure $P_c$ at which $\left(\partial P/\partial \rho\right)_T = 0$.
This \emph{critical point} marks the loss of thermodynamic stability and the onset of different physics.
The zero value of the bulk modulus $K$, or diverging compressibility $\kappa_T$, at the critical point is one example of the \emph{critical phenomena} that one encounters.

Saving the discussion of critical phenomena for later, let us address the consequences of \emph{thermodynamic instability}.
If $T < T_c$ and $P < P_c$ then there are certain values of $\rho$ where $\left(\partial_P/\partial \rho\right)_T < 0$ so the bulk modulus $K$ is negative.
Higher density regions are at lower pressure than the lower density region, so there is mass flow away from low density regions.
As a result, density fluctuations grow and the ultimate fate of such a fluid is \emph{phase-separation} into regions of low density and regions of high density.
However, density variations cannot grow forever: eventually, the high density and low density regions leave the unstable region of figure \ref{fig:vdw_discontinuous}(b) and enter stable regions of positive compressibility.
The difference between the higher and lower densities grows with the distance of the fluid from the critical point, characterized by a reduced temperature $t\equiv (T-T_c)/T_c < 0$; for large enough values of $|t|$, these two densities describe well-defined, distinguishable phases.
Eventually, these two fluid phases attain a \emph{phase-coexistent equilibrium}.

\begin{figure}
	\includegraphics[width=8.3cm]{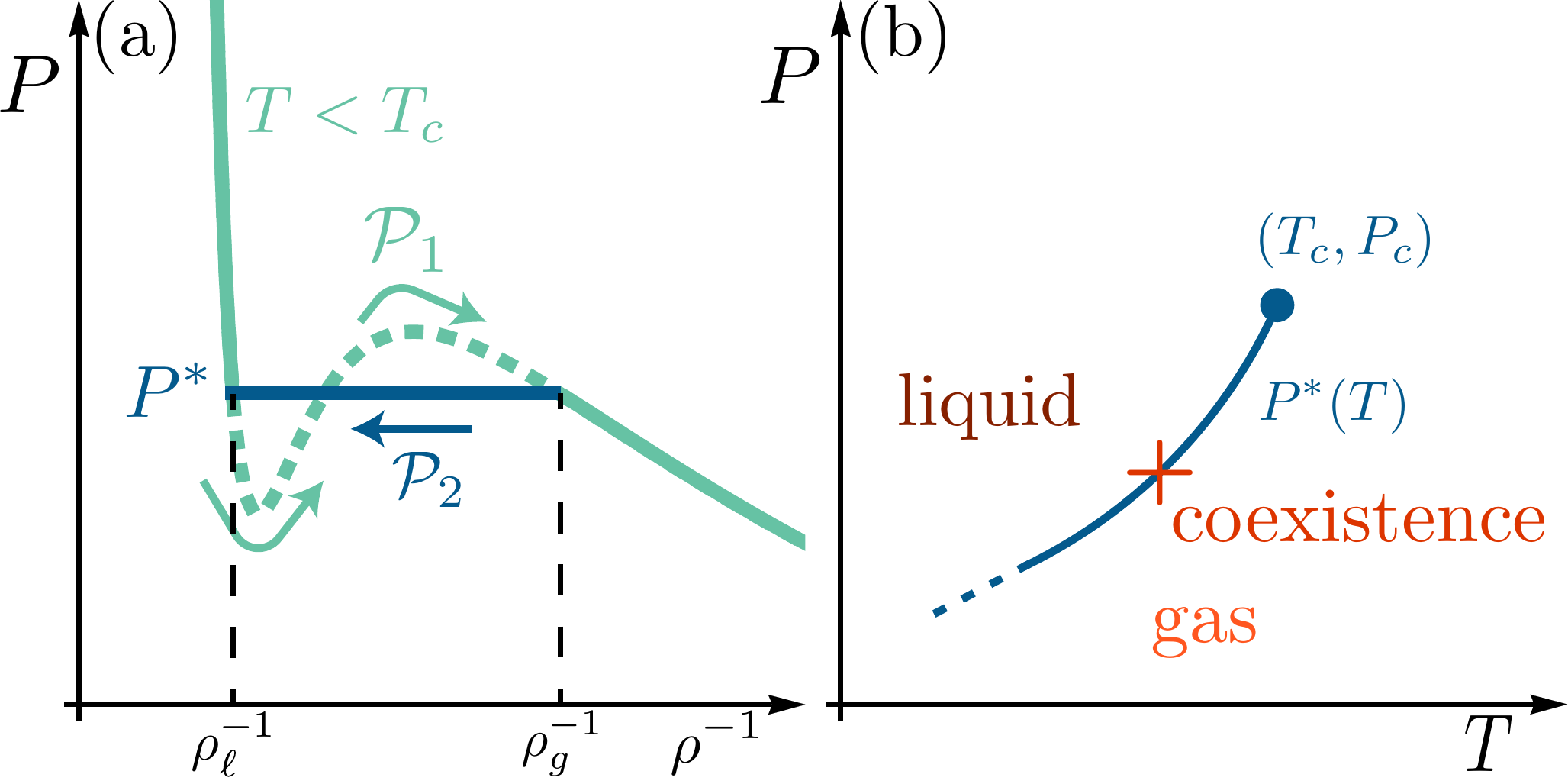}
	\caption{(a) An example isotherm below the critical temperature $T_c$ exhibiting an unstable $S$-shaped region ($\mathcal{P}_1$), along with the rectifying path ($\mathcal{P}_2$). This rectifying path gives a set of coexistent densities at a certain pressure $P^*$ at fixed temperature $T$. (b) Part of the phase diagram for a fluid described by the Van der Waals model. The locus of values of pressure and temperature $(T,P^*)$ that yield coexistence is shown and separates well-defined liquid and gas phases. This curve ends in a critical point $(T_c,P_c)$ where the fluid is single-phase.}
	\label{fig:maxwell_coex_vdw}
\end{figure}

Instead of using the Van der Waals equation of state to determine the pressure at a given density, now consider it as a way to determine density at a given pressure.
Much like the $\rho^{-1}(T)$ isobars in figure \ref{fig:vol_chi}(b), the $\rho^{-1}(P)$ isotherms are multi-valued graphs for $T<T_c$.
Thus, a high-density, liquid-phase fluid at $T < T_c$ undergoes a first-order phase transition to a low-density, gas-phase fluid for sufficiently low pressure.
However, we have shown that there should also be cases where the fluid is in a phase-coexistent equilibrium between liquid and gas phases.
In order for these phases to coexist in equilibrium, they must (i) have the same temperature $T$, (ii) the same pressure $P$, and (iii) the same chemical potential $\mu$; that is, they must be in thermal, mechanical, and chemical equilibrium.
While $T$ and $P$ are specified, $\mu$ must be determined.
We can take advantage of the Gibbs-Duhem relation $N\,{\rm d}\mu = - S\,{\rm d}T + V\,{\rm d}P$, which, at constant temperature, can be expressed as ${\rm d}\mu = \rho^{-1}\,{\rm d}P$ \cite{Callen1985}.
Therefore, chemical equilibrium is realized when
\begin{equation}\label{eq:chemical_potential_change}
\Delta \mu = \mu_{g} - \mu_{\ell} = \int_{\mathcal{P}_1} {\rm d}P\, \rho^{-1}(P) = 0 \, ,
\end{equation}
where $\mathcal{P}_1$ is the path along the isotherm, shown in figure \ref{fig:maxwell_coex_vdw}(a), that connects the point $(\rho^{-1}_{\ell}, P^*)$ to $(\rho^{-1}_{g}, P^*)$, where $\rho_{\ell}$ and $\rho_{g}$ are the respective densities of the liquid and gas phases.
Joining these two points by a constant-pressure line $\mathcal{P}_2$, we can define a loop $\mathcal{P} = \mathcal{P}_1 \cup \mathcal{P}_2$ as the union of these two paths; this is called a ``Van der Waals loop.''
Integrating equation (\ref{eq:chemical_potential_change}) by parts,
\begin{equation}\eqalign{
\int_{\mathcal{P}_1} {\rm d}P\, \rho^{-1}(P) &= P^*(\rho^{-1}_g - \rho^{-1}_{\ell}) - \int_{\mathcal{P}_1} {\rm d}(\rho^{-1})\, P \\
&\mkern-32mu= -\left[P^*\int_{\mathcal{P}_2} {\rm d}(\rho^{-1}) + \int_{\mathcal{P}_1} {\rm d}(\rho^{-1})\, P\right] \\
&\mkern-32mu= -\int_{\mathcal{P}}{\rm d}(\rho^{-1})\, P\, ,
}\end{equation}
we therefore find that coexistent phases are in equilibrium when the net area enclosed by the Van der Waals loop $\mathcal{P}$ is 0.
The line $\mathcal{P}_2$ that joins coexisting densities $\rho_{\ell}$ and $\rho_{g}$ gives the equilibrium pressure and replaces the $S$-shaped curve $\mathcal{P}_1$.
This ``Maxwell construction'' corrects multivalued isotherms in the Van der Waals equation of state \cite{Callen1985}.

Furthermore, in the two-dimensional $P\; vs.\; T$ phase diagram of fluids, we can identify a one-dimensional locus of coexistent equilibria consisting of the single pressure $P$ that yields coexistence for each isotherm $T$.
This \emph{coexistence curve} in the phase diagram terminates at the critical point $(T_c,P_c)$.
A fluid may be brought around the critical point without passing through the coexistence curve via appropriate temperature and pressure change protocols.
Whilst points on the coexistence curve describe coexistence between gas and liquid phases, points immediately to one side or the other are in the single phase region.
Passage through the coexistence curve results in a first-order phase transition, a discontinuous jump between high-density and low-density fluids.
Therefore, the only way to distinguish between gas and liquid phases of fluids is to pass through the coexistence curve.
In fact, \emph{the ability for a system to support coexisting densities in equilibrium is the defining feature of distinct phases that are separated by a first-order phase transition}.

\subsection{The common tangent construction in phase-separating systems}

The phase diagram \ref{fig:maxwell_coex_vdw}(b) shows a phase-coexistent region for a particular set of temperatures and pressures.
To land on this curve, one requires precise control over temperature and pressure, suggesting that coexistent phases are rarely realized.
As it turns out, one can readily achieve phase coexistence at constant temperature, volume, and number of particles $(T,V,N)$.

\begin{figure}
	\includegraphics[width=6.5cm]{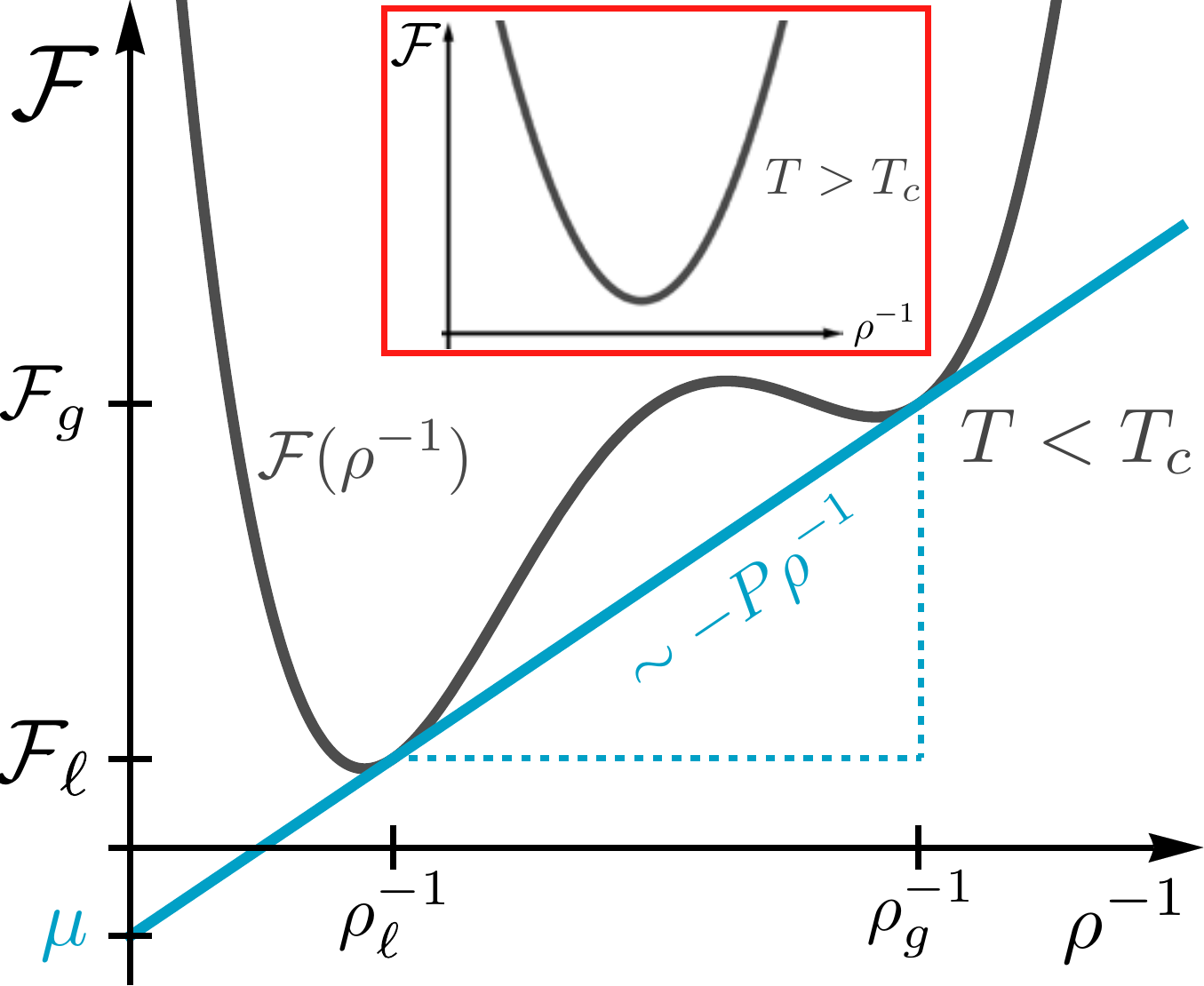}
	\caption{An example Helmholtz free energy density $\mathcal{F}(\rho^{-1})$ at temperature $T<T_c$ that possesses two local minima and a single local maximum is shown. The common tangent line has a slope of $-P$ and an intercept of $\mu$ and describes two equilibrium coexistent densities. Inset shows example Helmholtz free energy density at temperature $T>T_c$, where it exhibits a single minimum.}
	\label{fig:common_tangent_fig}
\end{figure}

Rather than working with the pressure-density equation of state, consider the Helmholtz free energy $F(T,V,N)$.
Since $F$ is a thermodynamic potential, it is a homogeneous first-order function in $V$ and $N$.
Therefore, we can write $F$ in terms of its density $\mathcal{F}$ as $F(T,V,N) = N\mathcal{F}(T,\rho^{-1})$, defined in terms of $N$.
If the free energy density $\mathcal{F}$ describes a thermodynamically stable system then the requirement for positive isothermal compressibility is satisfied for $(\partial^{2}\mathcal{F}/\partial(\rho^{-1})^2)_T > 0$ and there is a single free energy minimum $(\rho^{-1})^*$ for fixed values of $T$, as illustrated in the inset in figure \ref{fig:common_tangent_fig}.
However, for $T<T_c$, the free energy has a region of negative compressibility where $(\partial^{2}\mathcal{F}/\partial(\rho^{-1})^2)_T < 0$ and $\mathcal{F}$ is concave when plotted against $\rho^{-1}$.
In this case, the free energy supports two local minima, separated by a local maximum, as shown in figure \ref{fig:common_tangent_fig}.
Lacking a volume constraint, the system will seek to minimize the free energy so a local minimum that is not the global free energy minimum is considered meta-stable: eventually, given sufficient time, thermal fluctuations will drive the system to the global free energy minimum.
For example, it is possible to ``superheat'' a homogeneous liquid-phase fluid above the transition temperature for phase coexistence, keeping the fluid in its liquid phase, a metastable equilibrium, for a prolonged period of time.
To do this requires careful preparation, removing any possible nucleation sites for the gas phase from the liquid at, for  example, small pockets of trapped gas.
The introduction of a nucleation site, e.g., via disturbing the fluid, lowers the free energy barrier locally and allows a portion of the liquid to transition to the gas phase.
Without the introduction of a nucleation site from external influence, the superheated liquid nevertheless has a finite, albeit much longer, lifespan as a significantly large density fluctuation, driven by thermal fluctuations, will eventually provide a suitable nucleation site.
Since the gas-phase is of lower density than the liquid-phase, it occupies a larger volume than the same mass of liquid-phase fluid.
However, if the total volume of the fluid is constrained to remain constant, then even though the free energy associated with the gas is lower than that of the liquid, not all of the liquid can freely transition to the gas-phase.
Instead, a portion of the liquid can transition to the gas-phase, at the expense of increasing the density of the liquid phase, achieving a phase-coexistent equilibrium.

In order to determine the conditions for equilibrium phase coexistence at constant volume, recall that the equilibrium state minimizes the \emph{global} free energy $F$.
Let $\rho_{\ell}$ and $\rho_g$ be the densities of the liquid and gas phases, consisting of $N_{\ell}$ and $N_g$ particles, respectively.
By conservation of mass, the total number of particles in the container remains unchanged: $N_{\ell} + N_g = N$.
Therefore, we can define a fraction $f \equiv N_g/N$ of particles that are in the gas phase; by mass conservation, the fraction of particles that are in the liquid phase is $(1 - f)$.
Furthermore, the total volume of particles in the liquid and gas phases are given by $V_{\ell} = \rho_{\ell}^{-1}N_{\ell}$ and $V_{g} = \rho_g^{-1}N_{g}$.
Volume conservation requires that $V_{\ell} + V_g = V$; dividing by $N$, this conservation may be expressed as $f\rho_g^{-1} + (1 - f)\rho_{\ell}^{-1} = \rho^{-1}$, where $\rho\equiv N/V$ is the nominal density of the fluid, as if it were in a single phase.
Therefore, the fraction $f$ of particles in the gas-phase is given in terms of the equilibrium densities by
\begin{equation}\label{eq:fluid_lever}
\frac{N_g}{N} = f = \frac{\rho^{-1} - \rho_{\ell}^{-1}}{\rho_{g}^{-1} - \rho_{\ell}^{-1}} = \frac{V - V_{\ell}}{V_g - V_{\ell}}\, ,
\end{equation}
a result known as the \emph{Lever Rule}.

The phase-coexistent equilibrium is determined by minimizing the total Helmholtz free energy $F$, 
\begin{equation}
\frac{F}{N} = f\mathcal{F}_g + (1 - f)\mathcal{F}_{\ell} + \lambda\left(\frac{f}{\rho_g} + \frac{1 - f}{\rho_{\ell}} - \frac{1}{\rho}\right)\, ,
\end{equation}
which is the sum of free energy contributions from each phase, where $\mathcal{F}_g = F(\rho_g^{-1})/N_g$ and $\mathcal{F}_{\ell} = F(\rho_{\ell}^{-1})/N_{\ell}$, subject to the volume constraint, enforced by a Lagrange multiplier $\lambda$.
The minimization condition ${\rm d}F = 0$ requires that partial derivatives of the total free energy with respect to $\rho_g^{-1}$, $\rho_{\ell}^{-1}$, $f$, and $\lambda$ are all equal to 0.
This results in the Lever Rule (\ref{eq:fluid_lever}) along with three additional equilibrium equations, namely
\numparts\begin{eqnarray}
-\left(\frac{\partial \mathcal{F}_g}{\partial \rho_g^{-1}}\right)_{\rho_{\ell}^{-1},f,\lambda} &=& -\left(\frac{\partial \mathcal{F}}{\partial \rho^{-1}}\right)_{\rho^{-1}=\rho_g^{-1}} = \lambda \label{eq:fluid_tan_a} \, , \\[5pt]
-\left(\frac{\partial \mathcal{F}_{\ell}}{\partial \rho_{\ell}^{-1}}\right)_{\rho_{g}^{-1},f,\lambda} &=& -\left(\frac{\partial \mathcal{F}}{\partial \rho^{-1}}\right)_{\rho^{-1}=\rho_{\ell}^{-1}} = \lambda \label{eq:fluid_tan_b}\, , \\[5pt]
\mkern+64mu \mathcal{F}_g - \mathcal{F}_{\ell} &=& -\lambda\left(\rho_g^{-1}  - \rho_{\ell}^{-1}\right) \, . \label{eq:fluid_tan_c}
\end{eqnarray}\endnumparts
Consider an example free energy density $\mathcal{F}(\rho^{-1})$ at low enough temperature that it has two local minima, as plotted in figure \ref{fig:common_tangent_fig}.
The first two equilibrium equations (\ref{eq:fluid_tan_a},\ref{eq:fluid_tan_b}) show that the slopes of the free energy density $\mathcal{F}(\rho^{-1})$ at $\rho_g^{-1}$ and $\rho_{\ell}^{-1}$ are the same.
The third equation (\ref{eq:fluid_tan_c}), after taking into account that $\lambda$ is the slope of the lines that are tangent to the free energy at the equilibrium densities, shows that the two tangent lines overlap.
Thus, the equilibrium densities satisfy a \emph{common tangent construction}: the values $\rho_g^{-1}$ and $\rho_{\ell}^{-1}$ can be found graphically by drawing a straight line that is tangent to $\mathcal{F}(\rho^{-1})$ at two points.
As long as the free energy has two stable equilibria that are separated by an unstable equilibrium, this construction yields unique values for the equilibrium densities, and thus the fraction $f$ via the Lever Rule (\ref{eq:fluid_lever}).
Importantly, appreciate how the equilibrium densities are not at the local minima of $\mathcal{F}$.
Instead, $\rho_g^{-1}$ and $\rho_{\ell}^{-1}$ are close to those minima, as can be seen in figure \ref{fig:common_tangent_fig}.

There is a physical rationale behind the equilibrium equations.
Note that the pressure $P = -(\partial F/\partial V)_{T,N} = -(\partial \mathcal{F}/\partial \rho^{-1})_T$ so that the first two equations (\ref{eq:fluid_tan_a},\ref{eq:fluid_tan_b}) yields the interpretation of $\lambda$, a generalized force that maintains the volume constraint, as the pressure $P$ of the two phases, which are in mechanical equilibrium.
To interpret the third equation (\ref{eq:fluid_tan_c}), note that $\mathcal{F} - \lambda\rho^{-1} = (F - PV)/N = G/N$, where $G$ is the Gibbs free energy.
Therefore, we have that $G_g/N_g = G_{\ell}/N_{\ell}$, a balance of the Gibbs free energy density for each phase.
Noting that $G = \mu N$, this is also a balance between chemical potentials $\mu_g$ and $\mu_{\ell}$, so the two phases are in chemical equilibrium.
This result details a robust way to achieve equilibrium phase coexistence: as long as the temperature $T$ is low enough for the system to be thermodynamically unstable for certain values of the state functions, then there is equilibrium between coexistent phases at constant $N$ and $V$.

\subsection{A note on critical phenomena}

While we will not linger on the rich subject of physics near a critical point, a discussion of phase transitions requires at least a cursory mention of critical phenomena.
Until this point, we have focused on equilibrium thermodynamics in the vicinity of the coexistence curve.
Crossing this coexistence curve results in a first-order phase transition, which allows us to distinguish phases, such as the gas and liquid phase of a fluid.
Sitting on the coexistence curve, the system is not a single homogeneous phase but rather an admixture of two phases that are in equilibrium with each other.
The critical point is the endpoint of this coexistence curve and thus marks the onset of distinction between the two phases; equivalently, coexistence breaks down as this end of the curve is reached and the two phases lose distinction.

To study thermodynamics close to a critical point, it is helpful to define an \emph{order parameter} $\varphi$ that is zero when only a single phase exists (off of the coexistence curve) and is nonzero when multiple phases exist.
For a fluid, a choice of order parameter is the \emph{reduced density}, namely
\begin{equation}
\varphi = \frac{\rho - \rho_c}{\rho_c} \, ,
\end{equation}
where $\rho_c$ is the average density of the fluid at the critical point.
For the Van der Waals equation of state (\ref{eq:vdw}), the critical temperature is given by $T_c = 8 a \rho_0/27$ and the critical pressure is $a \rho_0^2/27$ so the critical density is $\rho_c = \rho_0/3$.
For small negative values of the reduced temperature $t = (T - T_c)/T_c$ along the critical isochore, where the fluid at the critical density $\rho_c$ is in unstable equilibrium, there are two new locally stable equilibria, one with $\rho > \rho_c$ and another with $\rho < \rho_c$; there is a positive and a negative value of $\varphi$.
Defining a reduced pressure $p = (P - P_c)/P_c$ and expanding the Van der Waals equation of state (\ref{eq:vdw}) yields a linear leading-order dependence of $p$ with $\varphi$ for small $\varphi$, that is $p \approx r\varphi$, where the coefficient $r \propto t$.
Noting that this implies that $\partial p/\partial \varphi \propto t$, we find stability via the Le Chatelier principle for $t>0$ and instability for $t<0$.
To recover the appearance of two new locally stable equilibria, there needs to be a dependence on $\varphi$ added to $p$ that yields positive bulk modulus for sufficiently large values of $|\varphi|$.
The simplest addition that stabilizes the reduced equation of state is a cubic $\varphi^3$ dependence, yielding $p = r\varphi + u\varphi^3$, where $u>0$.
We can integrate the pressure to find an approximate form of the free energy density $\mathcal{F}$ near the critical point, namely
\begin{equation}
\mathcal{F} \approx \frac{1}{2} r \varphi^2 + \frac{1}{4} u \varphi^4 \, .
\end{equation}
One immediate consequence of this approximation is that the equilibrium values of the new phases close to the critical point, i.e.~small $|t|$, are given by
\begin{equation}
\varphi = \pm\sqrt{\frac{|r|}{u}} \propto |t|^{1/2} \, ,
\end{equation}
which are symmetric about the unstable equilibrium $\varphi = 0$.
Note that this symmetry between the two phases holds only close to the critical point: further away, odd powers of $\varphi$ appear in the free energy.
Still, close to the critical point, we find that the separation in density between the liquid and gas phases, $(\rho_{\ell} - \rho_g)/\rho_c = 2|\varphi^*|$, where $\varphi^*$ is a minimum of $\mathcal{F}$, scales with $|t|^\beta$ with $\beta = 1/2$ along the critical isochore.
The exponent $\beta$ is one example of a \emph{critical exponent}.
There are a variety of critical exponents for different thermodynamic quantities, such as the isothermal compressibility $\kappa_T \sim |t|^{-\gamma}$.
Since $\kappa_T^{-1} \propto {\rm d}^2\mathcal{F}/{\rm d}\varphi^2$, we find that $\gamma = 1$ along the critical isochore.
These critical exponents are used to characterize the behavior of a system near a critical point.

The problem with the above analysis is that it does not yield good predictions for critical exponents that are measured in the lab.
The measured value of $\beta$ is actually close to $1/3$, but does not seem to be a rational number \cite{Goldenfeld1992}.
Even though the Van der Waals equation of state works well for describing equilibrium physics of liquid and gas phases of fluids for much of the phase diagram, it seems to fail near the critical point.
As it turns out, the root of the problem lies in the assumption that thermal fluctuations are not important.
The key assumption was that thermal fluctuations, characterized by a correlation length $\xi$, are important only over small length-scales.
To find the large length-scale physics, recall that in defining $\rho$ and other state functions as descriptions of the macroscopic state of a system, there was a coarse-graining length scale $\ell$ introduced over which the microscopic details, such as thermal fluctuations, were averaged over.
This coarse-graining length $\ell$ was taken to be at least as large as the correlation length $\xi$.
Since fluctuations are ignored in the resulting description of the thermodynamics, this is referred to as a \emph{mean-field theory}.
The predicted critical exponents are \emph{mean-field critical exponents}, which, owing to the simple structure of mean-field theories, are always rational numbers.

In order to correct mean-field theory, we need to properly incorporate this coarse-graining length scale and investigate corrections due to thermal fluctuations in the order parameter $\varphi$.
To do this, we can model a fluctuating order parameter near the critical point by a model free energy
\begin{equation}\label{eq:free energy-Landau}
F = \int {\rm d}^d x\left[\frac{1}{2} c |\nabla \varphi|^2 + \frac{1}{2}r \varphi^2 + \frac{1}{4} u \varphi^4\right] \, ,
\end{equation}
where the coefficient $c$ sets the energy cost of spatial variations in $\varphi$, and $d$ is the dimensionality of the space in which properties of the material vary.
Note that this is completely phenomenological: the presence of the gradient term $|\nabla \varphi|^2$ simply provides a positive free energy cost for spatial variation.
This term is necessary for the development of the coarse-grained model of the fluid as it describes a lower cutoff length for the wavelength of spatial fluctuations in the order parameter field $\varphi$.
To see this, let $\lambda$ represent the wavelength of a fluctuation in $\varphi$.
Then there is an energetic cost of this fluctuation that scales as $c |\varphi|^2/\lambda^2$ so that as the length scale $\lambda$ of the spatial variation in $\varphi$ decreases in size, the cost of this fluctuation grows.

To see how adjusting $c$ affects the length $\xi$ over which fluctuations $\delta\varphi(\mathbf{x})$ of $\varphi(\mathbf{x})$ about the equilibrium $\varphi^* \equiv \pm\sqrt{|r|/u}$ are correlated in space, we can determine the functional form of the fluctuation correlations, namely $\left<\delta\varphi(\mathbf{x})\delta\varphi(0)\right>$.
To leading order in fluctuations $\delta \varphi$, the change $\delta F$ in the free energy is given by
\begin{equation}\label{eq:free energy-fluctuation}
\delta F = F - F_0 = \int {\rm d}^d x\left[\frac{1}{2} c |\nabla \delta\varphi|^2 + r \delta\varphi^2\right] \, ,
\end{equation}
where $F_0 = -r^2 V/(4 u)$ is the free energy corresponding to the homogeneous equilibrium $\varphi^*$.
The probability that a particular fluctuation $\delta\varphi(\mathbf{x})$ is weighted by the Boltzmann factor $\exp(-\beta\delta F[\delta\varphi])$, where $\beta = (k_B T)^{-1}$.
Therefore, the fluctuation correlations are determined by
\begin{equation}\label{eq:fluctuation_correlation_integrals}
\left<\delta\varphi(\mathbf{x})\delta\varphi(0)\right> =  \frac{\int [{\rm d}\delta\varphi] e^{-\beta\delta F}\delta\varphi(\mathbf{x})\delta\varphi(0)}{\int [{\rm d}\delta\varphi] e^{-\beta\delta F}} \, ,
\end{equation}
where $\int [{\rm d}\delta\varphi]$ represents a sum over all possible fluctuations $\delta \varphi$ of the order parameter field.
Integrating the free energy fluctuation (\ref{eq:free energy-fluctuation}) by parts to yield
\begin{equation}
\delta F = \int {\rm d}^d x\,\delta\varphi\left[-\frac{1}{2} c \nabla^2 + r \right]\delta\varphi \, ,
\end{equation}
we recognize that the functional integrals in (\ref{eq:fluctuation_correlation_integrals}) have a Gaussian form and are thus simple to evaluate.
The fluctuation correlations are given by
\begin{equation}
\left<\delta\varphi(\mathbf{x})\delta\varphi(0)\right> = \beta^{-1}\left[-\frac{1}{2} c \nabla^2 + r \right]^{-1} \, ,
\end{equation}
and thus satisfy the Green's function equation \cite{ChaikinLubensky}
\begin{equation}
\beta\left[-\frac{1}{2} c \nabla^2 + r \right]\left<\delta\varphi(\mathbf{x})\delta\varphi(0)\right> = \delta^{(d)}(\mathbf{x}) \, ,
\end{equation}
where $\delta(\mathbf{x})$ is the Dirac delta function in $d$ dimensions.
Therefore, the position dependence of the fluctuation correlations is given by
\begin{equation}
\left<\delta\varphi(\mathbf{x})\delta\varphi(0)\right> \propto \frac{e^{-|\mathbf{x}|/\xi}}{\xi^{(d-3)/2}|\mathbf{x}|^{(d-2)/2}} \; ,
\end{equation}
where $\xi \equiv \sqrt{c/(2|r|)}$ defines the correlation length between thermal fluctuations in the coarse-grained field $\varphi$ \cite{Goldenfeld1992}.
As long as $|r| > 0$, the isothermal compressibility $\kappa_T \sim r^{-1}$ is finite, and we can always therefore coarse-grain to a length-scale $\ell$ larger than the correlation length $\xi$.
However, as $r \rightarrow 0^{-}$ on the approach to the critical point, this length-scale is ill-defined because $\xi$ diverges!
Therefore, fluctuations cannot be ignored and mean-field theory is destined to fail.
Indeed, this is confirmed in experiment via the phenomenon of ``critical opalescence'' \cite{Barrat2003}. 
As an otherwise transparent fluid, such as water, approaches the critical point, it turns opaque.
Whereas normally, the correlation length $\xi$ is shorter than the wavelength of visible light, on the approach to the critical point, it lengthens to the point that thermal fluctuations in the density of the fluid can scatter light.
The color of the fluid is a milky white, revealing that all visible wavelengths are scattered, so that fluctuations exist at many length-scales concurrently.
Furthermore, this opacity lingers even as $\xi$ increases in length closer to the critical point, confirming that fluctuations at visible wavelengths remain, even as $\xi$ moves into the infrared and beyond, eventually stopping at the macroscopic length-scale $L \sim V^{1/3}$ of the container.
Essentially, near the critical point, the fluctuations become scale-free.

Interestingly, while the mean-field theory predicts one set of critical exponents, in reality, critical exponents can vary from system-to-system.
However, there are certain, seemingly unrelated, systems that share sets of critical exponents.
For example, fluids, ferromagnets, and binary alloys all have approximately the same critical exponents \cite{Goldenfeld1992}.
This commonality of critical exponents amongst diverse systems means that their critical behavior is similar, even though the microscopic physics at play is very different, a phenomenon known as \emph{universality}.
Universality amongst systems is due to symmetry rather than microscopic physics.
The order parameter $\varphi$ that we introduced for fluids represents a density difference.
It works just as well for ferromagnets, which are described by magnetic dipoles that either point up or down; here, positive values of $\varphi$ correspond to an average magnetic dipole moment that is up and negative represents an average that is down.
Similarly, for binary mixtures consisting of species labeled $A$ and $B$, $\varphi$ represents the difference in densities of species $A$ and species $B$.
Regardless of the underlying microscopic physics, the form of the free energy at the critical point is identical, and the result is identical critical exponents, even when fluctuations are accounted for.
Systems represented by other types of order parameters typically lie in other universality classes.
The universality class of fluids, ferromagnets, and binary alloys is the three-dimensional Ising model, owing to the \emph{discrete} $+/-$ symmetry of the free energy $\mathcal{F}$, namely, $\mathcal{F}(-\varphi) = \mathcal{F}(+\varphi)$. 
If, for example, $\varphi$ was a complex order parameter instead of a real scalar and if $\mathcal{F}$ was invariant under \emph{continuous} transformation of the form $e^{i\theta}\varphi$, the corresponding critical phenomena would fall into the XY model universality class.
For example, many systems with a polar order parameter $\bm{\sigma}$ that exhibit continuous rotational symmetry, such as superfluids, certain superconductors, and hexatic liquid crystals, lie in the universality class described by the XY model.
The predictive power of the dimensionality of space, the dimensionality of the order parameter, and the symmetries of the system allow useful models of the behavior near the critical point via \emph{Landau theory}, where a simple free energy, such as (\ref{eq:free energy-Landau}), is constructed based on these considerations alone \cite{ChaikinLubensky,Goldenfeld1992}.

\subsection{Swelling-deswelling phase transition in polymer gels}

Whilst many polymer gels undergo continuous changes in their polymer volume fraction due to changes in solvent conditions, e.g., via changes in temperature, certain gels exhibit a seemingly discontinuous change in volume fraction $\phi$, jumping between a low-$\phi$ swollen state to a high-$\phi$ deswollen state.
As we have illustrated with our discussion about fluids, a discontinuous change in density in response to changing other state functions, e.g., temperature and pressure, indicates a first-order phase transition between a low-density and a high-density phase.
For fluids modeled by the Van der Waals equation of state, these are the gas and liquid phases, respectively.
Furthermore, little distinction between gas and liquid phases can be seen microscopically---unlike crystalline phases, there is no broken symmetry that distinguishes the two phases.
The only sure way to distinguish these two phases is passage through a coexistence curve in the phase diagram that either crosses through the discontinuous transition or ends in a state of equilibrium phase coexistence.
Therefore, we are led to conclude that polymer gels can have distinguishable swollen and deswollen phases.
However, unlike in the Van der Waals model of fluids, the Flory-Rehner model of polymer gels does not predict discontinuous change in $\phi$ for physically reasonable parameters.
Furthermore, within the formulation of the Flory-Rehner model for the osmotic pressure that has been presented thus far, not all values of $\chi$ yield a corresponding equilibrium value of $\phi$ when $\Pi$ is negative.
One way of achieving $\Pi < 0$ is by applying a mechanical pressure to the boundary of the gel, leading to solvent flow out of the gel via ``reverse osmosis.''
Interestingly, negative osmotic pressure states are typically thermodynamically unstable, favoring de-mixing of a solution into pure solute and pure solvent \cite{Powles1997}, i.e., phase-separation.
This lack of general applicability suggests that equation (\ref{eq:osmotic_pressure}) is an incomplete equation of state.

The Flory-Rehner model describes a rather simple picture of polymer gels in which the osmotic pressure $\Pi$ is expressed as two separate contributions: $\Pi_{\rm mix}$, which is due to the thermal motion of polymers amongst solvent molecules, and $\Pi_{\rm el}$, which is due to the elasticity of the polymer network.
For ionic gels, thermal motion of free counterions contribute $\Pi_{\rm ion}$ to the osmotic pressure.
This addition, which can be simply approximated as an ideal gas of counterions within the gel, is enough to theoretically obtain a discontinuous transition in the context of the Flory-Rehner model \cite{Onuki1993}.
Ionic gels are indeed known to exhibit a discontinuous transition between swollen and deswollen phases.
However, some neutral gels, such as pNIPAM, can also undergo a discontinuous transition yet do not have another obvious osmotic pressure contribution that is not captured within the Flory-Rehner model (\ref{eq:osmotic_pressure}).
Hence, as we have already emphasized, the Flory-Rehner model should be regarded as semi-empirical.
This, in part, is due to the important role of thermodynamic fluctuations as well as static inhomogeneities in the polymer network.
In the presence of a poor solvent, it has been shown \cite{Panyukov1996} that, beyond a straightforward renormalization of the elastic and osmotic contributions, the presence of network inhomogeneities can lead to phase-separation, either at high wavenumber (\emph{micro}phase separation) or at low wavenumber (\emph{macro}phase separation).

It is possible to extend the Flory-Rehner model such that it describes a phase transition.
This is accomplished by altering the mixing energy between polymer and solvent molecules, which is controlled by the Flory parameter $\chi$.
This term describes a two-body mean-field interaction between polymer and solvent molecules.
To see this, expand the mixing contribution $\Pi_{\rm mix}$ to the osmotic pressure in powers of the polymer volume fraction $\phi$:
\begin{equation}
\Pi_{\rm mix} = \frac{k_B T}{v}\left[\left(\frac{1}{2} - \chi\right)\phi^2 + \sum_{m = 3}^{\infty}\frac{\phi^m}{m}\right],
\end{equation}
where the sum is the remainder of the power series expansion of ${\rm ln}(1 - \phi)$.
This expansion has the form of a \emph{virial expansion} of the pressure $P$ of a fluid in terms of its density $\rho$, namely
\begin{equation}
P = k_BT\left[\rho + b_2 \rho^2 + b_3 \rho^3 +\dots\right],
\end{equation}
where the leading order term is the ideal gas contribution and the higher order terms are corrections due to interactions between particles, which become important with increasing density $\rho$ \cite{Kardar2007_particles}.
In particular, the virial coefficient $b_2$ captures the effect of two-body interactions.
For the Van der Waals equation of state, $b_2 = \rho_0^{-1} - a (k_B T)^{-1}$, which is positive for sufficiently high temperatures, meaning that two-body interactions contribute an additional pressure to the independent-particle ideal gas term.
This is much like the low-$\chi$ regime of polymer gels, which corresponds to the swollen phase.
However for low temperatures, the two-body terms contributes a negative pressure, which drives particles to condense to a liquid phase, much as the polymer gel deswells for high-$\chi$.
Note that at $\chi = 1/2$, the two-body term disappears in the Flory-Rehner model, describing a $\vartheta$-solvent; this is analogous to the Boyle temperature of the Van der Waals model $k_B T_b \equiv \rho_0/a$, for which $b_2 = 0$.

Whereas the virial expansion for the Van der Waals model has two parameters, $\rho_0$ and $a$, the virial expansion for the Flory-Rehner model only has one, $\chi$, which is taken to be independent of polymer volume fraction $\phi$.
However, measurements of $\chi$ have shown a nonlinear dependence on $\phi$ \cite{Flory1970,Hirotsu1987,Fernandez-Barbero2002,Lopez-Leon2007,AFN2009,Lietor-Santos2011,Nikolov2018}.
In general, the Flory parameter is a function of $\phi$ and can be expanded as a power series, namely
\begin{equation}\label{sub:floryParam_expansion}
\chi \rightarrow \chi_1 + \chi_2 \phi + \chi_3 \phi^2 + \dots
\end{equation}
and yields a more general virial expansion for the mixing contribution, 
\begin{equation}\label{eq:osm_mixing}
\Pi_{\rm mix} = \frac{k_B T}{v}\left[\left(\frac{1}{2} - \chi_1\right)\phi^2 + \left(\frac{1}{3} - \chi_2\right)\phi^3 + \dots\right] .
\end{equation}
Using this expansion, Erman and Flory \cite{ErmanFlory1986} have shown that a discontinuous transition as well as a critical point can be recovered by tuning $\chi_1$ and $\chi_2$, and ignoring all other terms, i.e., $\chi_{m>2} \equiv 0$.
In particular, acceptable fits to experimental swelling data can be found by fixing $\chi_2 > 1/3$ and varying $\chi_1$ with solvent quality, that is, taking $\chi_1$ to be a function of temperature only.
It is important to emphasize that this expansion is purely phenomenological and does not assign specific microscopic meaning to the values of $\chi_{m\geq 2}$ \cite{Hirotsu1994}.
In this phenomenological model of polymer gels, there are now two independent parameters, $\chi_1$ and $\chi_2$, in the virial expansion of the osmotic pressure, much like the two parameters of the Van der Waals model.

\begin{figure}
	\includegraphics[width=8.3cm]{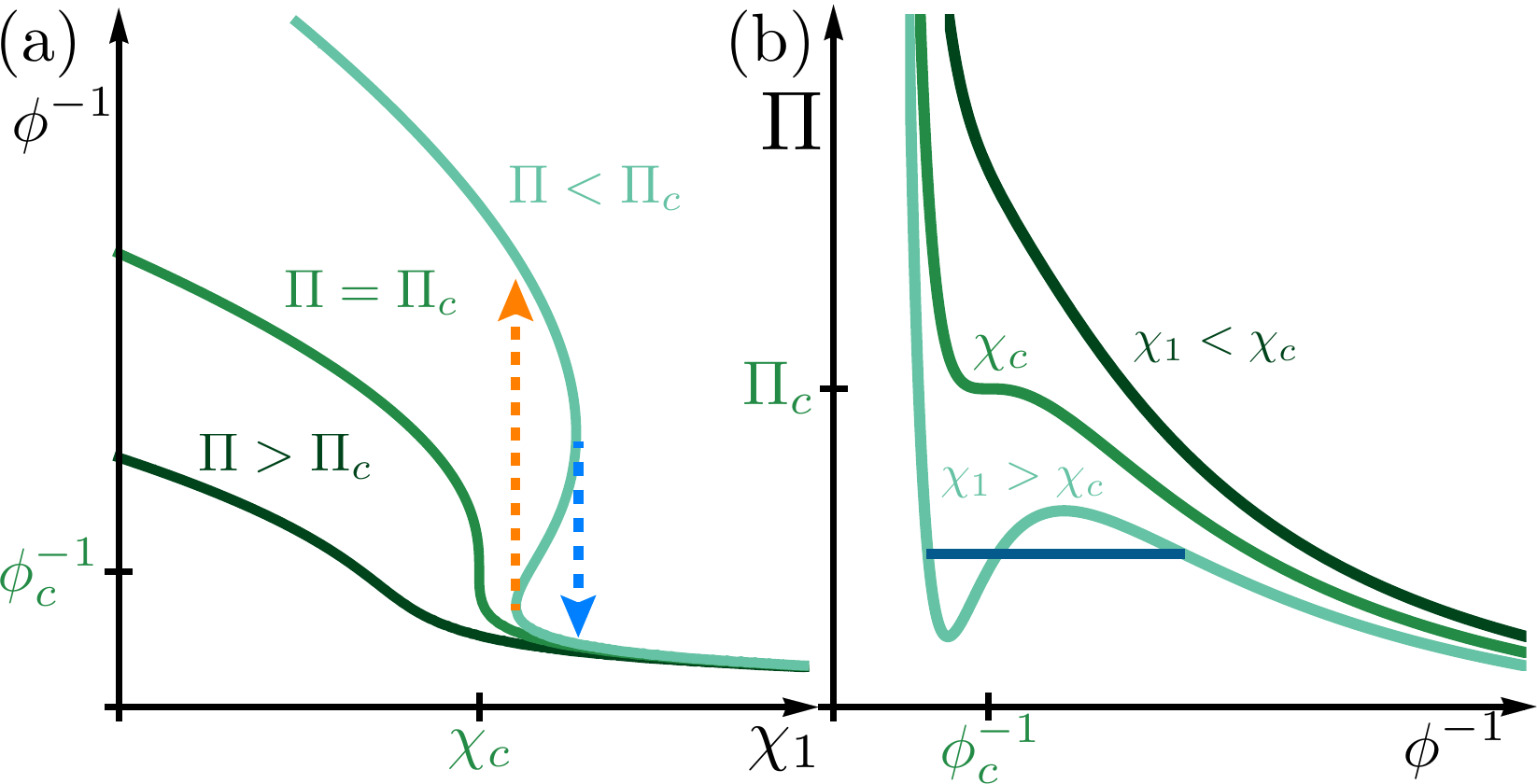}
	\caption{Plots of the modified Flory-Rehner equation of state with $\chi_2 = 0.56$ (see \cite{Hirotsu1994}). (a) Three osmotic isobars are shown as a function of $\chi_1$ with values of $\Pi$ greater than, equal to, and less than the critical osmotic pressure $\Pi$. Also shown are the critical value the Flory parameter $\chi_c$ and the critical polymer volume fraction $\phi_c$. Similar to the hysteresis seen in the Van der Waals model, there is hysteresis for $\Pi < \Pi_c$. (b) Three curves of constant $\chi_1$ are shown for $\chi_1 < \chi_c$, $\chi_1 = \chi_c$, and $\chi_1 > \chi_c$. The analogue of the Van der Waals loop is shown for the last, along with a rectifying line that describes coexistent volume fractions.}
	\label{fig:osm_press_disc}
\end{figure}

\begin{figure}
	\includegraphics[width=8.3cm]{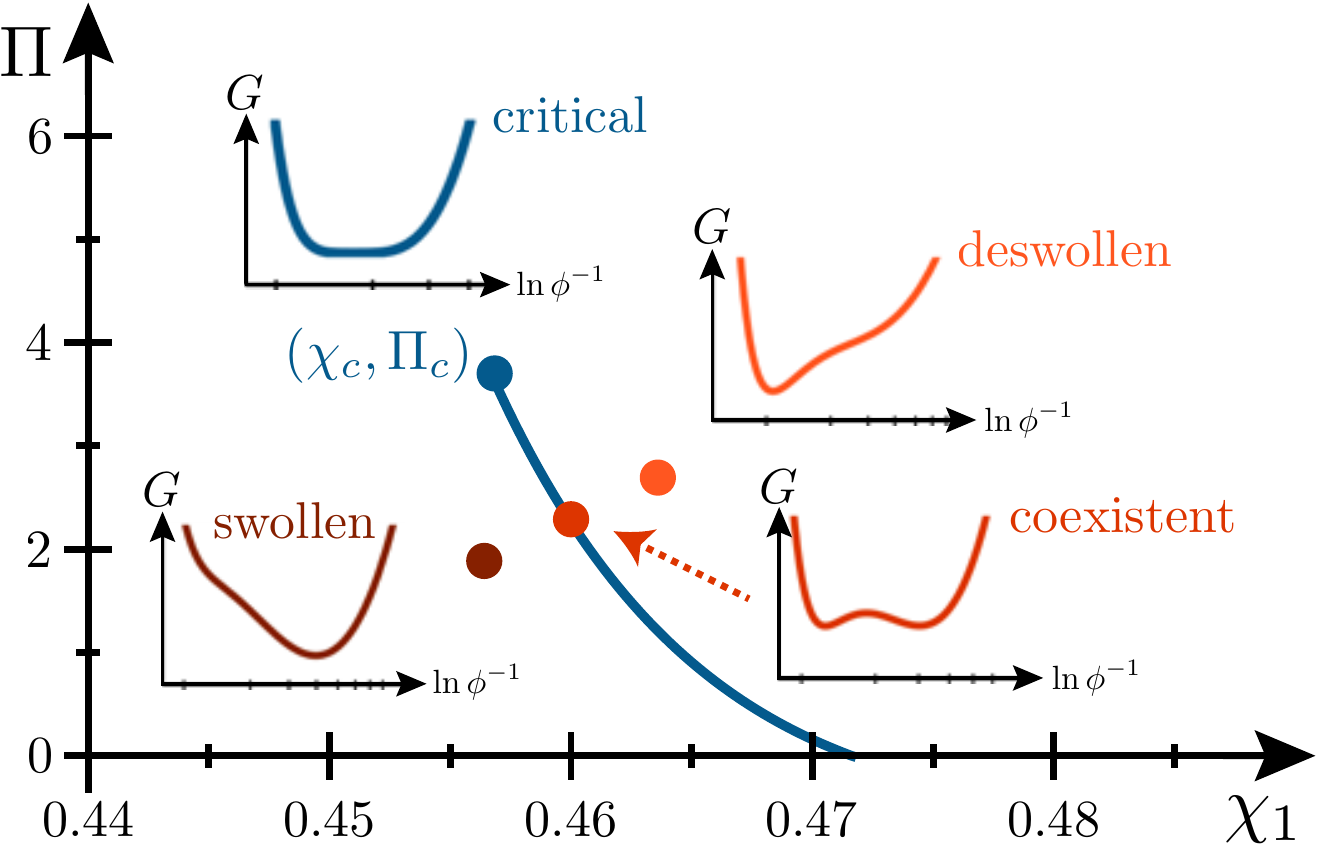}
	\caption{Phase diagram predicted by the Flory-Rehner model for a polymer gel with $\phi_0 = 0.1$, $\nu_0 v = 10^{-4}$, and $\chi_2 = 0.56$. The coexistence curve separates well-defined swollen and deswollen phases. Also shown is the Gibbs free energy $G(\phi^{-1})$ at the critical point, and along the coexistence curve, showing two local minima with the same value of $G$. Two plots of $G$ are shown at equal distances from the coexistence curve one in the swollen phase, the other in the deswollen phase, each displaying a single, well-defined absolute minimum.}
	\label{fig:gel_phase_diagram}
\end{figure}

Example osmotic isobars $(\Pi = {\rm const.})$ are shown in figure \ref{fig:osm_press_disc}(a) and curves of constant $\chi_1$, corresponding to isotherms, are shown in figure \ref{fig:osm_press_disc}(b).
Much like the analogous processes shown for the Van der Waals model in figure (\ref{fig:vdw_discontinuous}), there are continuous and discontinuous swelling processes, along with an identifiable critical point.
This critical point occurs at $\chi_1 = \chi_c$ and $\Pi = \Pi_c$, determined by the condition that the osmotic bulk modulus $K$ vanishes, i.e.,
\begin{equation}
\left(\frac{\partial \Pi}{\partial V}\right)_{\chi_1 = \chi_c} = -\frac{1}{v}\phi^2\left(\frac{\partial \Pi}{\partial \phi}\right)_{\chi_1 = \chi_c} = 0\, ,
\end{equation}
and that the osmotic pressure isotherm is flat at the critical point, i.e., 
\begin{equation}
\left(\frac{\partial^2 \Pi}{\partial V^2}\right)_{\chi_1=\chi_c} = \frac{1}{v^2}\phi^2\left(\frac{\partial}{\partial \phi}\phi^2\frac{\partial \Pi}{\partial \phi}\right)_{\chi_1=\chi_c} = 0\, .
\end{equation}
Note that these conditions are equivalent to requiring $(\partial \Pi/\partial \phi)_{\chi_c} = 0$ and $(\partial^2 \Pi/\partial \phi^2)_{\chi_c} = 0$.
Indeed, measurements of the bulk modulus $K$ near the phase transition show a dramatic softening when compared with the shear modulus $\mu$ of the gel \cite{Hirotsu1994,Sierra-Martin2011}.
The breakdown of thermodynamic stability can be rectified by the existence of equilibrium phase coexistence between swollen and deswollen regions at some constant value of $\Pi$ by the Maxwell construction, namely
\begin{equation}
\int_{\mathcal{P}_1} {\rm d}\Pi\, \phi^{-1}(\Pi) = 0 \, ,
\end{equation}
where $\mathcal{P}_1$ is portion of the $S$-shaped curve beginning and ending at the equilibrium value of $\Pi$ in the thermodynamically stable region, as shown in figure \ref{fig:osm_press_disc}(b).
Therefore, much like the phase diagram predicted by the Van der Waals model, there is a similar phase diagram for polymer gels described via the Erman-Flory model, with a coexistence curve with a terminal critical point, as shown in figure \ref{fig:gel_phase_diagram}.
The phase diagram of polymer gel swelling is similar to that of a fluid modeled by the Van der Waals equation of state.
Note that the coexistence curve is to the right of the critical point, whereas the curve in figure \ref{fig:maxwell_coex_vdw}(b) is to the left.
This is because the gas-like low-density phase, the swollen phase, occurs at low values of $\chi_1$, whereas the liquid-like high-density phase, the deswollen phase, occurs at higher values of $\chi_1$.
To understand the negative slope of the coexistence curve, we turn to the \emph{Clapeyron relation} \cite{Callen1985}, which relates this slope to the discontinuity change in volume and entropy that occurs when crossing the curve.
For fluids, the positive slope ${\rm d}P/{\rm d}T > 0$ shown in figure \ref{fig:maxwell_coex_vdw}(b) tells us that the increase in volume per particle that occurs when a liquid evaporates and becomes a gas accompanies a corresponding increase in entropy.
Conversely, the negative slope ${\rm d}\Pi/{\rm d}\chi_1 < 0$ for the gel coexistence curve indicates that there is a decrease in entropy as the gel passes from the deswollen phase to the swollen phase.
This is to be expected of elastomeric materials in general: an increase in volume stretches polymer chains, decreasing their configurational entropy.
Indeed, if one stretches a rubber band, decrease in the entropy of the polymer chains necessitates a passage of heat from the band into its surroundings, making it momentarily feel warm.

It is instructive to examine the behavior of the free energy near the coexistence curve and the critical point.
However, the Erman-Flory virial expansion has to first be incorporated into the mixing free energy $\Delta F_{\rm mix}$.
The power series expansion of the Flory parameter $\chi$ (\ref{sub:floryParam_expansion}) cannot be directly substituted into the mixing free energy as the calculated osmotic pressure $\Pi_{\rm mix}$ is inconsistent with that given in equation (\ref{eq:osm_mixing}).
Instead, start with the Erman-Flory mixing osmotic pressure in (\ref{eq:osm_mixing}) and integrate the relation $\Pi_{\rm mix} = \phi^2(\partial (\Delta\mathcal{F}_{\rm mix}/\phi)/\partial \phi)$ to find the mixing free energy density $\Delta\mathcal{F}_{\rm mix}$, up to an integration constant.
Since we require that the mixing free energy disappears when the gel is either purely polymer, $\phi = 1$, or in the limit where it is infinitely dilute, $\phi \rightarrow 0$, the integration constant is fixed, yielding
\begin{equation}\eqalign{
\Delta\mathcal{F}_{\rm mix} &= \frac{k_B T}{v}\bigg[(1-\phi){\rm ln}(1 - \phi)\\
&\mkern+150mu + \sum_{m = 1}^{\infty}\frac{\chi_m}{m}\phi(1-\phi^m)\bigg]\, ,
}\end{equation}
where the original form of the mixing free energy (\ref{eq:mixing_free energy}) is recovered if only $\chi_1$ is retained.
With this alteration to the total free energy $\Delta F$, there are values of $\chi_1$ and $\chi_2$ that cause $\Delta F$ to be a non-convex function of the polymer volume fraction $\phi$.
This has the implication that the osmotic equilibrium
\begin{equation}
\left(\frac{\partial \Delta F}{\partial \phi^{-1}}\right)_{\chi_1,\Pi} = -v\Pi \, .
\end{equation}
may be satisfied for multiple values of $\phi$.
Transforming to an analogue of the Gibbs free energy $G = \Delta F + \Pi V$, where $V = v\phi^{-1}$, the osmotic equilibrium condition is
\begin{equation}
\left(\frac{\partial G}{\partial \phi^{-1}}\right)_{\chi_1,\Pi} = \left(\frac{\partial \Delta F}{\partial \phi^{-1}}\right)_{\chi_1,\Pi} + v\Pi = 0
\end{equation}
so that coexistent equilibrium volume fractions correspond to local minimia of $G(\phi)$, as illustrated in figure \ref{fig:gel_phase_diagram}.
Plots of $G(\phi)$ are shown for sample values of $(\chi_1,\Pi)$ on the phase diagram \ref{fig:gel_phase_diagram}; note that the inverse polymer volume fraction $\phi^{-1}$ is plotted on a logarithmic scale, reflecting the large scale of the volume changes that occur.
At the critical point $(\chi_c,\Pi_c)$, the free energy minimum is broad, reflecting vanishing curvature $(\partial^2 G/\partial (\phi^{-1})^2)_{\chi_1,\Pi} = 0$, i.e., diverging isothermal compressibility.
Along the coexistence curve, two degenerate minima of $G$ emerge, corresponding to the two coexistent phases.
As predicted from Landau theory, via the model free energy (\ref{eq:free energy-Landau}), close to the critical point, these two minima emerge symmetrically from the equilibrium value of $\phi^{-1}$ at the critical point and their separation grows continuously as the parameters $(\chi_1,\Pi)$ are tuned along the coexistence curve; this is reflective of a continuous phase transition at the critical point.
Instead, if $(\chi_1,\Pi)$ are tuned transverse to the coexistence curve, the degeneracy of the free energy minima is lifted and there is an absolute minimum either for low $\phi^{-1}$ (deswollen phase) or for high $\phi^{-1}$ (swollen phase).
In this case, as the coexistence curve is reached, the free energy difference between the two minima goes to 0, marking the cross-over between the two phases.
This cross-over is a discontinuous jump between values of $\phi^{-1}$ that mark the local minima of $G$, indicating a first-order phase transition.
However, physically, the gel may not immediately switch to the new absolute minimum.
For example, if the gel is brought from the swollen phase across the coexistence curve to the deswollen phase, then the swollen phase still has a local free energy minimum; it is metastable.
There is a free energy barrier for the gel to leave this metastable equilibrium and attain the global free energy minimum.
Given sufficient time, thermal fluctuations will cause the gel to surmount this barrier.
However, practically, the gel remains in the swollen phase until $(\chi_1,\Pi)$ is tuned sufficiently far from the coexistence curve so that the free energy barrier disappears.
Therefore, the observed values $(\chi_1^*, \Pi^*)$ at which the transition occurs are not typically on the coexistence curve, but rather to the right of it.
Similarly, for the reverse process of starting from the deswollen phase and passing to the swollen phase, the transition typically occurs to the left of the coexistence curve.
These states that linger past the phase coexistence transition are the polymer gel analogies of the superheated liquid and supercooled gas states of a fluid.
This explains and generalizes the hysteresis that is shown in the osmotic isobars of figure \ref{fig:osm_press_disc}.

\subsection{Fluctuations and criticality in polymer gels}

\begin{figure}
	\includegraphics[width=7cm]{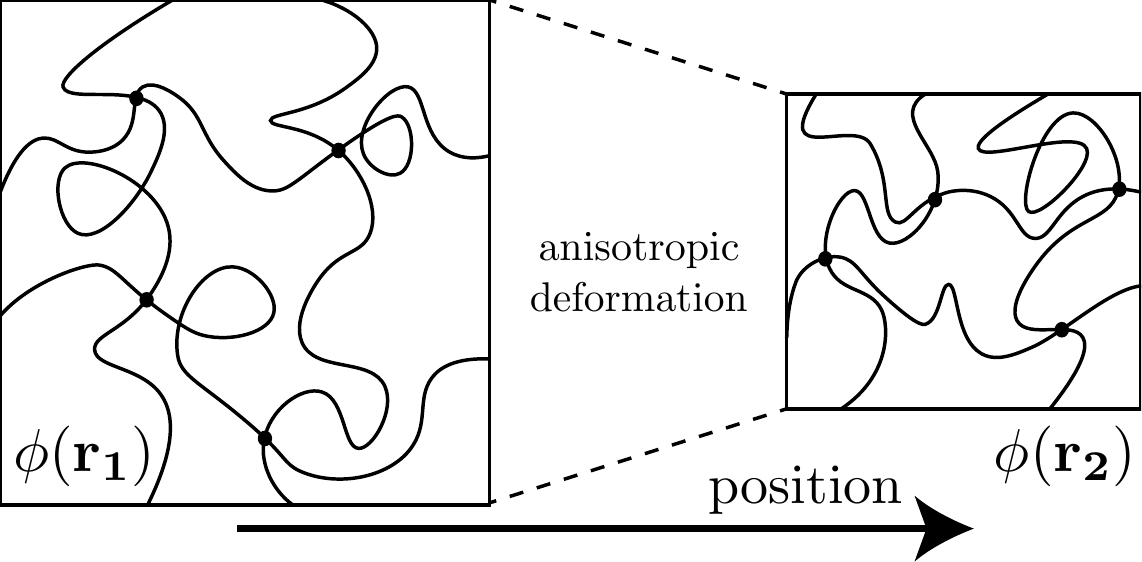}
	\caption{Cartoon of a polymer network with different polymer volume fractions $\phi$ at points $\mathbf{r_1}$ and $\mathbf{r_2}$. In order for the polymer network to interpolate continuously in size between these two points, there must be anisotropic deformation.}
	\label{fig:inhomogeneous_swell}
\end{figure}

So far, we have discussed how the swelling thermodynamics of polymer gels share features in common with fluids.
In discussing isotropic gels that undergo homogeneous changes in state, we have ignored the rigidity of these materials.
At the critical point and along the coexistence curve, however, the gel is inhomogeneous, either undergoing large thermal fluctuations or possessing two equilibrium volume fractions.
Here, the rigidity of the gel proves to have a profound effect on the equilibrium thermodynamics: since the polymer network is maintained by permanent chemical cross-links, its connectivity should not change under deformation.
Therefore, unless the gel is torn, the polymer network should remain contiguous even while supporting coexistent phases, as shown in figure \ref{fig:inhomogeneous_swell}.
Spatial inhomogeneities in $\phi$ therefore result in anisotropic stretching of the polymer network.
The phase diagram in figure \ref{fig:gel_phase_diagram} is thus a simplistic picture of swelling thermodynamics, starting from the placement of the critical point.
The remainder of this Topical Review is concerned with the consequences of rigidity for the swelling phase behavior.

We begin with the fate of thermodynamic stability and how elastic effects alter the critical point of the gel.
Consider again the picture presented by Le Chatelier's principle, where spatial variation in density at positive compressibility results in a spatial variation in pressure, which causes mass-flow that corrects the spatial variation.
Conversely, for negative compressibility, low-density regions have increased pressure compared with high-density regions, leading to runaway mass-flow that causes density variations to grow.
In an isotropic polymer gel, this runaway mass-flow that occurs for negative compressibility leads to spatial variation in the polymer volume fraction $\phi$.
In order to remain contiguous, the polymer network deforms inhomogeneously, which has a free energy cost that is not accounted for in the classical analysis presented so far.
In fact, \emph{the gel is stable for small, negative values of the compressibility}, as confirmed by light-scattering measurements \cite{Tanaka1977}.
It should be noted that materials with negative bulk modulus are generically unstable, since such a material will increase in volume under applied pressure.
However, since this type of instability is predicated on the absence of inhomogeneity, it is difficult to observe in bulk materials \cite{Moore2006,Lakes2008}.

In order to concretely formulate this discussion, let us construct the coarse-grained free energy $F$ of an inhomogeneous polymer gel, given a reference state $\mathcal{R}$ with homogeneous polymer volume fraction $\phi_0$.
This free energy is given by
\begin{equation}\label{eq:gel_inh}
F = \int_{\mathcal{R}} {\rm d}^3 r\, \left[\frac{1}{2}c\left|\bm{\nabla}\phi\right|^2 + \tilde{\mathcal{F}}(T,\Lambda,\phi)\right] \, ,
\end{equation}
where the first term represents the free energy cost of spatial variations in $\phi$ and is related to the correlation length $\xi$ of fluctuations in the volume fraction $\phi$, i.e., $c \sim \xi^2$.
The free energy density $\tilde{\mathcal{F}}$ is a density with respect to the reference state $\mathcal{R}$ rather than the current state of the gel; the two are related by $\tilde{\mathcal{F}} = (\phi_0/\phi)\mathcal{F}$.
Fluctuations in the state of the gel take points $\mathbf{r}$ in the reference state to points $\mathbf{R} = \mathbf{r} + \delta\mathbf{r}$ in the current state of the gel.
The homogeneous gel is stable if these small fluctuations always increase the total free energy.
Therefore, stability is determined by the second variations of the free energy $\delta^2 F$.

The fluctuations $\delta\mathbf{r}$ are represented by a displacement field $\mathbf{u}(\mathbf{r}) \equiv \mathbf{R}(\mathbf{r}) - \mathbf{r}$.
The deformation matrix can therefore be written as
\begin{equation}
\Lambda_{ij} = \partial_j R_i = \delta_{ij} + \partial_j u_i \, ,
\end{equation}
from which the polymer volume fraction $\phi = \phi_0/({\rm det}\, \Lambda)$ can be approximated as
\begin{equation}
\phi \approx \phi_0\left(1 - \partial_i u_i\right) \, ,
\end{equation}
so the variation of the volume fraction is given by $\delta \phi = -\phi_0 \partial_i u_i$.
Therefore, the second variation of the free energy is given by
\begin{equation}
\delta^2 F = \frac{1}{2}\int_{\mathcal{R}}{\rm d}^3 r\big[c\phi_0^2(\partial_{ij}u_j)^2 + E_{ijkl}(\partial_iu_j)(\partial_ku_l) \big] \, ,
\end{equation}
where $E_{ijkl}$ is the elasticity tensor, found by expanding $\tilde{\mathcal{F}}$ to second order in $\partial_i u_j$, and is thus a function of temperature $T$ and volume fraction $\phi_0$.
As the reference configuration corresponds to a homogeneous, isotropic gel, the elasticity tensor has the form
\begin{equation}
E_{ijkl} = \mu(\delta_{ik}\delta_{jl} + \delta_{il}\delta_{jk}) + \left(K - \frac{2}{3}\mu\right)\delta_{ij}\delta_{kl} \, ,
\end{equation}
where $\mu$ is the shear modulus and $K$ is the bulk modulus.
Note that the bulk modulus is given by $K = \phi_0^2 (\partial^2\tilde{\mathcal{F}}/\partial \phi^2)|_{\phi_0}$, which can adopt negative values when the mixing part of the free energy is near a local maximum  with respect to $\phi_0$.
If we take the approximation that the gel occupies all of space, then it is useful to use the Fourier representation of the displacement field,
\begin{equation}
u_i(\mathbf{r}) = \int_{-\infty}^{\infty}\frac{{\rm d}^3 k}{(2\pi)^3} e^{i\mathbf{k}\cdot\mathbf{r}}u_{i,\mathbf{k}} \, ,
\end{equation}
and the second variation in the free energy becomes: 
\begin{equation}\eqalign{
\delta^2 F = &\frac{1}{2}\int\frac{{\rm d}^3 k}{(2\pi)^3} u_{i,\mathbf{k}}u_{j,-\mathbf{k}}\bigg[c\phi_0^2|\mathbf{k}|^2k_ik_j \\
&\mkern+64mu + \mu |\mathbf{k}|^2\delta_{ij} + \left(K + \frac{1}{3}\mu\right)k_ik_j\bigg] \, .
}\end{equation}
We can then decompose the displacement field $\mathbf{u}$ into a longitudinal part $u^{\ell}_{\mathbf{k}}\hat{\mathbf{k}}$ and a transverse part $\mathbf{u}^t_{\mathbf{k}}\times\hat{\mathbf{k}}$, which yields two independent contributions to $\delta^2F$, namely that due to longitudinal fluctuations,
\begin{equation}
\delta^2 F^{\ell} = \frac{1}{2}\int\frac{{\rm d}^3 k}{(2\pi)^3} u_{\mathbf{k}}^{\ell}u_{-\mathbf{k}}^{\ell}|\mathbf{k}|^2\bigg[c\phi_0^2|\mathbf{k}|^2 + K + \frac{4}{3}\mu\bigg] \, ,
\end{equation}
and that due to transverse fluctuations,
\begin{equation}
\delta^2 F^{t} = \frac{1}{2}\int\frac{{\rm d}^3 k}{(2\pi)^3} \mathbf{u}_{\mathbf{k}}^{t}\cdot\mathbf{u}_{-\mathbf{k}}^{t}|\mathbf{k}|^2\mu \, .
\end{equation}
There is a correlation length $\xi_{\ell} \equiv [c \phi_0^2/(K + 4\mu/3)]^{1/2}$ associated with longitudinal fluctuations, whereas the transverse fluctuations do not have an associated length scale.
Since the longitudinal fluctuations correspond to fluctuations in the polymer volume fraction $\phi$, the longitudinal fluctuation correlation length $\xi_{\ell}$ also describes the polymer volume fraction correlation length.
Therefore, we can conclude that whereas the critical point for fluids is at vanishing bulk modulus $K = 0$, \emph{the critical point for polymer gels is at vanishing longitudinal modulus} $K + 4\mu/3 = 0$ \cite{Sekimoto1993}.
Note that for the classical rubber elasticity used in the Flory-Rehner model, the shear modulus $\mu$ is always positive.
However, as we have shown, the osmotic bulk modulus $K = \phi \partial \Pi/\partial \phi$ can be negative.
Whereas a negative bulk modulus marks the loss of stability for fluids and homogeneous elastic solids, polymer gels lose thermodynamic stability only when $K < -4\mu/3$.
The stabilizing effect of the shear rigidity near the critical point is an effect of inhomogeneity produced by thermal fluctuations.
It is remarkable that the critical behavior of polymer gels, amorphous solids primarily composed of liquid, should share some fundamental similarities with exotic solid state alloys that undergo structural phase transitions, such as ferroelastic and martensitic transformations, and are also capable of exhibiting negative elastic moduli \cite{Salje1991}.

As we have shown, there are features of the swelling behavior of gels that can be understood in analogy with the phase behavior of fluids.
However, as we have alluded to, there are key differences due to effects of shear rigidity.
While there is loss of stability for longitudinal fluctuations when $K + 4\mu/3 \leq 0$, it should be remembered that the transverse fluctuations retain stability since $\mu > 0$ \cite{Onuki1988}.
The fact that critical fluctuations in the gel are not seen at $K = 0$ means that the gel cannot be simply modeled by the Landau theory used for fluids, summarized in equation (\ref{eq:free energy-Landau}).
Instead, as shown by Golubovi\'c and Lubensky in \cite{Lubensky1989}, there is an additional term of the form $\mu(\int{\rm d}^3 x \phi)^2$ that must be added to the model free energy.
Since this term is non-local, meaning that it cannot be folded into a free energy density, it yields a departure from the Landau theory of simple isotropic fluids.
Furthermore, it is long-ranged, meaning that the equilibrium value of the field $\phi(\mathbf{x})$ depends on the value of $\phi$ at all other points in the gel.
More concretely, we can see the effect of this long-range interaction when one considers equilibrium phase coexistence between swollen and deswollen phases of gels, where the $\emph{gel shape}$ plays a role in determining the equilibrium values of $\phi$.
This is a notable departure from the behavior that we expect from fluids, where the density of the fluid at any point in its bulk is independent of the shape of the container.\footnote{The density of a fluid, as well as its fluctuations, do depend on shape when one considers finite samples and points near the boundary of the fluid \cite{Cardy1996}. However, for fluids, which have only local interactions, these boundary effects decay as points in the bulk are considered, and are therefore distinctly different than the shape-dependence that is seen in gels.}

\section{\label{sec:arrested_deswelling} Arrested deswelling: an exotic way to phase-separate} 

Whilst the transition between swollen and deswollen gels is in many ways similar to the transition between gas and liquid phases, shear rigidity alters some key characteristics of the transition.
As demonstrated in the previous section, the definition of the swelling critical point for gels is modified due to the stabilizing effect that shear rigidity has on thermal fluctuations.
Shear rigidity also has profound consequences for the phase-coexistent equilibria of polymer gels, as the mechanical equilibrium condition requires balance of a generally anisotropic stress tensor.
Thus, particularly in the case of macrophase separation, where macroscopic domains of swollen and deswollen gel give rise to such an anisotropic stress distribution, the result is a deformation of the gel at similarly long length scales.
As the response of a solid to an applied stress distribution depends on the shape of the solid, we expect that the conditions and configurations of the phase coexistent equilibrium states depend on the shape of the gel.

\begin{figure}
\centering
\includegraphics[width=8.3cm]{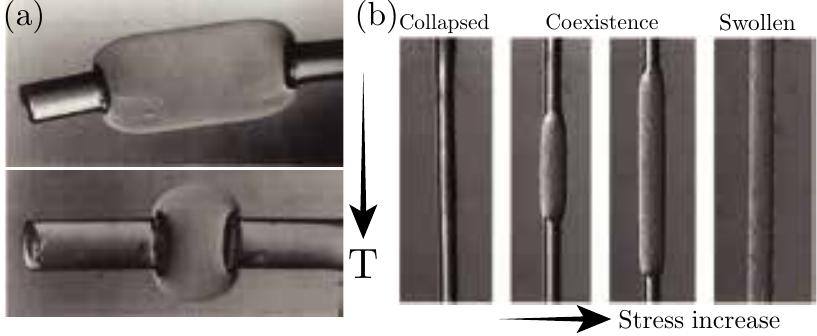}
\caption{\label{fig:cylinder} (a) Phase coexistence in ionized pNIPAM cylindrical gel at two different temperatures close to the transition temperature. The sample in the bottom image is $0.7^{\circ}{\rm C}$ warmer than the top panel. The diameter of the swollen portion is around 3 mm. Figure reproduced from \cite{Hirotsu1994}. (b) Stress-induced swelling and phase coexistence in neutral cylindrical gels. Figure reproduced from \cite{Suzuki1999}.}
\end{figure}

Experiments on cylindrical samples of an \emph{ionic polymer gel} have demonstrated phase coexistence at constant ambient osmotic pressure \cite{Hirotsu1994}.
The mass of swollen compared to deswollen gel in a single sample is controlled by temperature, which is similar to coexistent gas and liquid phases.
However, there is indeed a noticeable dependence on shape.
If an unconstrained cylinder-shaped gel, initially in the swollen phase, is brought to the coexistence regime, the deswollen phase nucleates at the two ends, as shown in figure \ref{fig:cylinder}(a); similarly, a swollen phase will grow from the ends of a deswollen cylinder.
However, if such a gel is instead stretched uniaxially, the new phase nucleates from the center, as shown in figure \ref{fig:cylinder}(b).

Demonstrating phase coexistence in \emph{neutral gels} has proven elusive due to the relatively narrow temperature range of equilibrium phase coexistence, spanning less than $0.1^{\circ}{\rm C}$, at constant osmotic pressure $\Pi$ \cite{Hirotsu1994}.
However, it has been shown that this temperature range can be broadened through the application of mechanical stress.
For example, equilibrium phase coexistence has been demonstrated in cylindrical samples of neutral gels that are stretched via a mechanical constraint applied to the ends of the gel \cite{Sekimoto1989,Onuki1989,Suzuki1999}, as shown in figure \ref{fig:cylinder} (b).
This is another realization of the effect of shear rigidity: since the equation of state for the osmotic pressure depends on the deformation matrix $\Lambda$, osmotic isobars are affected by anisotropic deformation.

Phase coexistence has been carefully achieved at constant osmotic pressure.
However, phase coexistence at \emph{constant volume} remains unexplored, likely because it is potentially challenging to fabricate a volume-constraining material grafted onto the boundary of the gel.
One alternative approach is to take advantage of the rich deswelling kinetics of polymer gels.
Matsuo and Tanaka \cite{Matsuo1988} studied the equilibration of 0.1 to 1 millimeter-radius spherical samples of neutral pNIPAM under heating and cooling through the first-order phase transition at zero osmotic pressure.
In their experiments, they observed that spheres, initially swollen at low temperature, that are \emph{rapidly heated} past the deswelling transition temperature at $\sim\mkern-4mu 32^{\circ}$C have a two-step deswelling process.
Immediately after the rapid heating process, there is some deswelling over the first few seconds, after which deswelling halts for tens of seconds.
During this pause in deswelling, referred to as the ``plateau period,'' there is thin skin of deswollen-phase gel at the boundary of the sample that is effectively impermeable to the solvent.
Towards the end of the plateau period, the deswollen skin appears thinner in some regions and thicker in others.
These thicker regions form a network-like structure of edges and vertices, with the thinner regions as the faces, reminiscent in morphology to a foam, that spans the surface of the gel.
The resulting inhomogeneous stress distribution about the skin causes the thinner-skinned regions to balloon outward, whereas the thicker-skinned regions are creased inward.
As a result of the ballooning, the thinner-skinned regions experience a large extensional strain tangential to the surface of the gel, facilitating the passage of solvent out of the gel.
This allows the gel to resume deswelling, eventually equilibrating and becoming a spherical deswollen gel.
Similar experiments on cylindrical \cite{Matsuo1992_cyl,Suzuki1999_JChemPhys,Bai2000,Boudaoud2003mechanical} and toroidal \cite{Chang2018} samples of pNIPAM gel reveal even richer phenomena.
In addition to the eventual formation of similar balloon-shapes on the surface of the gel near the end of the plateau period, which lasts for minutes in the experiments on toroidal gels, there is also a dramatic shape change, where the torus adopts, in some cases, a saddle-like or ``Pringle\textsuperscript{TM}''-like morphology.

\begin{figure}
	\includegraphics[width=8.3cm]{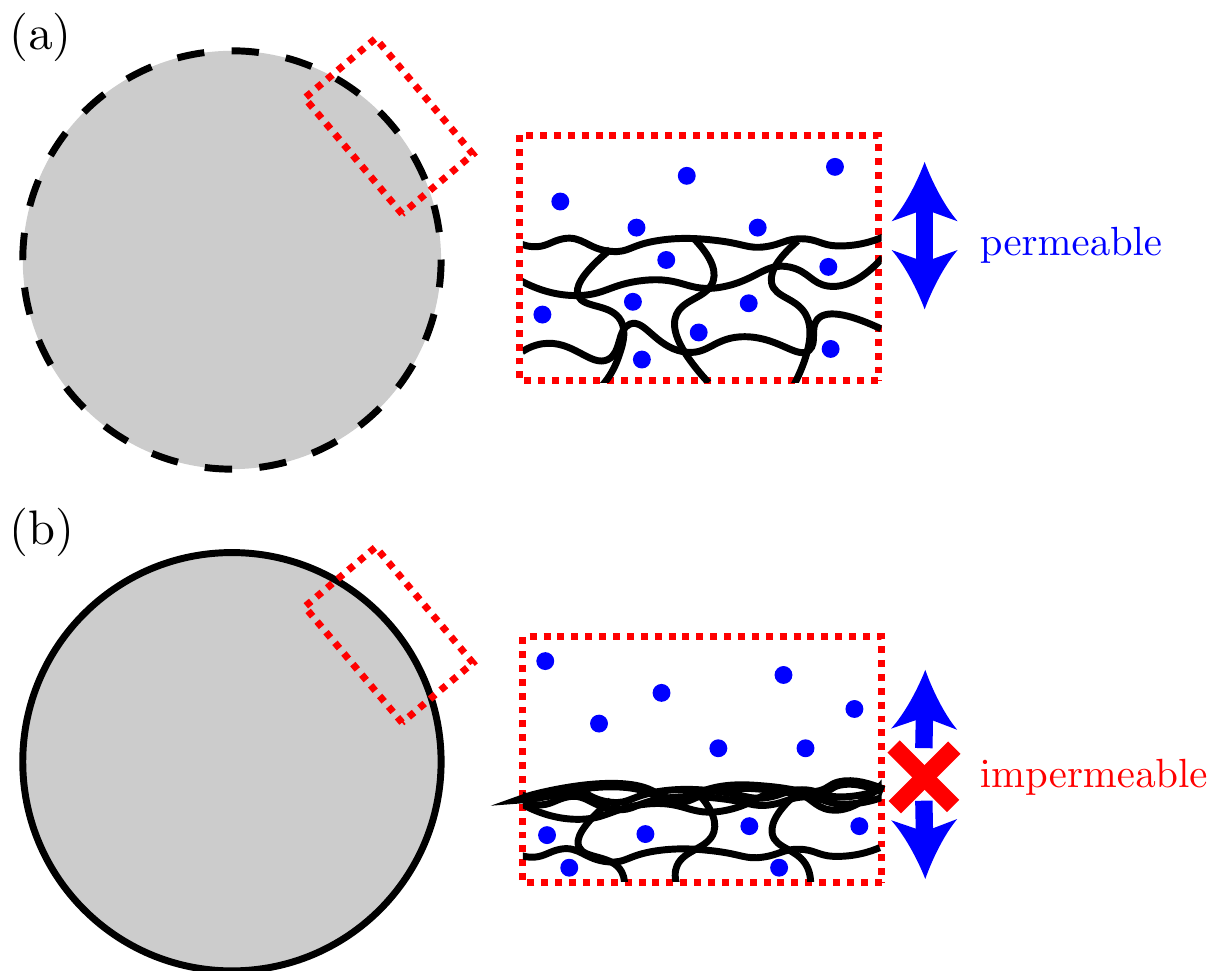}
	\caption{A swollen-phase polymer gel sphere shown (a) in equilibrium with the surrounding solvent bath due to its ability to pass solvent through its permeable boundary and (b) after rapid heating, where solvent exchange is cut off due to the presence of a dense, impermeable, thin, deswollen skin. In (b), due to the inability to exchange solvent, the swollen interior of the gel is out of equilibrium with the surrounding solvent.}
	\label{fig:impermeable}
\end{figure}

The plateau period after rapid heating is a prolonged time during which the volume of the polymer gel is essentially unchanging.
During this time, the interior of the gel is trapped in the swollen phase, even though it is at a temperature where the deswollen phase is an absolute minimum of the free energy.
If we ignore the small initial solvent loss in forming the very thin deswollen skin, this interior is effectively under a constant-volume constraint and is in a state that is far from the free energy minimum.
Much like a fluid in similar conditions, we expect that the gel will phase-separate, forming a high polymer volume fraction region at the expense of also forming a low polymer volume fraction region.
Since the plateau period is a relatively long-lasting part of the deswelling kinetics of the polymer gel, it is reasonable to assume that the phase-separation will approach an equilibrium phase-coexistent state before solvent starts to leak out at an appreciable rate.
Equilibrium phase coexistence in this situation is made possible because the thin deswollen skin is effectively impermeable, so the swollen interior is out of chemical equilibrium with the solvent bath, as illustrated in figure \ref{fig:impermeable}.
Thus, the rapid heating accomplishes (i) a quench across the first-order phase transition, which takes the gel far from  global equilibrium, (ii) the introduction of a volume constraint for the swollen interior, and (iii) a prolonged period during which the swollen interior can reach a phase separated state.


\subsection{Revisiting the common tangent construction \label{sec:sphere}}

To determine the phase coexistent equilibria for polymer gels of some given geometry, we need to minimize the total deformation free energy $\Delta F$ under the constraint that the total volume $V$ remains constant.
For convenience, take the reference configuration $\mathcal{R}$ to be the swollen gel immediately before phase-separation, where the gel is at a homogeneous polymer volume fraction $\phi_0$.
We adopt the following general form for the free energy density $\Delta\tilde{\mathcal{F}}$ of the gel in the reference state,
\begin{equation}
\Delta\tilde{\mathcal{F}}(T,\Lambda,\phi;\mathbf{r})  = \frac{1}{2}\mu_0 {\rm tr}\,\Lambda^T\Lambda + \Delta\hat{\mathcal{F}}(T,\phi;\mathbf{r})  \, ,
\end{equation}
where $\mu_0$ is the shear modulus corresponding to the reference state. 
Note that  the first term is a general form for the elastic contribution, originating from classical theory of rubber elasticity \cite{Treloar1975}, appearing in the Flory-Rehner model, as well as more sophisticated descriptions \cite{Panyukov1996}.
The free energy density $\Delta\hat{\mathcal{F}}(T,\phi)$ is the remaining part of the free energy density that only depends on the polymer volume fraction $\phi(\mathbf{r})$ at each point in space.
Note that in expressing $\Delta\tilde{\mathcal{F}}$ in terms of the Flory-Rehner theory, $\Delta\hat{\mathcal{F}}$ contains the mixing part of the free energy, as well as the additional terms of the Flory-Wall network entropy describing translational degrees of freedom of the chains; the shear modulus is given by $\mu_0 = n^0_{ch} k_B T$, where $n^0_{ch}$ is the chain density in the reference state of the gel.
The total free energy $\Delta F$, given by
\begin{equation}
\Delta F = \int_{\mathcal{R}}{\rm d}^3 r\left[\Delta\tilde{\mathcal{F}}(T,\Lambda,\phi;\mathbf{r}) + p\left(\frac{\phi_0}{\phi(\mathbf{r})} - 1\right)\right]\, ,
\end{equation}
incorporates a constant-volume constraint, enforced by the Lagrange multiplier $p$.
Since we seek a description of macroscopic phase-separation, we do not include the free energy cost of microscopic variation, $c|\bm{\nabla}\phi|^2$, as in equation (\ref{eq:gel_inh}); this amounts to neglecting thermal fluctuations of the state functions, which is justified as long as the gel is sufficiently far from the critical point.

Now consider phase-separation that forms a solvent-poor region with polymer volume fraction $\phi_p$ at the cost of forming a solvent-rich region with polymer volume fraction $\phi_r$.
Starting from a swollen phase of $0 < \phi_0 \ll 1$, we expect that a quench deep into the deswollen phase will result in $\phi_p^{-1} \ll \phi_0^{-1} \lsim \phi_r^{-1}$.
Furthermore, the solvent-poor and solvent-rich regions are distinct phases assuming that the gel is far enough from the critical point.
Therefore, we expect that $\phi_p$ and $\phi_r$ deviate very little from their representative values in their respective regions.
Ignoring these deviations, the volume conservation condition is given by
\begin{equation}
f\left(\frac{\phi_0}{\phi_p} - 1\right) + (1 - f)\left(\frac{\phi_0}{\phi_r} - 1\right) = 0 \, ,
\end{equation}
where $f$ is the fraction of gel in the reference configuration that will be solvent-poor.
We can solve for the fraction $f$, yielding
\begin{equation}\label{eq:gel_lever}
f = \frac{\phi_r^{-1} - \phi_0^{-1}}{\phi_r^{-1} - \phi_p^{-1}} \, ,
\end{equation}
which is identical to the Lever Rule of phase coexistence in fluids, equation (\ref{eq:fluid_lever}).

\begin{figure*}
	\includegraphics[width=\textwidth]{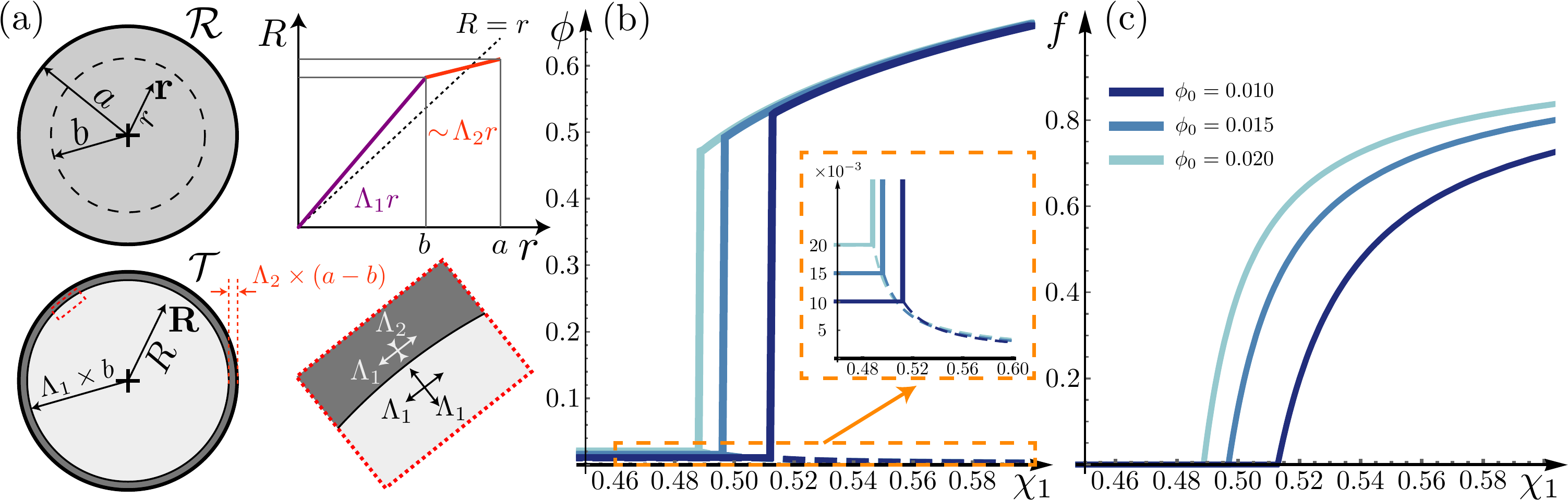}
	\caption{(a) A sphere immediately after rapid heating with a deswollen skin is shown in the reference state $\mathcal{R}$, where the interior is at a homogeneous polymer volume fraction $\phi_0$ and after phase-separation, in target state $\mathcal{T}$, where the solvent-rich spherical portion is at $\phi_r <\phi_0$ and the solvent-poor spherical shell is at $\phi_p>\phi_0$. In the reference state, points are described in spherical coordinates by $\mathbf{r} = r\mathbf{\hat{r}}$, where the outer radius of the sphere is at $r = a$ and the location of the phase-interface is $r = b$. Note that the schematic is not to scale; we expect $(a-b)\ll b$. In the target configuration, points are given by $\mathbf{R} = R\mathbf{\hat{R}}$. A simple linear approximation to $R(r)$ is also shown with slopes $\Lambda_1 > 1$ and $\Lambda_2 < 1$. Continuity of this function ensures that the components of the deformation matrix tangent to the phase interface, $\Lambda_{\Theta\theta}$ and $\Lambda_{\Phi\phi}$, are given by $\Lambda_1$, as shown in the inset to the lower right. (b) Equilibrium values of $\phi$ for $\phi_0 = 10, 15, 20 \times 10^{-3}$ as a function of $\chi_1$. For small enough values of $\chi_1$, there is no phase-separation, whereas above transition values of $\chi_1$, there are coexistent values of $\phi$ that are in equilibrium with each other; values of $\phi_p$ are easily seen (solid). The inset shows a blow-up of the small-$\phi$ region, showing equilibrium values of $\phi_r$ (dashed). (c) Values of the fraction $f$ occupied by solvent-poor gel, as predicted by the lever rule, equation (\ref{eq:gel_lever}), as a function of $\chi_1$. Note that at the transition values of $\chi_1$, $f$ increases continuously from 0.}
	\label{fig:sphere_equilibrium}
\end{figure*}

Now consider a spherical gel, such as that in the experiment of Matsuo and Tanaka \cite{Matsuo1988}.
We use spherical coordinates $(r,\theta,\phi)$ in the reference configuration and $(R,\Theta,\Phi)$ in the target configuration, such that reference configuration points are $\mathbf{r}  = r\hat{\mathbf{r}}$ and target configuration points are $\mathbf{R} = R\hat{\mathbf{R}}$.
Due to the symmetry of the sphere, we expect that the phase-coexistent equilibrium maintains the spherical symmetry; broken symmetry may arise from instability of this equilibrium.
Therefore, points $\mathbf{R}$ in the phase coexistent state depend only on the radial coordinate $r$ in the reference state; the deformed sphere is expressed entirely in terms of $R(r)$, where $R$ is the radial coordinate of the phase coexistent configuration.
Consequently, deformations maintain conformal symmetry so that we can take $\Theta = \theta$ and $\Phi = \phi$ and $\hat{\mathbf{R}} = \hat{\mathbf{r}} = (\sin\theta\cos\phi,\sin\theta\sin\phi,\cos\theta)$.
The transpose of the deformation matrix $\Lambda$, as expanded in spherical coordinates, is given by
\begin{equation}\eqalign{
\Lambda^T &=  \left(\hat{\mathbf{r}}\frac{\partial}{\partial r} + \frac{\hat{\bm{\theta}}}{r}\frac{\partial}{\partial \theta} + \frac{\hat{\bm{\phi}}}{r \sin\theta}\frac{\partial}{\partial \phi}\right)\otimes R(r)\hat{\mathbf{R}}(r,\theta,\phi) \\
&= \hat{\mathbf{r}}\otimes\hat{\mathbf{R}}\frac{{\rm d}R}{{\rm d}r} + (\hat{\bm{\theta}}\otimes\hat{\bm{\Theta}}+\hat{\bm{\phi}}\otimes\hat{\bm{\Phi}})\frac{R}{r} \, ,
}\end{equation}
with components,
\begin{equation}
\Lambda_{Rr} = \frac{{\rm d}R}{{\rm d}r};\; \Lambda_{\Theta\theta} = \Lambda_{\Phi\phi} = \frac{R}{r}\, .
\end{equation}
In the reference geometry, let $a$ be the outer radius of the sphere and let $b$ represent the radius of the interface between solvent-rich and solvent-poor regions, as in figure \ref{fig:sphere_equilibrium}(a).
Given that the sphere starts swollen and that in the experiments, there is a thin deswollen-phase skin on the boundary of the sphere, we will assume that the solvent-poor region grows from the deswollen-phase skin, inward.
Therefore we will take the region $r < b$ to be solvent-rich and the region $b < r < a$ to be solvent-poor.
In general, $R(r)$ is a continuous function since the gel must remain connected throughout the deformation.
However, it is expected that there is a change in the behavior of ${\rm d}R/{\rm d}r$ at the interface $r = b$ because whereas the solvent-rich layer should increase in radius, implying that $R(b) > b$, the solvent-poor layer should become thinner after deformation, so $a - b > R(a) - R(b)$, or $R(b) - b > R(a) - a$.
Therefore, we expect that ${\rm d}R/{\rm d}r$ is typically greater than 1 for $r < b$ and is typically less than 1 for $b \leq r < a$.
If the solvent-poor region is taken to be much smaller than the solvent-rich region, i.e., $f \ll 1$, then $(a-b)/b \ll 1$, and $R(r)$ adopts a piecewise-linear functional form,
\begin{equation}
R(r) \approx \left\{\begin{array}{cc}
\Lambda_1 r & r < b \\
\Lambda_1 b + \Lambda_2 (r - b) & b \leq r < a
\end{array}\right.
\end{equation}
where $\Lambda_1 > 1$, reflecting the further swelling of the swollen interior, and $0 < \Lambda_2 < 1$, reflecting the deswelling of the shell region.
The deformation matrix in the solvent rich region is given by $\Lambda_{ij} \approx \Lambda_1 \delta_{ij}$; since ${\rm det}\,\Lambda = \phi_0/\phi$, we find that $\Lambda_1 = (\phi_0/\phi_r)^{1/3}$.
In the solvent-poor region, we find that the deformation matrix is both inhomogeneous and anisotropic, with $\Lambda_{Rr} \approx \Lambda_2$ and $\Lambda_{\Theta\theta} = \Lambda_{\Phi\phi} \approx \Lambda_1$, with $r$-dependence appearing at higher order in $(a-b)/b$.
Therefore, we have that $\Lambda_1^2 \Lambda_2 = \phi_0/\phi_p$, giving the result $\Lambda_2 = (\phi_0\phi_r^2)^{1/3}/\phi_p$.
The total deformation free energy for the sphere $\Delta F_{\rm sphere}$, consisting of contributions from the swollen core and deswollen shell, is given by
\begin{equation}\label{eq:sphere_coex}\eqalign{
\frac{\Delta F_{\rm sphere}}{V} \approx &f\left[\frac{1}{2}\mu_0\phi_0^{2/3}\left(\frac{\phi_r^{4/3}}{\phi_p^2} + \frac{2}{\phi_r^{2/3}}\right) + \Delta\hat{\mathcal{F}}(\phi_p)\right] \\
&\mkern-40mu+(1 - f)\left[\frac{3}{2}\mu_0\frac{\phi_0^{2/3}}{\phi_r^{2/3}} + \Delta\hat{\mathcal{F}}(\phi_r)\right] \\
&\mkern-40mu+ p\left[f\left(\frac{\phi_0}{\phi_p} - 1\right) + (1 - f)\left(\frac{\phi_0}{\phi_r} - 1\right)\right] \, .
}\end{equation}
Minimizing $F_{\rm sphere}$ with respect to $\phi_r$, $\phi_p$, $p$, and $f$ yields three equilibrium equations, namely
\numparts\begin{eqnarray}
-\frac{\mu_0 \phi_r^{4/3}}{\phi_0^{1/3}\phi_p} + \frac{\phi_p^2}{\phi_0}\frac{\partial \Delta \hat{\mathcal{F}}}{\partial\phi}\bigg|_{\phi_p} \mkern+30mu= p \, , \label{eq:sphere_tan_a} \\
-\frac{\mu_0\phi_r^{1/3}}{\phi_0^{1/3}}\left[1 - \frac{2f}{3(1 - f)}\left(\frac{\phi_r^2}{\phi_p^2} - 1\right)\right] \nonumber\\
\mkern+110mu + \frac{\phi_r^2}{\phi_0}\frac{\partial \Delta \hat{\mathcal{F}}}{\partial\phi}\bigg|_{\phi_r} = p \, , \label{eq:sphere_tan_b} \\
\frac{\mu_0 \phi_0^{2/3}}{2\phi_r^{2/3}}\left[\frac{\phi_r^2}{\phi_p^2} - 1\right] + \Delta\hat{\mathcal{F}}(\phi_p) - \Delta\hat{\mathcal{F}}(\phi_r) \nonumber \\
\mkern+220mu = - p\left[\frac{\phi_0}{\phi_p} - \frac{\phi_0}{\phi_r}\right] \, ,\label{eq:sphere_tan_c}
\end{eqnarray}\endnumparts
as well as the lever rule.
However, the result cannot be cast as a common tangent construction, as evidenced by the dependence of $p$ on $f$ in equation (\ref{eq:sphere_tan_b}).
Ultimately, this results from the anisotropic stress of the solvent-poor region, which comes about through the \emph{coherency strain}: as the polymer network remains contiguous, the components of the deformation matrix that are tangent to the phase interface, namely $\Lambda_{\Theta\theta}$ and $\Lambda_{\Phi\phi}$, must be continuous through the interface.
Therefore, the deformation matrix for the solvent-poor region shares two of its three components with the solvent-rich region; the third component, $\Lambda_{Rr}$, which is normal to the phase interface, is discontinuous and therefore takes on two independent values for the two regions.
Note that this is alleviated for effectively one-dimensional gels: a gel that is constrained to undergo uniaxial deformation normal to the phase interface does not experience coherency strain at the interface and therefore the phase-coexistent equilibrium obeys the common tangent construction \cite{Doi2009}.

For a concrete solution, we use the Flory-Rehner model to provide a concrete form for $\hat{\mathcal{F}}$ and minimize the free energy in equation (\ref{eq:sphere_coex}) numerically.
As shown in figure \ref{fig:sphere_equilibrium}(b), coexistence between separate swollen and deswollen phases happens above certain transition values of $\chi_1$, depending on values of $\phi_0$.
For values of $\chi_1$ greater than the transition points, the equilibrium volume fraction bifurcates into a high-$\phi$ branch, corresponding to solvent-poor gel of volume fraction $\phi_p$, as well as a low-$\phi$ branch, corresponding to solvent-rich gel of volume fraction $\phi_r$.
Whereas there is a \emph{discontinuous} jump from the initial volume fraction $\phi_0$ to solvent-poor gel $\phi_p$, the solvent-rich volume faction decreases \emph{continuously} from $\phi_0$ above the transition.
The reason for this is attributed to the manner at which phase-separation occurs, namely a form of heterogeneous nucleation.
Since the solvent-poor gel grows from the boundary of the gel, inward, the fraction $f$ of solvent-poor gel grows continuously, as shown in figure \ref{fig:sphere_equilibrium}(c).
As the growth of solvent-poor gel comes at the cost of diluting the solvent-rich region, the continuous growth of the solvent-poor shell results in continuous solvent addition to the core-region; therefore $\phi_r$ decreases continuously as $f$ and $\phi_p$ grow.
This continuous growth of the solvent-poor region is consistent with predictions seen elsewhere, in the context of solvent-poor phase growing around a hole \cite{Kuroki1994}.
Note that the transition value of $\chi_1$ shifts to higher values as $\phi_0$ is set to higher values.
This is because for hight values of $\phi_0$, there is less interaction between solvent and polymer and therefore the total energetic cost of maintaining the homogeneous phase is less; this energetic cost increases for lower values of $\phi_0$.


\section{\label{sec:shape_change} Large shape change via thermodynamics} 


The characteristic swelling behavior of polymer gels makes them practical for many applications.
Their ability to absorb and retain solvent, enables them to remove unwanted liquids, such as water, making them attractive for cleaning applications \cite{Yu2016}.
At the same time, the ability to expel a liquid given certain stimuli has led to drug-delivery applications \cite{Peppas1997,Langer2004,Hoare2008}.
Furthermore, as we have outlined, the equilibrium swelling thermodynamics of polymer gels is well modeled by the Flory-Rehner model, given certain fitting parameters.
However, as we have discussed in the previous section, there are interesting consequences that emerge when a polymer gel is brought to a state in which having a single, homogeneous polymer volume fraction $\phi$ corresponds to a thermodynamically unstable situation, leading to a phase-coexistent equilibrium.
Moreover, due to the mechanism of skin formation after a rapid quench from the swollen phase to the deswollen phase, this situation is attainable if the volume phase transition is approached rapidly.
Thus, while the phase-coexistent state remains relatively unexplored, understanding it is, potentially, of considerable practical importance as it dramatically alters the expected behavior of a gel.
While in many cases, the deswelling-arrested, phase-coexistent state of a polymer gel may be an unwanted, troublesome feature of the polymer gel thermodynamics that must be avoided, it is possible that phase coexistence and the accompanying shape change can be useful, and seen as a potential design feature.
This is not without precedent, as exemplified by \emph{extreme mechanics}, in which \emph{mechanical instability}, traditionally regarded as a nuisance to be avoided, is harnessed in the design of novel material properties and responses to achieve shape change that would otherwise be unattainable.

\subsection{Preliminaries: Extreme mechanics}


Unlike many other rigid materials, the immutable network topology of polymer gels and other elastomeric materials enable them to undergo large elastic deformations.
As swollen polymer gels are primarily composed of solvent by weight, they are very soft materials that are comparable to soft biological systems.
Indeed, changes in the swelling state of a polymer gel can be used to mimic the \emph{growth} of soft tissues.
This combination of elasticity and swelling enables experiments on growth-induced instabilities as seen in nature.
In particular, swelling of polymer gels or other materials that are tethered to a rigid substrate exhibit a wide range of surface patterns, ranging from wrinkles, resembling fingerprints, to folds and creases, much like the sulci of brain tissue \cite{Trujillo2008,Breid2011,Hohlfeld2011,Kang2010,Tanaka1987mechanical,Sekimoto1987,Fogle2013}.
Interestingly, the appearance of these patterns are seemingly chosen at random, given the underlying symmetry of the substrate.
Such \emph{spontaneous symmetry breaking} of gels and other elastic materials under some sort of mechanical constraint is typically the result of mechanical instability \cite{Mora2006,Dervaux2012}.
Furthermore, the result of such pattern formation often yields new, and perhaps \emph{a priori} unexpected mechanical response \cite{Shim2012,Lazarus2015,Hillel2017,Rafsanjani2017,Holmes2019}.

\begin{figure}
\centering
\includegraphics[width=7.5cm]{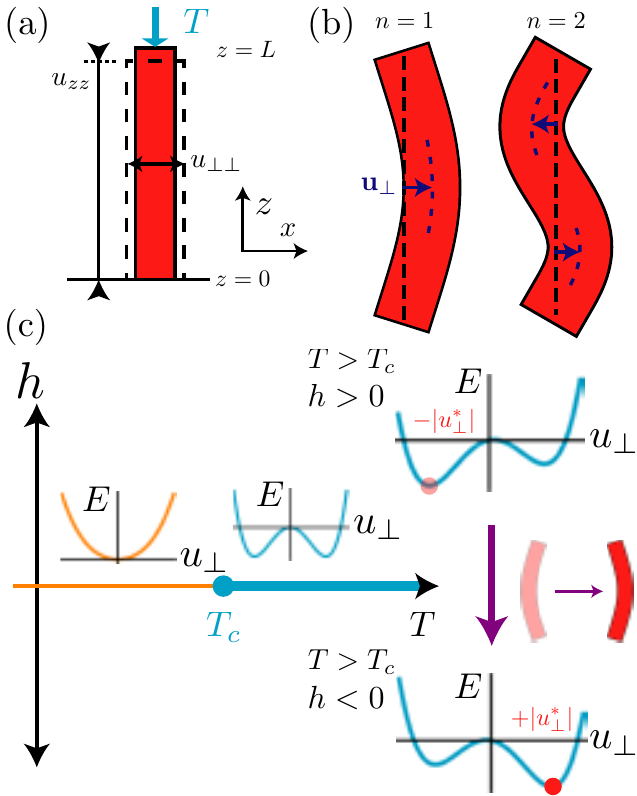}
\caption{\label{fig:euler}(a) An elastic rod of length $L$ is compressed at its end by a force $T$, resulting in a compression of its length by $u_{zz}$ and a dilation of its length by $u_{\perp\perp} = -\nu u_{zz}$. (b) Transverse deflections of the centerline of the rod by $\mathbf{u}_{\perp}$ describe bending deformations, the result of buckling. (c) Left: Portion of the ``phase diagram'' of the compressed rod about the critical force $T_c$ needed for buckling, along with depictions of the Landau energy at $h = 0$. Right: Depiction of the energy $E_{\rm Landau}$ as a function of transverse force $h$ for $T>T_c$, where the rod exhibits a discontinuous ``snap-through'' between bending directions.}
\end{figure}

Perhaps the most famous example of mechanical instability, due to its analytical tractability and appearance in structural engineering and nature, is \emph{Euler buckling}.
For concreteness, we will consider a three-dimensional material that is described by an isotropic linear elastic energy
\begin{equation}
E_{\rm el} = \frac{E}{2(1 + \nu)}\int {\rm d}^3 r\left[u_{ij}u_{ij} + \frac{\nu}{1 - 2\nu}u_{kk}^2\right] \, ,
\end{equation}
where $E$ is the Young's modulus, $\nu$ is the Poisson ratio, and $u_{ij} \equiv (\partial_i u_j + \partial_j u_i + \partial_i \mathbf{u}\cdot \partial_j \mathbf{u})/2$ is the finite symmetric strain tensor, corresponding to the displacement field $\mathbf{u}(\mathbf{r})$.
The nonlinear terms in the strain tensor, which are usually neglected, are retained here in order to properly compute the second variation of the elastic energy when determining stability.
If this material is formed into a slender column of length $L$ and circular cross-section radius $a \ll L$ that is under a compressive stress $\sigma_{zz} = T/(\pi a^2)$, as illustrated in figure \ref{fig:euler}(a), then the total energy is given by
\begin{equation}
E = E_{\rm el} + T \Delta L \, ,
\end{equation}
where $\Delta L$ is the a change in length.
Assuming that the rod is symmetric about the $z$-axis, the deformation will also be axisymmetric so that the strain will have longitudinal $u_{zz}$ and transverse $u_{\perp\perp}$ parts that, in the slender rod approximation, can be approximated as constants, i.e., $u_{zz} = \Delta L/L$ and $u_{\perp\perp} = \Delta a/a$.
The total energy can therefore be written as
\begin{equation}\eqalign{
E \approx \frac{E \pi a^2 L}{2(1 + \nu)(1 - 2\nu)}\big[&2 u_{\perp\perp}^2 + 4 \nu u_{\perp\perp}u_{zz} \\
&+ (1 - \nu)u_{zz}^2\big] + T L u_{zz} \, .
}\end{equation}
The new, stressed equilibrium of the rod is found by minimizing the total energy with respect to $u_{\perp\perp}$ and $u_{zz}$, yielding equilibrium strains
\numparts\begin{eqnarray}
u_{\perp\perp} &= -\nu u_{zz} \, ,\\
u_{zz} &= -\frac{T}{E \pi a^2} \, ,
\end{eqnarray}\endnumparts
which shows that for compression $(T>0)$, the rod decreases its length and for most elastic materials $(\nu>0)$, it increases its width proportionally.
This shortening of the length of the rod details the expected change in mechanical \emph{equilibrium} of the rod from an unstressed configuration to a stressed configuration.
However, it is not guaranteed that the stressed equilibrium is stable to deformations of the rod that break its symmetry \footnote{We will assume, however, that the unstressed equilibrium is stable. For polymer gels, as we have shown, this is a nontrivial assumption that depends on temperature and composition.}.
Indeed, it has been long known that if one overloads a column, the column may bend, as illustrated in figure \ref{fig:euler}(b).
To probe the stability of the rod with respect to deflections that bend the rod, we require a description of the elasticity of bending deformations.
It can be shown (see, e.g., \cite{LandauLifshitz1986}) that if the centerline of the rod is deflected from a straight configuration to a curved configuration, where it acquires a non-zero curvature $\kappa(z)$, the associated elastic energy cost is
\begin{equation}
E_{\rm bend} = \frac{1}{2}\int_0^L{\rm d}z\, B \kappa^2(z)  \, ,
\end{equation}
where $B$ is the bending modulus of the rod, a composite of the Young's modulus $E$ and the second moment of area of the cross-section of the rod; for rods of circular cross-section, $B = (\pi/4) E a^4$.
Stability of the straight configuration against transverse deflections $\mathbf{u}_{\perp}$ of the rod is ensured when the second variation of the total energy, $\delta^2 E$, with respect to these deflections is positive.
Expanding the energy $E$ about the stressed equilibrium, that is, where $\Delta L \approx -LT/(E\pi a^2) + \delta(\Delta L) + \delta^2(\Delta L)$, the second variation of the energy is given by
\begin{equation}
\delta^2 E = \frac{1}{2} \int_0^L {\rm d}z\, B (\delta \kappa)^2 - T \delta^2 (\Delta L) \, ,
\end{equation}
which can then be expressed in terms of the transverse deflections $\mathbf{u}_{\perp}$.
This is done by noting that the shape of the rod is given by $\bm{\gamma}(z) = z\hat{\mathbf{z}} + \mathbf{u}_{\perp}(z)$ so that the curvature change $\delta \kappa$ after deformation is given by
\begin{equation}
\delta \kappa = \delta\frac{|\partial_z\bm{\gamma}\times\partial_{zz}\bm{\gamma}|}{|\partial_z\bm{\gamma}|^{3/2}} \approx \left|\frac{{\rm d}^2 \mathbf{u}_{\perp}}{{\rm d}z^2}\right|
\end{equation}
and the second length variation $\delta^2 L$ after deformation is given by
\begin{equation}
\delta^2(\Delta L) = \delta^2\int_0^L  \!\!\!{\rm d}z (|\partial_z \bm{\gamma}|-1) \approx \frac{1}{2}\int_0^L  \!\!\!{\rm d}z\left|\frac{{\rm d}^2\mathbf{u}_{\perp}}{{\rm d}z^2}\right|^2.
\end{equation}
Therefore, the second variation of the energy is
\begin{equation}
\delta^2 E = \frac{1}{2}\int_0^L {\rm d}z\left[B \left|\frac{{\rm d}^2 \mathbf{u}_{\perp}}{{\rm d} z^2}\right|^2 - T \left|\frac{{\rm d}\mathbf{u}_{\perp}}{{\rm d}z}\right|^2 \right] \, .
\end{equation}
In general, for axisymmetric rods in three-dimensions, $\mathbf{u}_{\perp}$ can point anywhere transverse to the axis of symmetry (here, the $z$-axis) and the buckled, bent configurations spontaneously break a continuous symmetry, namely, rotations about the $z$-axis.
For simplicity, we will restrict our attention to symmetry-breaking deflections that keep the rod configuration confined to a plane, namely the $xz$-plane, so that $\mathbf{u}_{\perp} = u_{\perp}\mathbf{\hat{x}}$.
Now, if we assume that the rod is fixed at its two ends such that $u_{\perp}(0) = u_{\perp}(L) = 0$ then we can expand the deflection field as a sum of sine functions, namely
\begin{equation}
u_{\perp} = \sum_{n > 0} u_{\perp}^{(n)} \sin \frac{n \pi z}{L} \, ,
\end{equation} 
and the second variation of the energy is given by
\begin{equation}
\delta^2 E = \sum_{n>0} \frac{(n \pi)^2}{4 L}\left[\frac{B n^2 \pi^2}{L^2} - T\right]\left(u_{\perp}^{(n)}\right)^2 \, .
\end{equation}
Note that the form of the bending energy used assumes a slowly varying curvature along the length of the rod, so that the second variation of the energy, as shown above, is only valid when the wavenumber of the bending mode, $(n \pi)/L$ is much less than $1/a$.
Therefore, for low compressive force $T$, the energy remains positive for small transverse deflections.
However, for $T > T^{(n)}_c$, where the critical compressive force is $T^{(n)}_c = B n^2 \pi^2/ L^2$, the total energy may decrease by bending, forming $n/2$ wavelengths.
Hence, for $T > T_c \equiv T^{(1)}_c$, the rod is susceptible to buckling, with the critical compressive force $T_c$ marking the onset of the instability.

\subsection{Extreme thermodynamics}


There is an analogy that can be drawn between buckled rods and thermodynamic phases \cite{Mikulinsky1995}.
In particular, \emph{the onset of mechanical instability resembles the onset of thermodynamic instability at a critical point}.
Indeed, we can even take the Landau phenomenological approach and model the equilibrium of the bent rod, restricted to the lowest wavenumber mode, here $n=1$, via a model elastic energy of the form
\begin{equation}\label{eq:euler_landau}
E_{\rm Landau} = -\frac{\pi^2(T_c - T)}{4 L}u_{\perp}^2 + \frac{s}{4}u_{\perp}^4 + h u_{\perp} \, ,
\end{equation}
where $s>0$ is a parameter that stabilizes the amplitude of the deflection and $h$ is an external force that couples to the deflection.
This simple model recovers the spontaneous symmetry breaking due to Euler buckling at $h = 0$ [see figure \ref{fig:euler}(c)] but also is able to model the action due to an external force $h$ that is applied to the center of the rod (at $z = L/2$) oriented in the $\mathbf{\hat{x}}$ direction.
Interestingly, whereas $u_{\perp} \propto h$ at $T < T_c$, the response is much more complicated at $T > T_c$, where there is a discontinuous \emph{snap-through} from one curvature to the opposite, as depicted in figure \ref{fig:euler}(c).
This snap-through appears as the mechanical analogy of a first-order phase transition.
Interestingly, whereas the rod supports transverse elastic waves in both its straight and buckled configurations, the vibrational frequency of such waves goes to zero as the buckling threshold is reached, due to loss of elastic stability.
This is an elastic analogy of the ``critical slowing down'' of the relaxation time of perturbations to thermodynamic systems near a critical point \cite{Bobnar2011,Gomez2016}.

Euler buckling and snap-throughs of a slender rod are examples of how bifurcations and limit points in the equilibrium phase diagram of an elastic body exhibit parallels with phase transitions \cite{Mikulinsky1995}.
Indeed, as pointed out in a recent review by Douglas Holmes \cite{Holmes2019}, the analogy is useful for understanding mechanical instability for a wide variety of systems.
Additionally, Landau theory has proven to be a useful tool for translating mechanical instability into the language of phase transitions, providing additional insight, for example, into spin arrangements in magnetic dots \cite{Savelev2004}.
In the case of 2D elastic sheets that undergo a wrinkling instability, a similar transfer of ideas has revealed the smectic-like behavior of the wrinkle patterns \cite{Hillel2017}.

It is thus natural to inquire as to how far these analogies between mechanics and thermodynamics may be taken.
In particular, if mechanical instability can be used to design new material response, can we then harness thermodynamic instability in a similar way?
Can a case be made for \emph{extreme thermodynamics}, in which materials, such as polymer gels, are tuned near a point of thermodynamic instability in a manner that yields new, interesting behavior?

\begin{figure*}
\centering
\includegraphics[width=.8\textwidth]{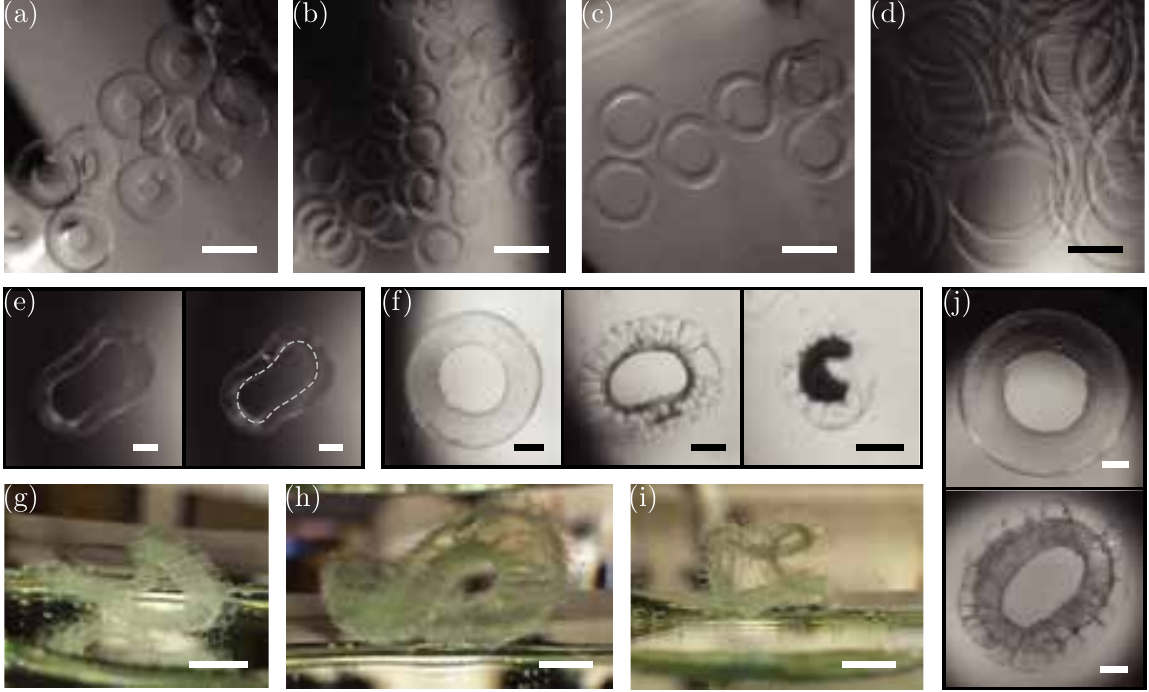}
\caption{\label{fig:bubbleTori} (a, b, c, d) Images of swollen toroidal polymer gels of various size and aspect ratio (defined as the ratio $R/a$ of the ring radius $R$ to the tube radius $a$), with increasing slenderness going from (a) to (d). Many toroidal gels of identical size are placed inside a capped vial, laid on its side on top of a circular platform used to make the tori, and imaged from below using a CCD camera. (e-j) Experimental images of pNIPAM toroidal gels after fast heating. (e) Buckling evolution of toroidal gel at 2.0 (left) and 2.5 mins (right) after heating. The dashed line in the second image outlines the position of the inner surface of the buckled tori at earlier time (2.0 mins). (f) Longer time evolution of buckling toroidal gels at (from left to right) 18 s, 3 mins, and 10 mins after heating. The inner handle portion of the ring darkens while the outer portion remains transparent, demonstrating the coexistence of solvent-rich and solvent-poor regions during deswelling. (g, h, i) Photograph of various buckled tori showing (i) out-of-plane buckling (h) ``Pringle\textsuperscript{TM}''-like morphology, and (i) folding. These photographs were taken from the side with a DSLR camera. (j) Low aspect ratio ($R/a = 2.7$, $a$ = 0.9 mm) toroidal gel immediately (top) and 12 mins (bottom) after rapid heating showing ballooning crease patterns and short-wavelength bamboo patterns (radial lines). The two distinct types of pattern reside on either half (top and bottom part, respectively) of the toroid. The scale bars in (a-d, g-i) represent 2 mm, and the ones in (e, f, j) represent 1mm. The dark areas images (a-f,j) are shadows from needles used to make the tori or circular objects placed in the light path to aid visualization of transparent gels.}
\end{figure*}

Thermodynamic instability, like mechanical instability, signals the development of multiple free energy minima, as opposed to a single, as well as the possibility of spontaneous symmetry breaking.
A hallmark of the appearance of multiple metastable equilibrium states is the ability to support coexistent phases within the same sample.
As we have discussed, phase coexistence can be achieved in polymer gels via rapid heating from the swollen phase to the deswollen phase.
In experiments on spheres, such phase coexistence leads to the formation of surface crease patterns; experiments on tori reveal additional buckling behavior of the toroidal shape, as shown in figure \ref{fig:bubbleTori}(e-i).
The source of these patterns is an internal stress distribution due to separation into coexistent swollen and deswollen phases.
Therefore, one might regard the appearance of the patterns as mechanical buckling due to internal stress generated by the allocation of solvent in the gel \cite{Chang2018}.

However, there is a feedback: the arrangement of the coexistent phases, or the distribution of the solvent within the gel, depends on the mechanical stress distribution.
This is illustrated by the example of the phase-coexistent equilibria of a sphere in section \ref{sec:sphere}, in which anisotropic deformations of the polymer network due to coherency strain between the two interfaces altered the equilibrium polymer volume fractions of each of the two phases.
As we shall demonstrate in the next section, this coherency strain has a real effect on the distribution of solvent in polymer gel tori, leading to internal stresses that cause buckling.
We can therefore conclude that such buckling is the result of phase coexistence, and thus the presence of thermodynamic instability, showing that thermodynamic instability can be harnessed to produce new behavior (buckling and pattern formation) that is distinct from the normal behavior in the thermodynamically stable regime (swelling and deswelling).

\subsection{The case of toroidal polymer gels}

\begin{figure}
\centering
\includegraphics[width=\columnwidth]{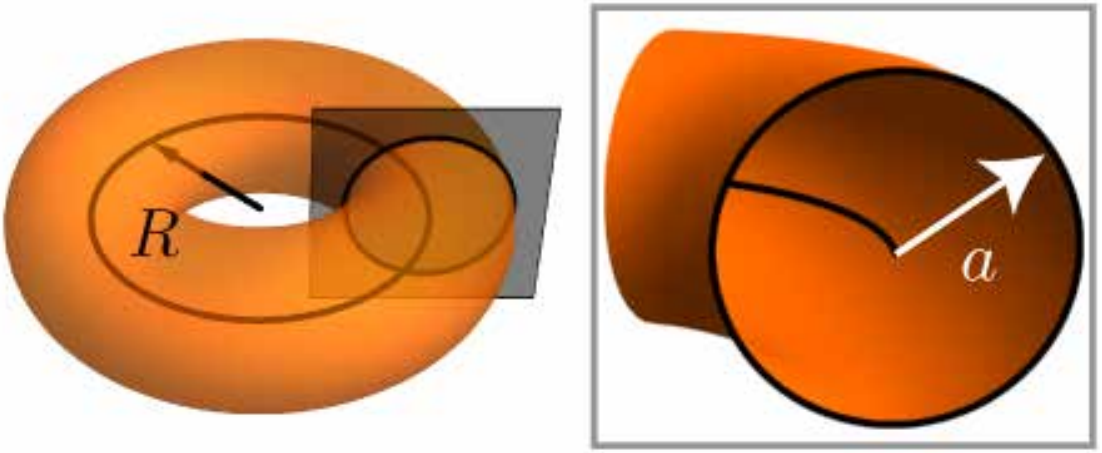}
\caption{\label{fig:toroid} Left: Depiction of the toroidal geometry considered, where $1/R$ is the ``ring curvature,'' with $R$ the radius of the ring passing through the center of the circular cross-section of the torus. Right: A slice through the circular cross-section, where $a$ is the ``tube radius,'' i.e., the radius of the boundary of the torus as measured from the center ring.}
\end{figure}


Tori are characterized by two length scales; it is convenient to use the tube radius $a$ and the ring radius $R$, as depicted in figure \ref{fig:toroid}.
In order to address phase coexistence in a toroidal sample of polymer gel as well as the buckling instability in a way that is analytically tractable, we will take the slender rod approximation $a \ll L$ that was used to study Euler buckling.
This is equivalent to working in the limit where $\kappa a \ll 1$, where $\kappa = 1/R$ is the initial, fabricated curvature of the toroidal centerline.
We can therefore approximate the toroid as a slender cylinder of radius $a$ and length $L = 2\pi R$, whose ends are identified, ensuring that it still has the topology of a solid torus.
Curvature is then incorporated as a perturbative correction of size $\kappa a$ to this \emph{flat limit} of the torus.

\begin{figure}
\centering
\includegraphics[width=8.3cm]{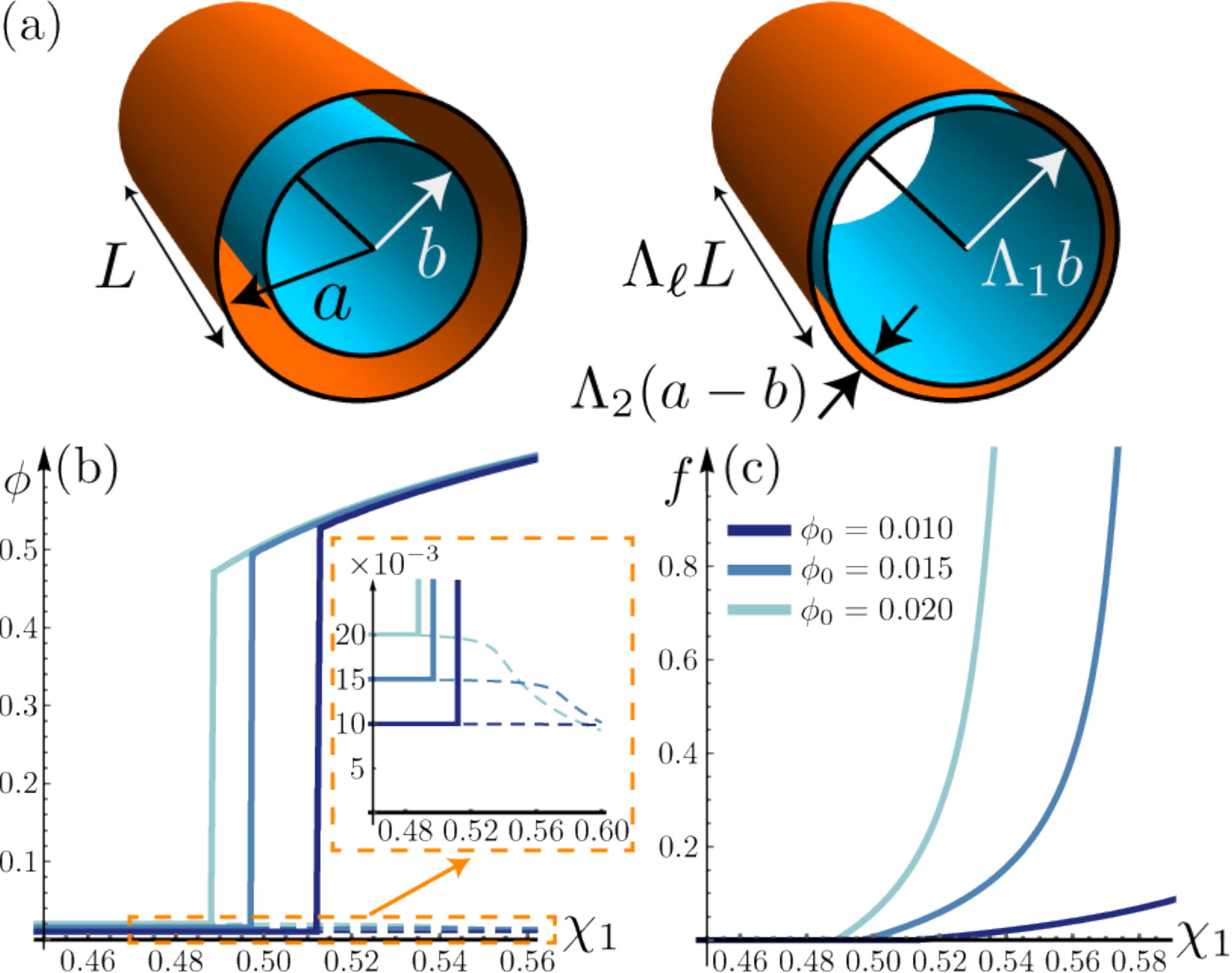}
\caption{\label{fig:flat_torus} Phase coexistent equilibrium of the flat toroid, as depicted by a cylinder of length $L$ with periodic boundary conditions at $z = 0,L$. (a) Left: Reference configuration, where $a$ is the radius of the outer boundary of the gel and $b$ is the interface boundary; Right: Target configuration, where the length dilation is $\Lambda_{\ell}$, $\Lambda_1$ is the isotropic dilation of the solvent-rich core, and the solvent-poor shell shrinks in thickness by $\Lambda_2$. (b) Equilibrium volume fraction, as a function of $\chi_1$, as predicted by the Flory-Rehner model; after the transition, the solvent-poor volume fraction $\phi_p$ is shown as the solid curve whereas the solvent-rich volume fraction $\phi_r$ is dashed, as highlighted in the inset. (c) Mass fraction $f$ of the gel corresponding to solvent-poor gel as a function of $\chi_1$.}
\end{figure}

Under similar conditions to the sphere, phase-coexistent equilibria of the flat limit of the torus can be treated in a similar manner.
We utilize the same approximations as those used in section \ref{sec:sphere}.
Proceeding in cylindrical coordinates, the reference configuration is parametrized by $(\rho,\theta,z)$ and the phase-separated target configuration by $(P,\Theta,Z)$.
In the limit in which the deswollen shell is thin, we can approximate the radial coordinate $P$ as a piecewise, linear function of $\rho$, such that $P(\rho) \approx \Lambda_1 \rho$ for $\rho < b$ and $P(\rho) \approx \Lambda_1 b + \Lambda_2 (\rho - b)$ for $b \leq \rho < a$, where $\Lambda_1$ and $\Lambda_2$ represent re-scaling of points transverse to the $z$-axis.
Next, translational symmetry along the $z$-axis ensures that the deformation matrix is not a function of $z$.
Since points $z = 0$ and $z = L$ are identified under the periodic boundary conditions of the torus, translational symmetry requires that $Z = \Lambda_{\ell} z$, where $\Lambda_{\ell}$ describes the longitudinal stretch or compression of the torus.
Therefore, in the solvent-rich core, the polymer volume fraction $\phi_r = \phi_0/(\Lambda_1^2\Lambda_{\ell})$ and in the solvent-poor shell, $\phi_p = \phi_0/(\Lambda_1\Lambda_2\Lambda_{\ell})$.
Accordingly, the total free energy $F^0_{\rm torus}$ of the phase-separated, unbent torus is given by
\begin{equation}\label{eq:torus_coex}\eqalign{
\frac{F^0_{\rm torus}}{V} \approx &f\,\left[\frac{1}{2} \,\mu_0\, \left(\Lambda_{\ell}^2 + \phi_0\,\frac{1 + (\phi_r/\phi_p)^2}{\phi_r\,\Lambda_{\ell}}\right) + \hat{\mathcal{F}}(\phi_p)\right] \\
&\mkern-64mu+(1 - f)\left[\frac{1}{2}\, \mu_0\, \left(\Lambda_{\ell}^2 + 2\,\frac{\phi_0}{\phi_r\, \Lambda_{\ell}}\right) + \hat{\mathcal{F}}(\phi_r)\right] \\
&\mkern-18mu+ p\left[f\left(\frac{\phi_0}{\phi_p} - 1\right) + (1 - f)\left(\frac{\phi_0}{\phi_r} - 1\right)\right] \, .
}\end{equation}
Values of $\phi_p$, $\phi_r$, and $f$ that minimize this free energy are plotted in figure \ref{fig:flat_torus}, assuming the Flory-Rehner model.
Comparing the results with figure \ref{fig:sphere_equilibrium}(b,c), notice there are some marked differences between the variation in $\phi_r$ and $f$ with $\chi_1$ between a sphere and a torus.
For example, whereas $\phi_r$ for the sphere steadily decreases with $\chi_1$, there is not much change in $\phi_r$ for the flat torus until larger values of $\chi_1$.
This is due to the additional length-change degree of freedom of the flat torus: for sufficiently small values of $\chi_1$, the deformation is mainly of the cross-section, so that $\Lambda_{\ell} \approx 1$, and involves a large shrinking of the thin shell with a modest stretch of the core, resulting in $\phi_p \sim \Lambda_2^{-1}$ and $\phi_r$ roughly constant; for larger values of $\chi_1$, deformations of the length become more important as the shell radius change becomes costly, so that $\phi_p \sim \phi_r \sim \Lambda_{\ell}^{-1}$, shown in the inset of figure \ref{fig:flat_torus}(b).
There is also a drastic difference in $f(\chi_1)$ for the two geometries, where the fraction of the solvent-poor phase appears to level-off with increasing $\chi_1$ for the sphere but rapidly grows for the toroid.
This is also due to the additional degree of freedom present in the toroid and there is eventually a similar level-off of $f$ for higher values of $\chi_1$ (not shown), due to the activation of the length change.

\begin{figure}
\centering
\includegraphics[width=6cm]{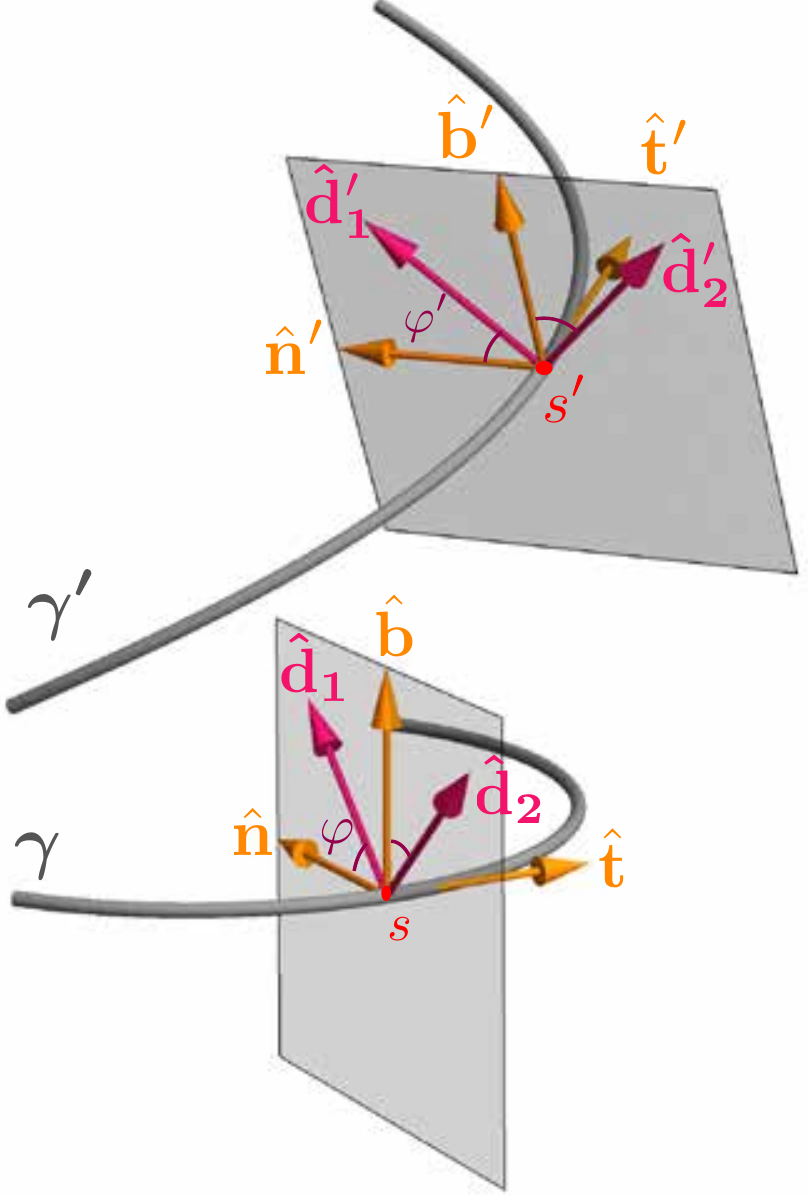}
\caption{\label{fig:centerline_frame} A centerline $\bm{\gamma}$, where the Frenet-Serret frame $\{\mathbf{\hat{t}},\mathbf{\hat{n}},\mathbf{\hat{b}}\}$ is shown at point $\bm{\gamma}(s)$ along the curve. A section of the transverse subspace, spanned by $\{\mathbf{\hat{n}},\mathbf{\hat{b}}\}$ is highlighted, along with the transverse material frame $\{\mathbf{\hat{d}_1},\mathbf{\hat{d}_2}\}$, defined by the rotation angle $\varphi$. Also shown: example deformed centerline $\bm{\gamma'}$, with associated Frenet-Serret and material frames.}
\end{figure}

Next, we consider the effects of curvature on the phase-coexistent equilibria.
The curvature of a toroidal centerline lifts the rotational symmetry of the flat, cylinder-like torus.
Consequently, we must revise the assumption that the interface between the solvent-rich core and the solvent-poor shell is axisymmetric.
Furthermore, this also breaks the assumption of axisymmetric deformation, requiring a more general representation of the deformation matrix $\Lambda$.
In order to address this asymmetry, we move from the cylindrical coordinate system to a more general one in which the straight $z$-axis is replaced with a closed parametric curve $\bm{\gamma}(s)$, where $s \in [0,L)$ is the arclength parameter of the centerline, defined such that $|\partial_s\bm{\gamma}| = 1$.
As long as the centerline is everywhere curved, we can uniquely construct the Frenet-Serret frame $\{\mathbf{\hat{t}},\mathbf{\hat{n}},\mathbf{\hat{b}}\}$, where $\mathbf{\hat{t}}(s) \equiv \partial_s \bm{\gamma}$ is the unit tangent vector, $\mathbf{\hat{n}}(s) \equiv (\partial_s \mathbf{\hat{t}})/|\partial_s \mathbf{\hat{t}}|$ is the unit normal vector, and $\mathbf{\hat{b}}(s) \equiv \mathbf{\hat{t}}(s)\times\mathbf{\hat{n}}(s)$ is the unit binormal vector, which completes an orthonormal triad at all points on the centerline; this frame is illustrated in figure \ref{fig:centerline_frame}.
The rotation rate of this frame along the centerline depends on the curvature $\kappa$ and the torsion $\tau$ of the centerline, via
\begin{equation}
\partial_s\left(\begin{array}{c}\mathbf{\hat{t}} \\ \mathbf{\hat{n}} \\ \mathbf{\hat{b}}\end{array}\right) =
\left(\begin{array}{ccc}0 & \kappa & 0 \\ -\kappa & 0 & \tau \\ 0 & -\tau & 0 \end{array}\right)\left(\begin{array}{c}\mathbf{\hat{t}} \\ \mathbf{\hat{n}} \\ \mathbf{\hat{b}}\end{array}\right) \; ,
\end{equation}
a set of geometric relations known as the Frenet-Serret equations \cite{Pressley2010}.
For the fabricated, planar torus of constant curvature $\kappa$, there is no torsion, $\tau \equiv 0$.
However, to study buckling-type deformations of the torus, where the ring can adopt non-planar deformations, we require this full geometry of curves where $\tau \neq 0$.
Furthermore, the cross-section of the gel can, in general, \emph{twist} independently of the centerline.
We therefore define a material frame $\{\mathbf{\hat{d}_1},\mathbf{\hat{d}_2},\mathbf{\hat{d}_3}\}$ that is \emph{adapted} to the centerline and can therefore be expressed in terms of the Frenet-Serret frame.
It is convenient to choose
\numparts\begin{eqnarray}
\mathbf{\hat{d}_1}(s) &= \mathbf{\hat{n}}(s)\cos\varphi(s) + \mathbf{\hat{b}}(s)\sin\varphi(s) \, , \label{eq:material_frame_a}\\
\mathbf{\hat{d}_2}(s) &= -\mathbf{\hat{n}}(s)\sin\varphi(s) + \mathbf{\hat{b}}(s)\cos\varphi(s) \, , \label{eq:material_frame_b}\\
\mathbf{\hat{d}_3}(s) &= \mathbf{\hat{t}}(s) \, , \label{eq:material_frame_c}
\end{eqnarray}\endnumparts
where $\{\mathbf{\hat{d}_1},\mathbf{\hat{d}_2}\}$ define the \emph{transverse frame} to the centerline and $\varphi(s)$ is the angle of rotation from the Frenet-Serret frame [see figure \ref{fig:centerline_frame}].
There is a more general set of equations describing the rotation rate of the material frame, namely
\begin{equation}
\partial_s \mathbf{\hat{d}_{\mu}} = \bm{\omega}(s)\times\mathbf{\hat{d}_{\mu}}(s) \, ,
\end{equation}
where $\mu \in\{1,2,3\}$ and $\bm{\omega}(s)$ is known as the \emph{Darboux vector} and is given by 
\begin{equation}
\bm{\omega} = \kappa \sin\varphi \mathbf{\hat{d}_1} + \kappa \cos\varphi \mathbf{\hat{d}_2} + (\tau + \partial_s\varphi)\mathbf{\hat{d}_3} \, .
\end{equation}
Therefore, we can represent points $\mathbf{r}$ near an arbitrarily curved centerline as
\begin{equation}
\mathbf{r}(s,x_1,x_2) = \bm{\gamma}(s) + x_1 \mathbf{\hat{d}_1}(s) + x_2\mathbf{\hat{d}_2}(s) \, ,
\end{equation}
where $x_1,x_2$ give coordinates in the cross-section.

\begin{figure}
\centering
\includegraphics[width=8.2cm]{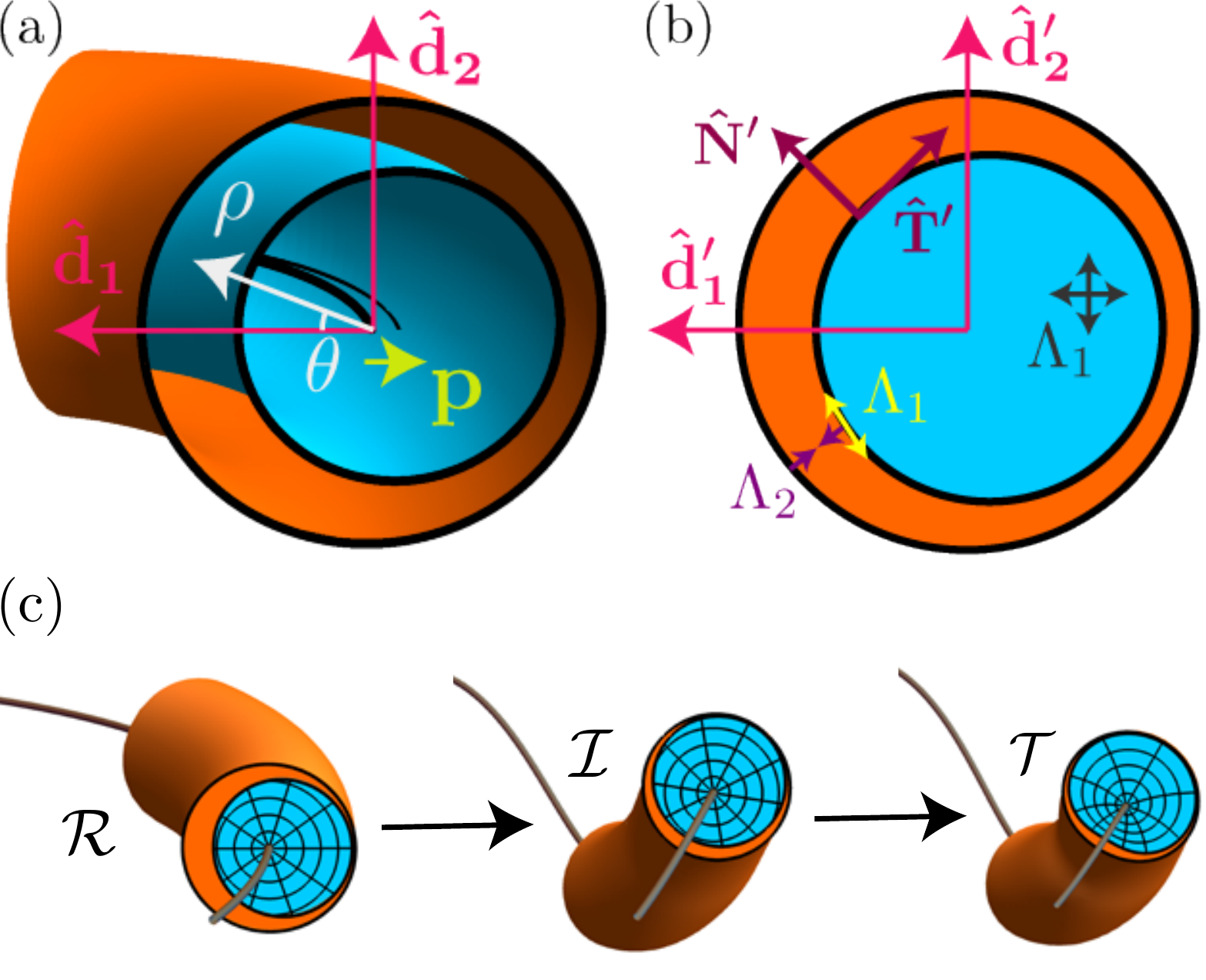}
\caption{\label{fig:deformed_toroid} (a) Slice-through of the reference configuration of the curved toroid with polarized arrangement of the coexistent phases, where $\mathbf{p}$ is the polarization vector, defining an offset between the centerline and the center of the solvent-rich region. Polar coordinates $(\rho,\theta)$ are also shown, in reference to the transverse material frame $\{\mathbf{\hat{d}_1},\mathbf{\hat{d}_2}\}$. (b) Cross-section of the target configuration, highlighting the isotropic dilation of the solvent-rich core by $\Lambda_1$, as well as the local tangent $\mathbf{\hat{T}'}$ and normal $\mathbf{\hat{N}'}$ to the phase-interface. The solvent-poor shell is stretched by $\Lambda_1$ tangentially to the interface and compressed by $\Lambda_2$ normal to the interface. (c) Schematic of the deformation process where the reference configuration $\mathcal{R}$ is deformed to the intermediate configuration $\mathcal{I}$ by a change in curve shape and macroscopic deformation of the cross-section due to phase separation. Finally, residual internal stress is relaxed by through microscopic deformations, leading to configuration $\mathcal{T}$.}
\end{figure}

We will therefore represent the reference configuration $\mathcal{R}$ of the torus with centerline $\bm{\gamma}$, frame $\{\mathbf{\hat{d}_1},\mathbf{\hat{d}_2},\mathbf{\hat{d}_3}\}$, and coordinates $(s,x_1,x_2)$.
The cross-section coordinates $(x_1,x_2)$ can be written in polar form $(\rho,\theta)$ via $x_1 = \rho \cos \theta$ and $x_2 = \rho \sin \theta$.
While $\rho = a$ is the boundary of the torus, the interface between the solvent-rich and solvent-poor regions is generally not axisymmetric but adopts a more general form $\rho = b(\theta)$ in these polar coordinates.
There are many shapes that the interface can adopt and there is evidence, namely the ballooning patterns that forms on the surface of toroidal gels [see figure \ref{fig:bubbleTori}(f-j)], that the deformation may have a relatively low wavelength along both the toroidal meridian (the $\bm{\hat{\theta}}$ direction) and the toroidal ring (the $\mathbf{\hat{d}_3}$ direction).
However, in examining the effect of curvature on the phase-coexistent equilibrium, we will focus on longest-wavelength alterations to the coexistence pattern that may arise due to broken axial symmetry.
Therefore, let the interface shape $b(\theta)$ adopt the simple form
\begin{equation}\label{eq:interface}
b(\theta) = b_0\left(1 + \mathbf{p}\cdot\bm{\hat{\rho}}(\theta)\right) \, ,
\end{equation}
which is a single-wavelength deformation of the interface in the cross-section.
As long as the amplitude $|\mathbf{p}|$ of this deformation is small, its overall effect is a simple translation of the circular interface such that it is no longer centered on the toroidal centerline as shown in figure \ref{fig:deformed_toroid}(a).
In effect, it describes a \emph{polarization} of the distribution of solvent mass within the cross-section of the toroid; we shall refer to $\mathbf{p}$ as the polarization vector.
It is possible to describe other moments of the solvent mass distribution, such as an inertia tensor, which corresponds to a two-wavelength deformation of the interface shape in the cross-section.
However, the coupling between curvature and polarization is the simplest, as dictated by symmetry; the curvature coupling to other moments occurs at higher powers of the free energy $F$ and thus may be neglected in the low-curvature regime.

In order to represent the equilibrium target configuration $\mathcal{T}$, let us return to the idea of incorporating the effect of curvature as a \emph{small} correction to the equilibrium phase coexistence in the flat torus.
To do this, we will add a step into the deformation process $\mathcal{R}\rightarrow\mathcal{T}$, and call this intermediate step $\mathcal{I}$.
In the intermediate configuration $\mathcal{I}$, the toroid consists of points $\mathbf{r'}$, given by 
\begin{equation}
\mathbf{r'}(s',x'_1,x'_2) = \bm{\gamma}'(s') + x'_1 \mathbf{\hat{d}'_1}(s') + x'_2\mathbf{\hat{d}'_2}(s')\; ,
\end{equation}
where $\bm{\gamma}'$ is the deflected centerline, e.g., due to buckling, with new frame  $\{\mathbf{\hat{d}'_1}, \mathbf{\hat{d}'_2}\}$ [see figure \ref{fig:centerline_frame}].
Furthermore, assume that in going from $\mathcal{R}\rightarrow\mathcal{I}$, the toroid undergoes phase-separation \emph{assuming the same deformation matrices as in the axisymmetric, flat torus limit}.
Therefore, in the solvent-rich region,
\begin{equation}
\frac{\partial x'_{\alpha}}{\partial x_{\beta}} = \Lambda_1 \delta_{\alpha\beta} \; {\rm for} \; \rho < b(\theta) \, ,
\end{equation}
where $x'_1 = \rho'\cos\theta'$ and $x'_2 = \rho'\sin\theta'$ yields the polar representation $(\rho',\theta')$ of the cross-section coordinates in $\mathcal{I}$.
The anisotropic deformation of the solvent-poor shell is complicated by a more general, thickness-varying shell; if the solvent-rich core maintains the above isotropic deformation then the continuity requirement across the phase interface results in a deformation $\Lambda_1$ \emph{tangential} to the interface and $\Lambda_2$ \emph{normal} to the interface in the cross-section.
In the reference configuration $\mathcal{R}$, the interface tangent is given by $\mathbf{\hat{T}} \equiv \partial_{\theta}(b\bm{\hat{\rho}})/|\partial_{\theta}(b\bm{\hat{\rho}})|$ and normal by $\mathbf{\hat{N}} \equiv  \mathbf{\hat{T}} \times \mathbf{\hat{d}_3}$; to leading order in $p = |\mathbf{p}|$, defined in equation (\ref{eq:interface}), these are given by
\numparts\begin{eqnarray}
\mathbf{\hat{T}} &= \bm{\hat{\theta}} + (\bm{\hat{\theta}}\cdot\mathbf{p})\bm{\hat{\rho}} + \mathcal{O}(p^2) \, , \\
\mathbf{\hat{N}} &= \bm{\hat{\rho}} - (\bm{\hat{\theta}}\cdot\mathbf{p})\bm{\hat{\theta}} + \mathcal{O}(p^2) \, .
\end{eqnarray}\endnumparts
The interface tangent and normal, $\mathbf{\hat{T}'}$ and $\mathbf{\hat{N}'}$, in the intermediate configuration $\mathcal{I}$ have the same form due to the isotropic deformation of the core, replacing $\bm{\hat{\rho}}\mapsto \bm{\hat{\rho}'}$ and $\bm{\hat{\theta}}\mapsto \bm{\hat{\theta}'}$.
Therefore, in the solvent-poor shell,
\begin{equation}
\frac{\partial x'_{\alpha}}{\partial x_{\beta}} \approx \Lambda_1 \hat{T}'_{\alpha}\hat{T}_{\beta} + \Lambda_2 \hat{N}'_{\alpha}\hat{N}_{\beta} \; {\rm for} \; b(\theta) \leq \rho < a \, ,
\end{equation}
as depicted in figure \ref{fig:deformed_toroid}(b).
Let $\Lambda'$ be the deformation matrix field describing position-dependent changes in length that occur in going from $\mathcal{R}\rightarrow\mathcal{I}$.
Then
\begin{equation}
\Lambda'_{ij} \equiv \frac{\partial r'_i}{\partial r_j} = (\delta_{ik} + u^{\rm ext}_{ik})\Lambda^{0}_{kj}
\end{equation}
where $\Lambda^{0}$ is the deformation matrix corresponding to axisymmetric deformation, namely
\begin{equation}
\Lambda^{0}_{ij} = \left\{\begin{array}{c}
\Lambda_1 \left(\hat{d}'_{1,i}\hat{d}_{1,j} + \hat{d}'_{2,i}\hat{d}_{2,j}\right) + \Lambda_{\ell}\hat{d}'_{3,i}\hat{d}_{3,j} \\ [5pt]
\Lambda_1 \hat{T}'_{i}\hat{T}_{j} + \Lambda_2 \hat{N}'_{i}\hat{N}_{j} + \Lambda_{\ell}\hat{d}'_{3,i}\hat{d}_{3,j}
\end{array}\right. \, ,
\end{equation}
where the top holds for points in the solvent-rich core $\rho < b(\theta)$ and the bottom holds for points in the solvent-poor shell $b(\theta) \leq \rho < a$.
The other part $(\mathbbm{1} + u^{\rm ext})$ of the deformation matrix $\Lambda'$ describes small changes in length due to deformations of the centerline and the material frame.
These small changes are encoded in an ``external strain'' $u^{\rm ext}$, which is given by
\begin{equation}\eqalign{
u^{\rm ext} = &\varepsilon_{mn3}\bigg[\bigg(\Delta \omega_m + \frac{\Lambda_1 - 1}{\Lambda_1}\omega_m\bigg)x'_n \mathbf{\hat{d}_3'}\otimes\mathbf{\hat{d}_3'}\bigg) \\
&\mkern+26mu + \Delta \omega_3x'_m \mathbf{\hat{d}_n'}\otimes\mathbf{\hat{d}_3'} \bigg]
}\end{equation}
where $\Delta \omega_{\alpha} \equiv \omega'_{\alpha} - \omega_{\alpha}$ is the change in the Darboux vector, i.e., change in the curvature, torsion, and twist of the framed curve degrees of freedom of the gel.
Here, we have taken the approximation that the arclength $s'$ after phase-separation is proportional to the arclength $s$ before phase-separation and that this proportionality is the longitudinal deformation matrix $\Lambda_{\ell}$, describing the change in length of the gel.
Even though it can be expected that curvature and torsion of the toroidal centerline results in a renormalization of $\Lambda_{\ell}$, this effect should be small since changes in the length of the rod are described by deformations of the gel that respect axial symmetry; as curvature breaks axial symmetry, we can expect such renormalization to be a higher order effect than what we seek to describe.
The external strain also describes the non-axisymmetric changes in length that occur due to axisymmetric phase-separation when curvature is present; this effect disappears when $\Lambda_1 \rightarrow 1$, i.e., when there is no transverse deformation of the gel in the solvent-rich region. 

The intermediate configuration $\mathcal{I}$ is a poor representation of the phase-separated gel as it assumes the axisymmetric solution in a geometry without axial symmetry.
To rectify this, points $\mathbf{R}$ in the actual equilibrium state, that is, in the target configuration $\mathcal{T}$ are obtained from points $\mathbf{r'}$ in $\mathcal{I}$ via an ``internal'' displacement field $\mathbf{u}$ such that
\begin{equation}
\mathbf{R}(\mathbf{r'}) = \mathbf{r'} + \mathbf{u}(\mathbf{r'}) \, .
\end{equation}
This displacement field $\mathbf{u}$ represents a ``correction'' to the shape of the equilibrium gel, resulting in a small, non-axisymmetric strain that is added to the large axisymmetric deformation.
By including the intermediate configuration $\mathcal{I}$, we are able to factor the deformation matrix $\Lambda$ into an axisymmetric part $\Lambda^{0}$ and a part that describes non-axisymmetric deformations, namely
\begin{equation}\label{eq:def_decomposition}\eqalign{
\Lambda_{ij} &= \frac{\partial R_i}{\partial r_j} = \frac{\partial R_i}{\partial r'_k}\frac{\partial r'_k}{\partial r_j} = (\delta_{ik} + \partial_k'u_i)\Lambda'_{kj} \\
&\approx (\delta_{il} + \partial_l u_i + u^{\rm ext}_{il})\Lambda^0_{lj} \, ,
}\end{equation}
where we have retained only the leading order non-axisymmetric terms, $\partial_i u_j$ and $u^{\rm ext}_{il}$.
Whereas the external strain encodes the shape of the centerline of the toroid, the remaining strain $\partial_l u_i$ describes degrees of freedom that we can regard as ``internal'' to the gel and that are allowed to equilibrate given a certain centerline shape.
Thus, we call $\partial_i u_j \equiv u^{\rm int}_{ij}$ the ``internal strain.''
Finally, it is useful to fold the external and internal strain into a total strain $u \equiv u^{\rm ext} + u^{\rm int}$.

The total free energy $F$ of the gel is then given by
\begin{equation}\eqalign{
F &=  F_p + F_r \\
&+ p\left[\int_{p}\frac{{\rm d}^3 r}{V}\,{\rm det}\,\Lambda  + \int_{r}\frac{{\rm d}^3 r}{V}\,{\rm det}\,\Lambda - 1\right] \, ,
}\end{equation}
where $F_p$ and $F_r$ are the free-energies of the solvent-poor and solvent-rich region, respectively, given by
\begin{equation}
F_{p/r} = \int_{p/r} {\rm d}^3 r \left[\frac{1}{2}\mu_0{\rm tr}\,\Lambda^T\Lambda + \hat{\mathcal{F}}(\phi)\right] \, ,
\end{equation}
with $\int_p$ and $\int_r$ representing integration over solvent-poor and solvent-rich regions of the gel in the reference configuration $\mathcal{R}$.
The third term enforces the volume constraint, with $p$ the Lagrange multiplier.
Thus, we can compute the total free energy $F$, given the deformation matrix expressed in equation (\ref{eq:def_decomposition}) and using the relation $\phi = \phi_0/({\rm det}\, \Lambda)$.
However, the amount of solvent in each region of the gel, described by the polymer volume fraction $\phi(\mathbf{r})$, should be largely unaffected by the presence of a small, non-axisymmetric strain $u$; only the shape of the regions is affected by such strain.
This is a common assumption adopted in theories of elasticity: the ``elastic'' degrees of freedom are independent of any order parameter describing the phase of the material.
For example, in the elastic theory of nematic-phase liquid crystals, gradients in the director field do not greatly affect how ``nematic'' the material is---such elastic terms are considered \emph{transverse} to order parameter (see e.g., \cite{ChaikinLubensky}).
This assumption holds as long as the thermodynamic phase of the material is well-defined, failing near the critical point.
Similarly, even though the gel in study is in the coexistence region, it consists of two well-defined phases, assuming that it is far enough from the critical point.
Therefore, the polymer volume fraction $\phi$, which describes the phase of the gel, has two discrete values, $\phi_p$ and $\phi_r$.
Referring to the decomposition of the deformation matrix (\ref{eq:def_decomposition}), we therefore require that the strain $u$ satisfies the \emph{incompressibility constraint}, ${\rm det}(\mathbbm{1} + u) = 1$, everywhere within the gel.
If this holds, then the volume constraint is maintained independently of $u$.
Importantly, while the non-axisymmetric strain $u$ does not affect the volume fraction $\phi$ of the two phases, it does affect the spatial distribution of the two coexistent phases, which is described by the interface shape $b(\theta)$.
Any interface shape $b(\theta)$ that does not preserve the continuous rotational symmetry of the gel about its centerline axis results in a non-axisymmetric coherency strain, which is encoded in $u$.
We seek to determine the change $\Delta F = F - F^0_{\rm torus}$ in free energy due to changing the arrangement of phases in the presence of curvature.

To proceed, we furthermore require that only the symmetric part $\epsilon \equiv (u + u^T)/2$ of the strain $u$ appears in the free energy in order to properly describe the cost of elastic deformations.
Furthermore, in order to enforce the incompressibility constraint ${\rm det}(\mathbbm{1} + u) = 1$, the determinant can be expanded using the relation ${\rm det}\, A = {\rm exp}({\rm ln}\,A)$ for any matrix $A$, resulting in the condition
\begin{equation}
{\rm tr}\, \epsilon = \frac{1}{2}{\rm tr}(\epsilon^2) + \mathcal{O}(\epsilon^3) \, .
\end{equation}
As a result, the free energy change $\Delta F$ due to lifting the axial symmetry is given by
\begin{equation}\eqalign{
\Delta F &\approx \int_p {\rm d}^3 r\,\bigg[\mu_0(\Lambda_{\ell}^2 - \Lambda_2^2)\epsilon_{33} + \mu_0(\Lambda_1^2 - \Lambda_2^2)\epsilon_{TT}\\
&\mkern+80mu + \frac{1}{2}c_{\alpha\beta\gamma\delta}^p \epsilon_{\alpha\beta}\epsilon_{\gamma\delta}\bigg] \\
&+\int_r {\rm d}^3 r\, \bigg[\mu_0(\Lambda_{\ell}^2 - \Lambda_1^2)\epsilon_{33} + \frac{1}{2}c^r_{\alpha\beta\gamma\delta}\epsilon_{\alpha\beta}\epsilon_{\gamma\delta}\bigg] \, ,
}\end{equation}
where $c^p$ and $c^r$ are anisotropic elasticity tensors that characterize the linear elasticity of the phase-separated ``flat'' toroid (see \cite{dimitriyev} for details).
Integrals over the solvent-poor and solvent-rich regions are given by
\numparts\begin{eqnarray}
\int_p {\rm d}^3 r &\approx \int_0^L {\rm d}s \int_0^{2\pi} {\rm d}\theta\, b(\theta)(a - b(\theta)) \, ,\\
\int_r {\rm d}^3 r &\approx \int_0^L {\rm d}s \int_0^{2\pi} {\rm d}\theta\, \int_0^{b(\theta)} {\rm d}\rho \, \rho \, ,
\end{eqnarray}\endnumparts
where the portion of the Jacobian $|\partial \mathbf{r}/\partial(s,\rho,\theta)|$ that depends on curvature contributes a higher-order correction to the free energy; thus these integrals are over cylindrical regions of the gel. 
Next, the internal displacement field $\mathbf{u}$ is found by finding conditions under which the free energy is an extremum, $\delta \Delta F = 0$, at fixed centerline geometry and polarization $\mathbf{p}$, i.e.,
\begin{equation}
\left(\frac{\delta \Delta F}{\delta \mathbf{u}(\mathbf{r})}\right)_{\{\Delta\omega_{\mu},\omega_{\mu}\},\mathbf{p}} = 0 \, ,
\end{equation}
which yields, to leading order, a displacement field that is linear in $\Delta \omega$, $\omega$, and $b(\theta)$.
The result is an effective curve elastic energy 
\begin{equation}\label{eq:free energy-change-result}\eqalign{
\Delta F &\approx \frac{1}{2}\int_0^L \!\!\!{\rm d}s \bigg\{ B\!\!\!\sum_{m = 1,2}\!\!\!\left(\Delta \omega_m + \frac{\Lambda_1 - 1}{\Lambda_1}\omega_m\right)^2 \!\!\! \\
& \mkern-32mu +C \Delta\omega_3^2 - 2 k\!\!\! \sum_{m,n=1,2}\!\!\!\epsilon_{mn3}\left(\Delta \omega_m + \frac{\Lambda_1-1}{\Lambda_1}\omega_m\right)p_n \\
&\mkern-32mu - r p^2\bigg\} \, ,
}\end{equation}
where $B$ and $C$ are effective bending and twisting moduli, $k > 0$ is a constant that couples toroid curvature to interface shape, and $r > 0$ characterizes the free energy change due to polarization of the solvent distribution (see \cite{dimitriyev} for values of these parameters in terms of $\mu_0$, $\Lambda_1$, $\Lambda_2$, and $\Lambda_{\ell}$).
The positive value of $r$ implies that there is always a free energy decrease that can be achieved for a polarized solvent distribution.
While this is perhaps unexpected, note that a polarized solvent distribution results in the formation of a solvent-poor shell with a thinner region and a thicker region, resulting in a non-uniform stress transmitted across the phase interface.
This stress results in a non-axisymmetric deformation of the solvent-rich core that, due to coherency strain, also stretches parts of the shell, leading to a local free energy density increase, whilst compressing other parts of the shell, leading to a local free energy decrease.
However, the part of the shell that is stretched is the thinner part, whereas the part that is compressed is thicker, leading to a net free energy decrease, as compared to the free energy cost of interfacial strain with a shell of uniform thickness.
As long as the toroid maintains this solvent-poor shell, however, the magnitude $p$ of the polarization is ultimately limited: for large $p$, the thin part of the shell is greatly stretched, resulting in a large elastic penalty.
Still, the polarization, and other higher moments of the solvent distribution, represent a \emph{coarsening process} where the initial form of the phase-separation into a coexistent axisymmetric core-shell geometry evolves over time.
Within the plateau period of the equilibration dynamics, the solvent distribution evolves to minimize the elastic free energy cost due to coherency strain.

\begin{figure}
\centering
\includegraphics[width=8.3cm]{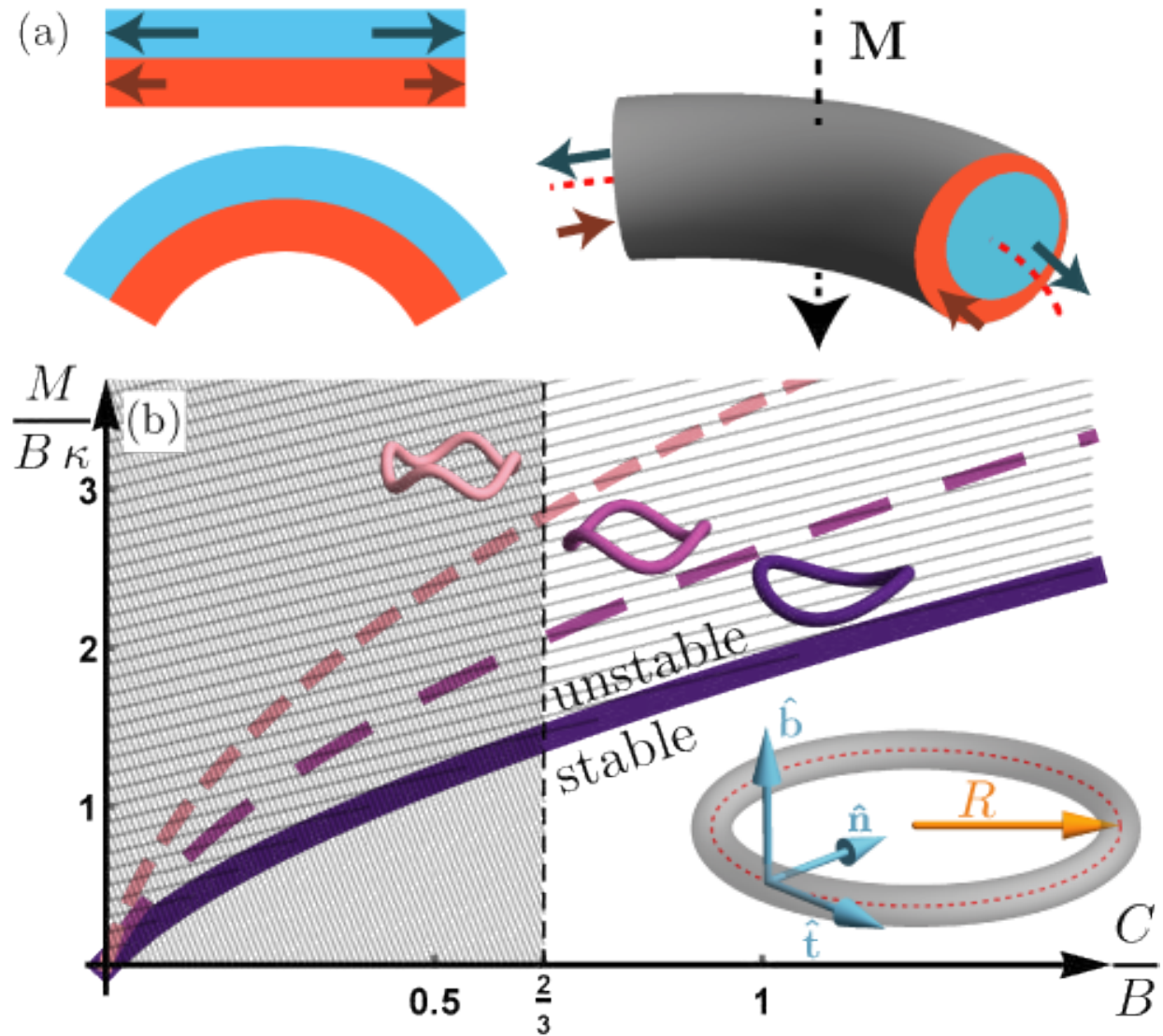}
\caption{\label{fig:buckling_torus}(a) Schematic of a bimetallic strip before (top left) and
after (bottom left) heating. A slice through the cross section of a
phase-separated toroid is shown on the right with centerline (dashed
red), polarized arrangement of solvent-rich
(blue) and solvent-poor (orange) regions, and the swelling moment $\mathbf{M}$. (b) Prediction of instability
from linear stability analysis in terms of dimensionless measures of
the swelling moment, $M/(B\kappa)$, and the ring rigidity, $C/B$. The inset
schematically shows the Frenet-Serret frame in an unperturbed ring,
as well as the “Pringling” and the next-two-lowest-order modes. Note
that for uniform incompressible tori with a circular cross section,
elasticity theory dictates that $C/B \approx 2/3$. Figure adapted from \cite{Chang2018}.}
\end{figure}

The curvature-polarization coupling term in (\ref{eq:free energy-change-result}) shows that the polarization direction is controlled by the curvature of the toroid, both the curvature ($\omega_1,\omega_2$) at which it was fabricated and the change in curvature ($\Delta\omega_1,\Delta\omega_2$) after deformation.
A toroid fabricated with initial curvature $\kappa = 1/R$ and without twist, so that the transverse material frame is defined by $\varphi = {\rm constant} \equiv 0$, has an initial Darboux vector $\bm{\omega} = \kappa \mathbf{\hat{b}}$, where $\mathbf{\hat{b}}$ is the binormal of the centerline of the torus prior to deformation.
The resulting change in free energy due to coupling between initial curvature and polarization is therefore
\begin{equation}
\Delta F = \frac{1}{2}\int_0^L{\rm d}s\,\left\{\dots + 2 k \frac{\Lambda_1 - 1}{\Lambda_1} \kappa (\mathbf{p}\cdot \mathbf{\hat{n}}) + \dots \right\} \, ,
\end{equation}
where $\mathbf{\hat{n}}$ is the normal of the centerline.
The free energy therefore decreases when $\mathbf{p}$ points along $-\mathbf{\hat{n}}$, describing a situation in which solvent mass is greater towards the outer radii of the torus and there is more polymer mass near the ``hole'' region; this is confirmed by experiment [see figure \ref{fig:bubbleTori}(f-j)].
This curvature-solvent distribution coupling has been observed in bent rubber and is due to the effect of internal stress on a material's ability to swell \cite{Nah2011}; swelling is promoted in regions under tension and impeded in regions under compression.
Furthermore, with such a polarization of the solvent distribution, the similar coupling with $\Delta \omega_m$ implies that free energy is decreased if the torus deforms such that curvature increases.
This coupling between bending deformations and the curvature of the centerline is due to an internal stress distribution, causing a torque about the centerline.
A simplified picture of the situation is shown in \ref{fig:buckling_torus}(a), where the solvent-rich region is under compressive stress due to lamination to the solvent-poor region; if axial symmetry is broken, then this stress is centered along a region that is offset from the centerline.
The effect is then similar to the classical problem of Timoshenko's heated bimetallic strip \cite{Timoshenko1925}, in which two metals with different thermal expansion coefficients are laminated together in a strip-like geometry; under heating, the constraint imposed by lamination results in a coherency strain that causes the strip to bend in the direction of the strip that expands less.
Here, the gel bends in order to compress the solvent-poor region whilst expanding the solvent-rich region, thus increasing the curvature of the centerline; this swelling version of the bimetallic strip is ubiquitous in shape-changing soft materials \cite{Reyssat2009,Douezan2011,Nah2011,Holmes2011,Gladman2016}.
We may therefore define a \emph{swelling moment} $M_m = -k\epsilon_{mn3}p_n$ that characterizes the local internal torque that is applied transverse to the centerline, bending it.
Since this moment is uniform around the central ring of the torus, it acts to uniformly increase the curvature $\kappa$.

However, a fundamental result of the differential geometry of curves (see e.g., \cite{Pressley2010}) implies that the curvature $\kappa$ of any planar closed curve, when integrated over the curve's arclength, remains constant.
Therefore, unless the magnitude and direction of the polarization changes, the only way for the toroid to deform such that the total curvature $\kappa$ increases is for the toroid to deform out-of-plane.
In order to bend out-of-plane, however, the deformation must overcome the cost of bending and twisting.
To study this buckling transition, we fix the magnitude and direction of the polarization $\mathbf{p}$ and consider only the elastic part of the free energy change, namely
\begin{equation}\eqalign{
\Delta F_{\rm el} &= \frac{1}{2}\int_0^L {\rm d}s\bigg\{B \sum_{m=1,2}\Delta \omega_m^2 + C \Delta \omega_3^2 \\
&+ 2 \sum_{m=1,2} M_m \Delta \omega_m\bigg\} \, ,
}\end{equation}
where we fix $\mathbf{p}$ along the $-\mathbf{\hat{n}}$ direction, which then fixes the swelling moment $\mathbf{M}$ along $-\mathbf{\hat{b}}$.

To determine the critical swelling moment $M_c$ needed to buckle the toroid from its planar configuration, parametrize a deformed centerline $\bm{\gamma'}$ by
\begin{equation}
\bm{\gamma'}(s) = \bm{\gamma}(s) + \zeta(s/R)\mathbf{\hat{b}} \, ,
\end{equation}
where $\bm{\gamma}(s)$ is the original centerline
\begin{equation}
\bm{\gamma}(s) = R\left(\cos(s/R),\sin(s/R),0\right) \, ,
\end{equation}
which has a corresponding Frenet-Serret basis
\numparts\begin{eqnarray}
\mathbf{\hat{t}} &= \left(-\sin(s/R), \cos(s/R), 0 \right) \, , \\
\mathbf{\hat{n}} &= -\left(\cos(s/R), \sin(s/R), 0 \right) \, , \\
\mathbf{\hat{b}} &= \left(0,0,1\right) \, ,
\end{eqnarray}\endnumparts
as pictured in the inset of \ref{fig:buckling_torus}(b).
To second order in the out-of-plane deflection $\zeta$, the Frenet-Serret basis of the deformed centerline is 
\numparts\begin{eqnarray}
\mathbf{\hat{t}'} &\approx [1 - \frac{1}{2}(\partial_s \zeta)^2]\mathbf{\hat{t}} + \partial_s\zeta\, \mathbf{\hat{b}} \, ,\\
\mathbf{\hat{n}'} &\approx - (\partial_s\zeta)(\partial_{ss}\zeta)\,\mathbf{\hat{t}} \nonumber\\
&\mkern+80mu+ [1 -\frac{1}{2}(\partial_{ss}\zeta)^2]\mathbf{\hat{n}} + \partial_{ss}\zeta\, \mathbf{\hat{b}} \, ,\\
\mathbf{\hat{b}'} &\approx -\partial_s\zeta\,\mathbf{\hat{t}} - \partial_{ss}\zeta\,\mathbf{\hat{n}} \nonumber\\
&\mkern+80mu+ [1 - \frac{1}{2}((\partial_s\zeta)^2 + (\partial_{ss}\zeta)^2)]\,\mathbf{\hat{b}} \, ,
\end{eqnarray}\endnumparts
from which we can define the deformed material frame $\{\mathbf{\hat{d}'_1},\mathbf{\hat{d}'_2}\}$ via a rotation by $\varphi'(s)$ as shown in (\ref{eq:material_frame_a}-\ref{eq:material_frame_c}).
To second order in $\zeta$ and $\varphi'$, the elastic part of the free energy change is given by
\begin{equation}\eqalign{
\Delta F_{\rm el} &\approx \frac{B}{2 R^2}\int_0^{2\pi R}\mkern-24mu {\rm d}s\, \bigg[\varphi'^{2} + \frac{C}{B}\,\left(\partial_{sss}\zeta + \partial_s \zeta + \partial_s \varphi' \right)^2\\
& - \frac{M}{B \kappa}\bigg((\partial_{ss}\zeta)^2 - 2(\partial_{s}\zeta)^2 -\varphi'^2\bigg) \bigg] \, ,
}\end{equation}
which may be simplified via the substitution\footnote{Since there is an ambiguity in how to define the material frame, the field $\varphi$ represents a gauge degree of freedom of the framed curve and this substitution is a gauge transformation.} $\varphi' = -\partial_{ss}\zeta - \partial_s\zeta + \tilde{\varphi}'$ resulting in a transformed form of the free energy change
\begin{equation}\eqalign{
\Delta F_{\rm el} &\approx \frac{B}{2 R^2}\int_0^{2\pi R}\mkern-24mu {\rm d}s\, \bigg[\left(1 + \frac{M}{B \kappa}\right)(\partial_{ss}\zeta + \partial_s\zeta - \tilde{\varphi})^2 \\
&+ \frac{C}{B}\,\left(\partial_s \tilde{\varphi} \right)^2 - \frac{M}{B \kappa}\bigg((\partial_{ss}\zeta)^2 - 2(\partial_{s}\zeta)^2\bigg) \bigg] \; .
}\end{equation}
Next, the two perturbing fields $\zeta(s)$ and $\tilde{\varphi}'(s)$ can be expanded in Fourier modes,
\numparts\begin{eqnarray}
\zeta &=  \sum_{n = -\infty}^{\infty} \hat{\zeta}_n e^{i n s/R}, \;\;\; \hat{\zeta}_{-n} = \hat{\zeta}_n^* \, , \\
\tilde{\varphi}'' &=  \sum_{n = -\infty}^{\infty} \hat{\tilde{\varphi}}'_n e^{i n s/R}, \;\;\; \hat{\tilde{\varphi}}'_{-n} = \hat{\tilde{\varphi}}_n'^* \, ,
\end{eqnarray}\endnumparts
which diagonalizes the free energy change, yielding a quadratic form
\begin{equation}
\Delta F_{\rm el} = \sum_{n=-\infty}^{\infty} (\hat{\zeta}_n\; \hat{\tilde{\varphi}}'_n)^{\dagger}\mathcal{A}_n(\hat{\zeta}_n\; \hat{\tilde{\varphi}}'_n) \, ,
\end{equation}
with
\begin{equation}
\mathcal{A}_n \! = \!\left(\!\!\begin{array}{cc}
(n^2 - 1)^2 + \frac{M}{B \kappa} & \left(1 + \frac{M}{B\kappa}\right)(n^2 - 1) \\ [5pt]
\left(1 + \frac{M}{B\kappa}\right)(n^2 - 1) & 1 + \frac{M}{B\kappa} + \frac{C}{B}n^2
\end{array}\!\!\right) .
\end{equation}
The planar torus is therefore unstable to buckling out-of-plane when the free energy change of a certain mode $n$ becomes negative.
This stability threshold occurs when ${\rm det}\,\mathcal{A}_n = 0$, i.e.,
\begin{equation}
\frac{C}{B}\bigg(\frac{M}{B\kappa} + (n^2 - 1)^2\bigg) - \frac{M}{B\kappa}\left(\frac{M}{B\kappa} + 1\right)(n^2 - 2) = 0 \; ,
\end{equation}
which results in a critical swelling moment $M^{(n)}_c$ for each mode $n$ as a function of bending modulus $B$, twisting modulus $C$, and initial curvature $\kappa$, as plotted in figure \ref{fig:buckling_torus}(b) \cite{Chang2018}.
For set values of bending and twisting moduli and initial curvature, there is a finite value of $M$, above which the planar toroid is unstable to buckling out of the plane.
The first mode that becomes unstable is the $n = 2$, corresponding to a saddle or Pringle\textsuperscript{TM}-like morphology; higher modes become unstable for larger values of $M$.
While the values of the effective bending and twisting moduli depend on factors such as the composition of the torus and the thickness of the shell, the ratio $C/B$ is well-approximated by the result for a uniform elastic rod with circular cross-section, namely $C/B \approx 1/(1 + \nu)$, where $\nu$ is the Poisson ratio of the elastic material.
Since the gel at fixed volume fraction is similar to an incompressible rubber, we take $\nu = 1/2$, yielding $C/B \approx 2/3$, which marks the lower limit of $C/B$ for rods of circular cross-section, as predicted by classical elasticity theory \cite{LandauLifshitz1986}.
Note that threshold value of $M$ for buckling decreases as the curvature $\kappa$ of the torus decreases at fixed bending modulus $B$.
This is completely analogous to the Euler buckling prediction of smaller critical compression $T_c$ for longer rods at fixed bending modulus.
Given a simple estimation of $M/(B\kappa)$, we have found that the predicted buckling threshold at $C/B \approx 2/3$ agrees with experiments, supporting the phase-separated ring model \cite{Chang2018}.

The form of the effective elastic energy $\Delta F$ given in equation (\ref{eq:free energy-change-result}) hints at a description of the phase-separated gel in terms of Landau theory, namely
\begin{equation}\eqalign{
\mathcal{L} = \frac{1}{2}\bigg[&B|\Delta \omega_m|^2 + C\Delta\omega_3^2 + C_p |\partial_s\mathbf{p}|^2 - r p^2 + \frac{u}{2}p^4 \\
&- 2 \epsilon_{mn3}(k_1\Delta \omega_m + k_2 \omega_m)p_n\bigg] \, ,
}\end{equation}
where $C_p$ and $u$ are positive coefficients that stabilize the spatial variations and magnitude of the polarization order parameter $\mathbf{p}$.
Note that in the case where the gel is constrained to lie straight, i.e.~where $\omega_m =  0$, the equilibrium configuration of the solvent polarization field $\mathbf{p}$ is ordered in a \emph{ferromagnetic} arrangement, with fixed magnitude and a spontaneously selected alignment direction transverse to the centerline of the gel.
Consequently, uniform rotations of this alignment direction do not increase the free energy and thus disturbances in the configuration of the gel, such a material inhomogeneity or an interruption in the gel's uniform shape, can easily cause long-wavelength modulation of the alignment direction.
These \emph{Nambu-Goldstone modes}, which in the context of the ferromagnetic order that we expect of the polarization field, are similar to \emph{spin waves} in ferromagnetic materials \cite{ChaikinLubensky,Kardar2007_fields}.
If the gel is then allowed to bend in response to the polarized solvent distribution, re-introducing the coupling between centerline shape and solvent distribution, the ferromagnetic order causes a uniform bending moment, resulting in a uniformly curved gel ring; again, the direction of the ring curvature is spontaneously selected, much like the magnetic field of a ferromagnetic material in the absence of an externally applied field.
As an aside, note that our discussion of toroidal gels carries through here: if the gel was initially formed in a ring shape, then the manufactured curvature acts as an applied field, aligning the solvent polarization vector in a preferred direction.
The spin-wave excitations of the polarization field have a rather interesting consequence for the shape of the gel.
A long-wavelength rotation of the polarization field causes a similarly long-wavelength rotation of the curvature direction.
This means that the shape of the centerline is no longer confined to a plane; it adopts a helical shape rather than a circular one.
In other words, the gel has a soft torsion mode so that the Nambu-Goldstone modes of this theory are perhaps better referred to as twist or torsion waves.

\begin{figure}
\centering
\includegraphics[width=8.2cm]{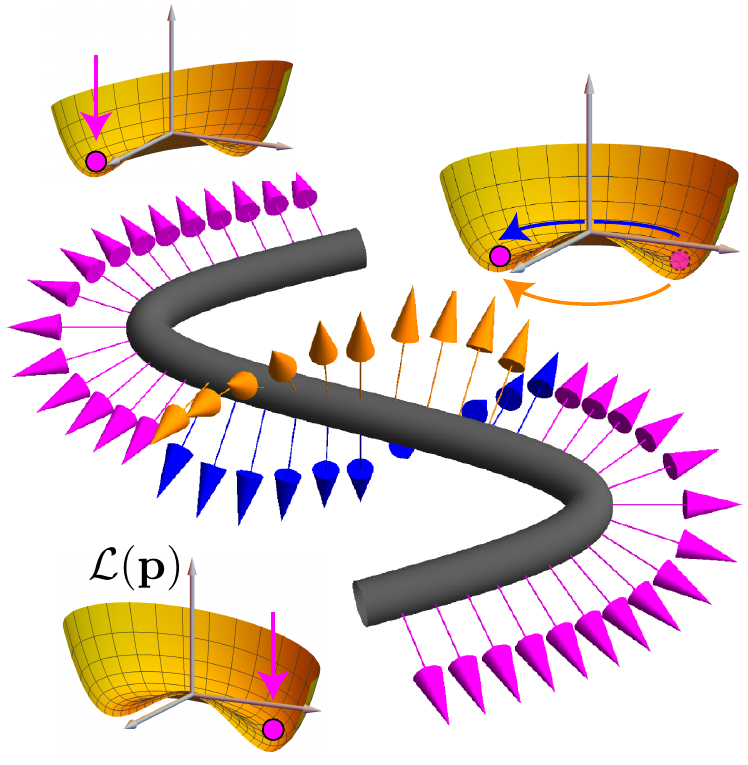}
\caption{\label{fig:s_shape} Depiction of configurations of the solvent polarization for an ``$S$''-shaped gel, as predicted by the Landau theory. Curvature acts as an external field, ``tilting'' the quartic potential that models the free energy of the polarization field $\mathbf{p}$. In the middle, where the curvature vanishes, the free energy density is rotationally symmetric, suggesting that the polarization field interpolates between its two orientations by twisting  either clockwise or counterclockwise.}
\end{figure}

Aside from a theoretical curiosity, this observation suggest some potentially interesting experiments.
Using polymer gel printing techniques \cite{Chang2018} it is possible to create gel samples with a variety of different shapes.
Continuing our analogy with ferromagnetic materials, a gel drawn in the shape of the letter ``$S$,'' as illustrated in figure \ref{fig:s_shape}, should, under rapid heating, adopt a polarization field that undergoes a reversal in direction due to the flip in the curvature direction.
Since the two arcs of the ``$S$''-shape are joined by a straight segment, our model suggests that the polarization should undergo a $\pi$ rotation, actuating an out-of-plane twist of the $S$-shape.
This is an investigation for the future.
For now, we conclude there are remarkable similarities between the solvent-stress coupling in the phase-separated polymer gel and magnetoelastic effects, which are currently being investigated for use in so-called ``shape-programmable magnetic soft matter'' \cite{Lum2016}, including recent work with flexible ferromagnetic rings \cite{Gaididei2019}.

\section{\label{sec:conclusion} Conclusions \& Outlook} 


We have discussed a small subset of the rich array of phenomena that polymer gels can exhibit.
Starting with a discussion of the derivation and assumptions that are present in the Flory-Rehner model of the equation of state for isotropic polymer gels, we have attempted to discuss some of the swelling behavior of gels in a manner that is model-agnostic, only using the Flory-Rehner model to illustrate certain predictions.
In particular, we have highlighted the swollen-deswollen phase transition, situating it within the classical theory of phase transitions of fluids.
Departing from the traditional discussion surrounding this topic, we have paid particular attention on how intuition derived from the phase behavior of fluids fails, particularly at the critical point and along the phase coexistence curve, due to shear rigidity.
In a spectacular departure from quasistatic processes, we have seen that rapid quenches across the first-order swelling transition can lead to arrested deswelling, trapping the gel in the coexistence region for a prolonged period, where it reaches an equilibrium state of coexisting phases.
We have shown that the shapes adopted by toroidal gels when forced to coexist in this manner are dramatically distinct from the shapes adopted in quasistatic processes.
By developing a description of the rapidly-heated gel that couples the spatial distribution of solvent within the gel to its elasticity, we have shown that the observed buckling arises from phase coexistence, and is thus linked to \emph{thermodynamic instability} of single-phase gel.
This demonstrates that, in the context of polymer gels, thermodynamic instability can be used to achieve a shape change that cannot be normally accessed in the thermodynamically stable regime.
There are other interesting changes in material properties that can occur due to thermodynamic instability, such as possible auxetic behavior \cite{Hirotsu1991} and microstructure formation \cite{Hirotsu1994,Onuki1999}.
We have conjectured that such instability may be used as part of material design, an idea that we describe as \emph{extreme thermodynamics}.

Recently, there has been interest in \emph{designing} equilibrium shape change in polymer materials \cite{Dias2011,Bakarich2015}.
In one approach, the swelling response of the bulk gel is tuned by spatially modulating the density of cross-links or the gel's chemistry, which impacts both the elasticity of the gel and its equilibrium volume fraction at constant temperature \cite{Mora2006,Guvendiren2009,Wu2013,Santangelo2017}.
Since different portions of the gel equilibrate to different volume fractions, the result is an internal stress distribution due to coherency strain that frustrates the original shape of the gel, leading to interesting shape change.
Contrasting this design of the equilibrium gel shape, it has been shown \cite{Pandey2013} that if the gel is taken out of equilibrium by exposing different parts to different temperatures or by only exposing part of the gel to solvent whilst keeping other parts dry, the gel undergoes a dramatic set of shape transformations.

Another approach that yields the ability to design complex geometries from initially planar gels has its roots in certain processes observed in nature \cite{Marder2003,Dervaux2009}.
It has been observed that pine cones are able to actuate shape change in response to changes in humidity, opening and closing their scales.
The reason for this is a bilayer structure built from plant tissue that yields an anisotropic response upon swelling \cite{Reyssat2009}.
Following this realization, it was found that certain seed pods \cite{Armon2011} split open from a flat state, forming two helical halves of opposite chirality, via an anisotropic shrinking process where two layers of tissue shrink in different directions, changing the intrinsic curvature of the seedpod.
This layered anisotropy has been adopted for use in a novel additive manufacturing technique \cite{Bakarich2015,Gladman2016}, where polymer gel, made anisotropic through the use of aligned cellulose fibers within the gel, is printed in layers of different swelling-direction.
When swelling is actuated by immersion in a solvent, these gel structures undergo morphological evolution that mimics natural processes, such as the opening of orchids.

We suggest that the shape changes accessed by rapidly heating gels through their phase transition may be considered in the context of these examples of designed shape change.
As we have shown, the solvent distribution within the torus that is brought to a state of phase-coexistent equilibrium is set by the curvature of the toroidal centerline.
While we have focused our attention on toroidal gels with a single curvature, the form of the solvent distribution polarization-curvature coupling in the free energy change (\ref{eq:free energy-change-result}) shows that the solvent polarization direction can be \emph{guided} by local curvature of the centerline.
We therefore conjecture that for more general polymer gel rings, where the curvature can vary continuously along the centerline, the solvent distribution will be polarized according to the local curvature direction.
The result is that after rapid heating, these rings should deform in a manner that \emph{increases} the magnitude of the local curvature. 
Moreover, in the case where the curvature direction undergoes a rapid reversal, such as in the letter ``$S$,'' we speculate that in order to interpolate between the opposite curvature directions, the polarization vector will \emph{rotate}, actuating a twist of the gel.
It is worth noting that the interesting deformations observed in experiments on tori require a simple actuation, namely rapid heating, without any prior patterning of the gel: the only feature that guides shape change is the initial shape of the gel.
Thus, a single toroid can undergo at least two very different types of shape change depending only on heating protocol, namely isotropic deswelling under slow heating and buckling under rapid heating.
This provides access to a much richer array of possible material responses that extends beyond the current regime of prescribed buckling, leading to a possibility of feedback between the material shape, its phase, and its response to applied stress.

Naturally, there is much more that can be explored regarding the physics of polymer gels.
For example, we have neglected the topic of polymer gel dynamics entirely.
Continuum hydrodynamic models have been developed, based on small deviations from equilibrium conditions for the gel \cite{Sekimoto1991,Tsutomu1995,Doi2009,Wahrmund2009,Bouklas2012,Nikolov2018}.
There have been descriptions of the coarsening of gels that have undergone spinodal decomposition \cite{Onuki1999} as well as of the development of surface patterns \cite{Suematsu1990,Maskawa1999,Boudaoud2003mechanical,Guvendiren2009}.
However, many of these models either represent important idealizations of the gels or are extremely complex, requiring considerable computational resources.
Furthermore, even robust hydrodynamic descriptions based on the Flory-Rehner model may not adequately describe the kinetics of the phase-transition, due to the limitations of the model.
In particular, the equilibration kinetics that arise in quench experiments on polymer gels represent a considerable challenge for dynamical studies, due to the multiple timescales involved.
For example, it is difficult to determine or predict the thickness of the solvent-poor skin as a function of time after the quench; this information is essential for a full understanding of the process as it determines the flow rate of solvent out of the swollen interior. 
It is not clear that the Flory-Rehner model, which provides a clear description of a homogeneous gel in or near equilibrium, is the appropriate equation of state for studying the dynamics in this regime; rather, a separate kinetic description may be necessary.
In addition, the development of patterns, such as balloon and bamboo-like structures that appeared in both cylindrical \cite{Tanaka1987mechanical,Boudaoud2003mechanical} and toroidal gels during deswelling, as shown in figure \ref{fig:bubbleTori}(j), is complicated by the necessity of a full nonlinear elasticity description of the gel that is able to incorporate large strains.

\section*{Acknowledgments}
This work was supported by the National Science Foundation (DMR-1609841).

\section*{References}
\bibliographystyle{iopart-num}
\bibliography{refs}

\end{document}